# ADVANCED MRI FOR CARDIAC ASSESSMENT IN MICE

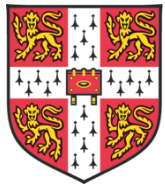

This thesis is submitted for the degree of
Doctor of Philosophy
*University of Cambridge*

Guido Buonincontri

Downing College

2013

Table of contents

All components of this thesis were carried out in accordance with the UK Animals (Scientific Procedures) Act, 1986, and with the approval of the University of Cambridge Ethical Review Panel.

This thesis is the result of my own work and includes nothing which is the outcome of work done in collaboration except where specifically indicated in the text. No part of this thesis has previously been submitted for any degree or diploma at any institution.

This thesis does not exceed 60,000 words in length (including tables, footnotes, bibliography and appendices).

# Acknowledgements

I am grateful to my supervisors Dr. Adrian Carpenter and Dr. Stephen Sawiak for their mentoring. This work owes a great deal to their commitment and dedication. I have really enjoyed the frequent discussions and the frankness with which they have shared their experience and passion for scientific research.

The biological component (surgical procedures, husbandry, etc.) of mouse experiments was performed by Dr. Thomas Krieg and Dr. Carmen Methner from the Department of Medicine, and I am grateful to both of them for their support during this thesis.

The transgenic R6/2 mice were provided by Prof. Jenny Morton and Dr. Nigel Wood from the Department of Physiology, Development and Neuroscience, I am grateful to both for their support.

Thanks to Dr. Robert Hawkes and others in the WBIC for technical advice and support.

Ultimately, thanks to my family and friends for their continuous support.

Financial support for my studentship was given by the medical research council (MRC).

# Abstract

Heart failure is a leading cause of mortality in the Western world. The mouse is a widely used model for a number of diseases, induced by genetic modification or surgical intervention. When performing experiments in mice, in vivo magnetic resonance imaging (MRI) can be used to evaluate heart anatomy and function at multiple levels.

Volumetric measurements of ventricle size at each phase of the heart cycle (obtained with cine MRI) can be used as a sensitive measure for heart failure. Cine MRI is applied here to genetic models of heart disease, including stress tests with pharmacological manipulation. As the long duration of the acquisition for functional assessment is a critical limitation, a new method to make cine MRI twelve times faster is presented and validated, maintaining standard accuracy. The method utilises a combination of compressed sensing and parallel imaging with a radial acquisition of k-space. Eddy-current induced artifacts, present when acquiring radially, are corrected retrospectively with a novel scheme.

Tissue viability can be measured with late gadolinium enhancement (LGE) imaging, where a contrast agent is utilised. After injection, the agent rapidly washes out of healthy tissue, though has a slower kinetic in infarcted regions. A novel method to perform LGE efficiently is presented and validated against histology. This method is applied to the evaluation of a new treatment for infarction.

Finally, a novel method to perform a multi-modal assessment of the mouse heart within a single exam is presented. The method includes cine and LGE MRI as well as a measure of mechanical strain in the myocardium with displacement encoding with stimulated echoes (DENSE) MRI and assessment of cellular metabolism with positron emission tomography (PET). This method is demonstrated in the evaluation of a new protective agent for infarction.

This thesis shows that MRI can significantly contribute to refinement and consequent reduction of animal experiments in experimental cardiology, offering high throughput, accuracy and versatility. Further, the techniques described in this thesis permit disease staging on an individual basis, using non-invasive technologies which are readily translatable to the clinical environment.

# List of abbreviations

AMI – Acute Myocardial Infarction
ASL – Arterial Spin Labelling
BOLD – Blood Oxygen Level Dependent
CANSEL – Cosine ANd Sine modulation to Eliminate artifacts
CHD – Coronary Heart Disease
CI – Confidence Interval
CNR – Contrast to Noise Ratio
CO – Cardiac Output
CT – Computed Tomography
CTPH – Chronic Thromboembolic Pulmonary Hypertension
DANTE – Delay Alternating with Nutation for Tailored Excitation
DENSE – Displacement ENcoded with Stimulated Echoes
DTI – Diffusion Tensor Imaging
ECG - ElectroCardioGram
EDV – End Diastolic Volume
EF – Ejection Fraction
EPI – Echo Planar Imaging
ESV – End Systolic Volume
FA – Flip Angle
FBP – Filtered Back Projection
FDG – FluoroDeoxyGlucose
FID – Free Induction Decay
FISP – Fourier Imaging with Steady state free Precession
FOV – Field Of View
GRAPPA – Generalized Autocalibrated Partially Parallel Acquisition
HARP – HARmonic Phase
HD – Huntington's Disease
HF – Heart Failure
HR – Heart Rate
IR – Inversion Recovery
I/R – Ischemia and Reperfusion
KO – Knock Out
LA – Left Atrium
LAD – Left Anterior Descending
LGE – Late Gadolinium Enhancement
LV – Left Ventricle
LVM – Left Ventricular Mass
MI – Myocardial Infarction
MRI – Magnetic Resonance Imaging
MRS – Magnetic Resonance Spectroscopy
NEX – Number of Excitations
NMR – Nuclear Magnetic Resonance
NUFFT – Non Uniform Fast Fourier Transform
PAH – Pulmonary Artery Hypertension
PET – Positron Emission Tomography

POCS – Projection Over Convex Sets
RA – Right Atrium
RF - RadioFrequency
RV – Right Ventricle
SENSE – SENSitivity Encoding
SNR – Signal to Noise Ratio
SPAMM – SPAtial Modulation of Magnetization
SPECT – Single Photon Emission Computed Tomography
SPIRiT – Iterative Self-consistent Parallel Iterative Reconstruction
SV – Stroke Volume
TE – Echo Time
TI – Inversion Time
TR – Repetition Time
TTC – Triphenyl Tetrazolium Chloride
WT – Wild Type

# Section 1: Introduction

# Chapter 1

# Introduction

Heart disease is the main cause of mortality in the United Kingdom (1). New therapies and basic mechanisms of heart disease are being actively studied, however is difficult to investigate these directly in patients. Research questions on disease and treatment are most commonly studied in animal models. Among these, mouse models have assumed a central role, including transgenic and interventional models (2). Evaluation of organ function as well as tissue viability and metabolism are required for useful translational studies, however direct assessment of the heart is difficult. Magnetic resonance imaging (MRI) is an excellent method to investigate the mouse heart *in vivo*, as it offers high anatomical accuracy and sensitivity. In this thesis methods for cardiac MRI in the mouse heart are implemented and expanded. In this chapter, the main motivations behind this work are introduced. Before outlining these, a brief introduction to the basic physiology and pathology of the heart is presented as a guide for future chapters.

## 1.1 The cardiovascular system

The cardiovascular system consists of blood, vessels and the heart, and is essential for the circulation of blood through the body in order to deliver oxygen and nutrients. Systemic circulation has the function of supplying organs with oxygenated blood, while pulmonary circulation has the function of cycling the blood through the lungs, where this is oxygenated. Part of the systemic circulation, the coronary circulation, has the role of supplying the heart with oxygenated blood (3).

Within the cardiovascular system, the heart is responsible for pumping the blood though the circulation, and can be divided in two sections: the left and the right heart. Each part is made of two contractile chambers, a ventricle and an atrium. The right ventricle (RV) pumps the blood into the pulmonary circulation, while the right atrium (RA) receives blood from the systemic circulation. The left ventricle (LV) pumps blood to the systemic circulation and the left atrium (LA) receives it from the pulmonary circulation (see Figure 2).

Sequential contraction of atria and ventricles is caused by an action potential travelling through the muscle. Electro-mechanical coupling promotes the interaction between actin and myosin filaments, causing the contraction of myocytes (3). As the electrical pulses travel through the myocardium, the generated electrical potential can be measured from the surface of the body with an electrocardiogram (ECG, see Figure 1). With an ECG the main phases of the heart cycle can be discriminated, as the measured potential is proportional to the number of depolarizing muscle cells. The signal in the ECG is dominated by the depolarization of the ventricles. Contraction of the ventricles is called systole while relaxation, when chambers are loaded again with fresh blood, is called diastole.

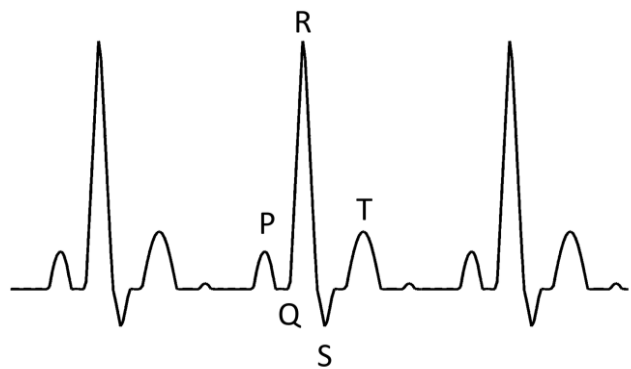

**Figure 1 Electrocardiogram.**

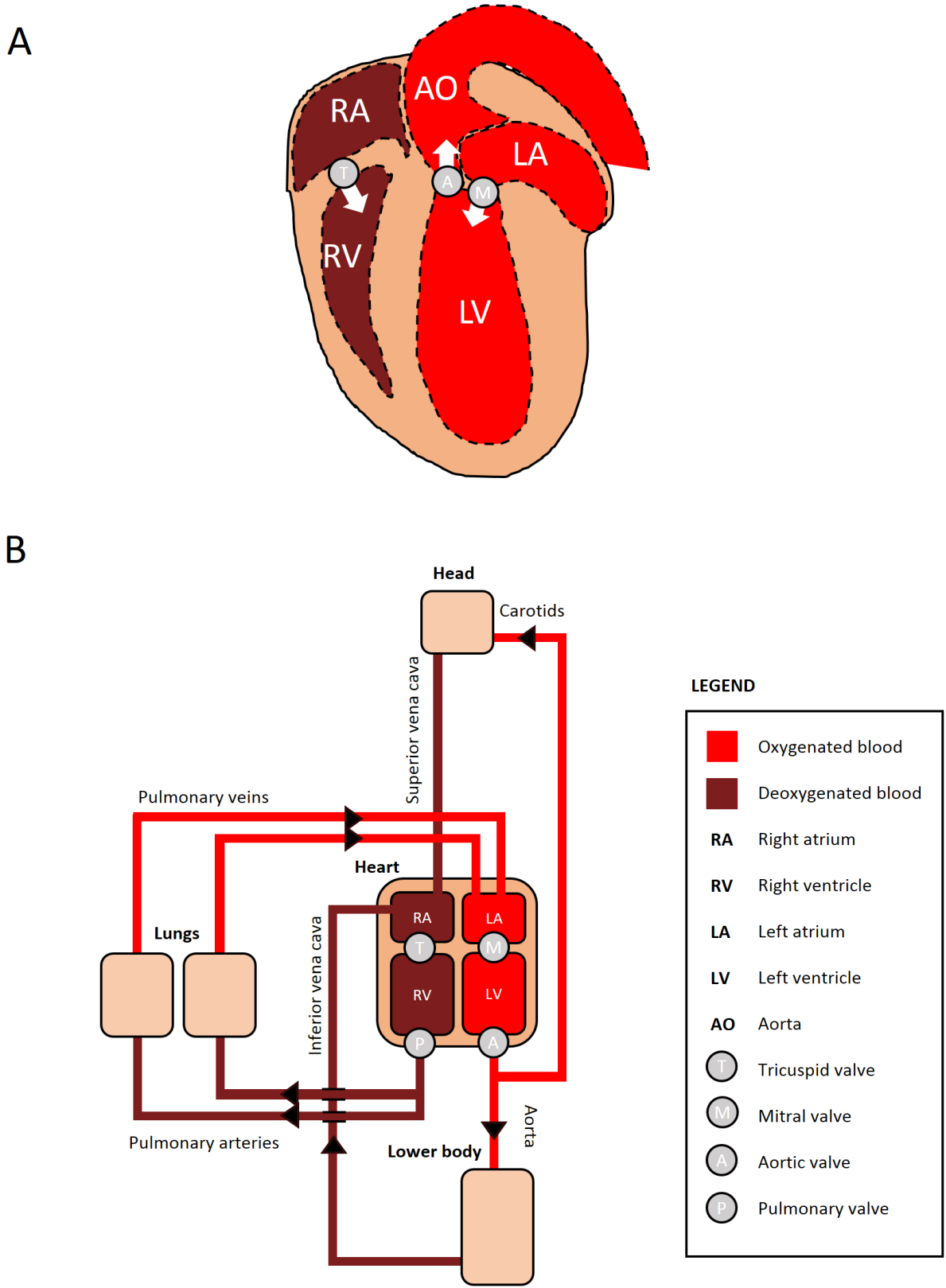

**Figure 2 Anatomy of the circulatory system. a) A four chamber view of the mouse heart. b) The circulatory system of a mouse.**

## *1.2 Heart failure*

Heart failure occurs when the heart is not capable of pumping sufficient oxygenated blood through the circulation with enough pressure. This condition leads to electrophysiological abnormalities and sudden death if no intervention is performed. In patients, heart failure is most commonly a result of coronary heart disease (CHD). In CHD, a blood clot forms in a coronary artery, occluding the vessel and depriving the heart tissue from blood. CHD is the biggest killer in the United Kingdom, causing over 82,000 deaths per year. It is estimated that one in five men and one in eight women die from the disease (1, 4). Current therapies for addressing coronary artery disease consist of reperfusion of the occluded artery during the first hours after the onset of chest pain. Interventions such as reperfusion, aiming at reducing necrosis in infarction are called cardioprotective. The utilisation of cardioprotective pharmaceuticals together with reperfusion therapy is an active area of research, and several strategies are currently under investigation in clinical trials (5).

CHD is not the only cause of heart failure. Heart failure can be also a result of high blood pressure, cardiomyopathy, genetic defects, damaged valves, as well as a complication of other diseases (6). The role of single genes and proteins in cell and organ physiology is of interest for the understanding of these diseases and to develop new treatments. However, it is difficult to perform these studies directly, and animal models are most often used to simplify complex diseases into manageable research questions.

## *1.3 Mouse models of heart disease*

Mouse models have emerged as the most utilised model for heart disease for several reasons. First, the mouse heart is relatively similar to the human heart, which makes it a sensible model when assessing disease. Taken aside obvious size differences and the anatomy of the venous pole of the heart, the two species have followed similar evolutionary paths (7). Secondly, mice are housed at a low cost compared with bigger animals, and disease course is quicker, meaning that the time-course of the experiments is accelerated and results can be obtained faster. In addition, there is a vast literature on mouse genome structure and modification. The role of different genes in heart function and disease can be studied in mice by selectively knocking out genes, or introducing genes known to generate disease in patients. A great number of relevant transgenic and knockout (KO) mice have been developed over the years for the investigation of mechanisms of heart disease as well as new treatments. Further, cardiomyocite-specific models and inducible knockouts can be used to answer specific questions in experimental cardiology (2).

In addition to genetically modified models, interventional models are utilised for the occlusion of the coronaries characterising CHD. Coronary obstruction can be surgically induced in a controlled environment to obtain acute myocardial infarction and the subsequent development of heart failure. When a transient coronary obstruction is followed by a restoration of blood flow, the model mimics a typical situation where a patient undergoes reperfusion therapy (8).

## 1.4 Assessing the mouse heart invasively

Traditionally, assessment of heart function in mice uses invasive or terminal procedures, such as *in vivo* catheterization of the heart (9) or the *ex vivo* working heart model (10). These techniques can offer assessment of heart function in controlled conditions, however they are either terminal procedures, or can induce injury to the system that they are measuring due to the severity of the procedure.

For heart tissue, mouse models are commonly investigated using histopathological techniques. Wall thickness, as well as tissue viability and cell metabolism have been traditionally studied on the laboratory bench with *ex vivo* techniques. One of the most widely used staining agents is triphenyl tetrazolium chloride (TTC), which is commonly used in post-mortem identification of myocardial infarctions. As shown in Figure 3, healthy viable heart muscle stain deep red, while infarcted areas appear brighter. Volumetric measurements can be used to derive infarct size.

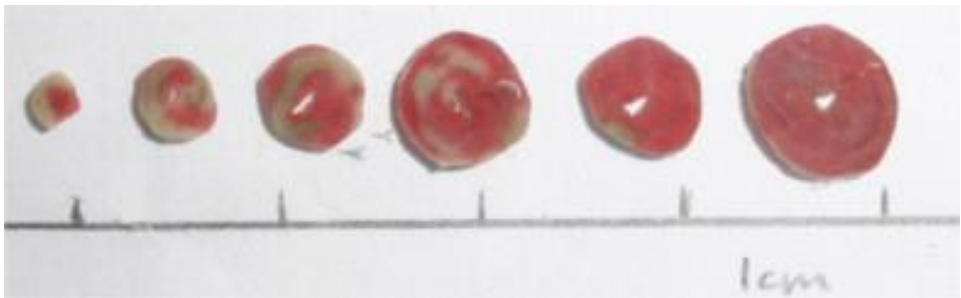

**Figure 3 Slices from TTC staining in a mouse model of acute myocardial infarction. TTC stains in the presence of dehydrogenase enzymes and the non-infarcted myocardium appears brick red. Performed by Dr Carmen Methner (Department of Medicine, Cambridge).**

The last ten years have seen a wide adoption of clinical methods adapted to the preclinical environment such as ultrasound, computed tomography, and magnetic resonance imaging (MRI).

## 1.5 Structure of this thesis

In Section 1, MRI in mouse models of heart disease is introduced in the context of experimental cardiology. Chapter 2 introduces the theoretical background of MRI, focussing on the techniques utilised within this thesis, while Chapter 3 puts the work of this thesis in the context of the relevant literature.

In Section 2, standard methods and improvements of current techniques to measure heart function with MRI are discussed. Chapter 4 describes a protocol to perform MRI of the mouse heart, describing common difficulties and pitfalls. The protocol described in Chapter 4 is part of the paper "PET and MRI in mouse models of myocardial infarction"(11). The protocol is demonstrated in a mouse model of cardiomyocyte-specific Complex I KO and in the R6/2 mouse model of Huntington's disease (HD). The experiment in 4.6 is part of the paper: "Direct evidence of progressive cardiac dysfunction in a transgenic mouse model of Huntington's disease"(12). Experiments in 4.5 and 4.7 are part of separate works currently submitted for publication. In particular, 4.7 describes an application of cine MRI in the R6/2 mouse model of HD during inotropic stress testing with dobutamine. In this experiment, a limitation of the MRI protocol was encountered, represented by the long acquisition times. To accelerate acquisition, Chapter 5 explores methods to reconstruct cine-MRI acquisitions from undersampled data, implementing compressed sensing combined with parallel imaging utilising self-gated MRI with rectilinear and radial acquisition schemes. As radial MRI presented eddy-current induced artifacts, these were recovered retrospectively with a novel method, described in Chapter 6.

Methods developed within this thesis for assessment of tissue properties in-vivo are reported in Section 2. A novel method to measure infarct size *in vivo* that correlates well with TTC staining has been included in Chapter 7, based on the publication "A fast protocol for infarct quantification in mice" (13). The method is also demonstrated in the investigation of a new cardioprotective pharmaceutical in 7.5, which is part of a work currently submitted for publication.

Although complex cell metabolism experiments remain difficult to be performed *in vivo*, tracer techniques such as positron emission tomography (PET) can be used to probe specific biochemical pathways non-invasively. In Chapter 8 a new protocol for performing multi-modality imaging in a single examination is presented, based on the publication "PET/MRI assessment of the infarcted mouse heart" (14), while the data reported in 8.5 demonstrate this method in the assessment of a new pharmaceutical, part of the publication "Riociguat reduces infarct size and post-infarct heart failure in mouse hearts: Insights from MRI/PET imaging" - *PLoS ONE*, in press.

## 1.6  Chapter summary

Heart failure is one of the major causes of mortality in the United Kingdom. Mouse models can be utilised to model the effect of different genes on ischaemic heart disease and for the investigation of new interventions. As well as models for infarction, mice serve as models for a number of cardiovascular diseases, such as genetic heart disease, hypertension and cardiomyopathies. The last decades have seen wide adoption of these models in preclinical research. In order to obtain meaningful translational studies, non-invasive *in vivo* techniques can be utilised in mice. MRI is a translational tool which has high anatomical accuracy and can measure heart function as well as viability. The next chapter introduces the theoretical background of MRI, focussing on the technologies utilised throughout this thesis.

# Chapter 2

# Magnetic resonance imaging

Magnetic resonance imaging (MRI) is an imaging technique based on the nuclear magnetic resonance (NMR) phenomenon. The MRI signal is commonly generated by the spin of the hydrogen proton, mostly found biologically in water molecules. Unlike high-energy techniques such as computed tomography (CT), where contrast is generated by differences in material attenuation, MRI uses the interaction of spins with the local environment. This provides alternative soft-tissue contrast mechanisms depending on how the system is excited.

In this chapter, the essential theoretical background of MRI is outlined, focussing on the terminology and technology that will be used throughout this thesis (for a more detailed introduction to MRI see (15, 16)).

To perform MRI, a magnet is used to generate a static magnetic field, gradient coils are used to vary this static field spatially for localization. Transmitter radiofrequency coils excite the sample, while receiving radiofrequency coils measure the change in magnetic field generated by the sample using Faraday induction. The signal is then digitized in a spectrometer. The main steps required to obtain an image are briefly reviewed here.

## 2.1 Basic principles

The NMR phenomenon can be observed in spin systems within a static magnetic field. In medical systems the static magnetic field is usually generated by a superconducting magnet. By convention, the direction of the magnetic field is put along the **z** coordinate of a Cartesian frame of reference. The field, commonly labelled **B₀**, causes the spin of protons to align in a parallel or antiparallel direction. These two states of the proton nucleus are commonly referred to as spin-up state and spin-down state, with the following energies:

Spin down: $E_\downarrow = \frac{1}{2}\gamma\hbar B_0$

Spin up: $E_\uparrow = -\frac{1}{2}\gamma\hbar B_0$ .

The energy gap is proportional to B₀ (with $\gamma$ the gyromagnetic constant of the proton and $\hbar$ Planck's constant). At room temperature the thermal equilibrium is given by the Boltzmann distribution, with $N_\uparrow$ and $N_\downarrow$ respectively the number of spins in the up and in the down state:

$$\frac{N_\uparrow}{N_\downarrow} = exp\left(\frac{\Delta E}{k_B T}\right) \xrightarrow{\Delta E \ll k_B T} 1 + \frac{\gamma\hbar B_0}{k_B T}$$

The slight excess of spins in the up state with respect to the down state (about one part in $10^6$ per Tesla at room temperature), gives rise to a macroscopic magnetic moment, accounting for the whole NMR signal. This equation explains the reason behind the adoption of high magnetic fields in clinical MRI (1T and above): a higher B₀ generates more signal.

In a static magnetic field, a magnetic moment precesses at the Larmor angular frequency about the axis of the static magnetic field:

$$\omega_0 = \gamma B_0$$

the value of $\gamma/2\pi$ for an hydrogen nucleus is ~42 MHz/Tesla, but in the stationary state in which the magnetic moment is aligned with **z**, no precession will be observed.

In NMR, radiofrequency pulses are utilised to flip the direction of the magnetization in the sample. If an oscillating magnetic field $\mathbf{B_1}$ in a direction perpendicular to $\mathbf{B_0}$ is applied, the magnetization will precess about the direction of the vector sum of $\mathbf{B_0}$ and $\mathbf{B_1}$.

A practical way of thinking of this system is the so-called rotating frame of reference (17): a frame of reference that rotates about the z axis at the frequency $\omega_0$.

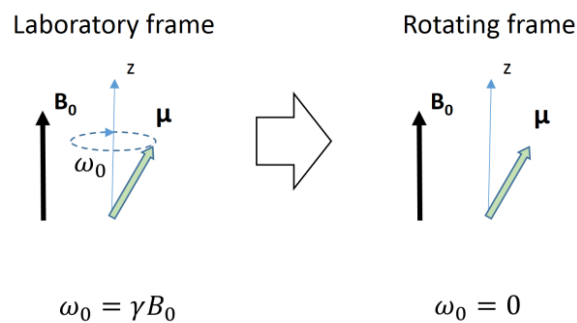

**Figure 4** In the rotating frame free precession at the Larmor frequency is eliminated.

In this frame of reference no free precession will be observed, and a radiofrequency electromagnetic pulse, circularly polarised in plane at the frequency $\omega_0$ will look like a steady vector $\mathbf{B_1}$. If a radiofrequency pulse is applied along the x-direction of this frame of reference, the magnetization will rotate about the x-axis with an angular frequency $\omega_1 = \gamma B_1$ while the pulse is present (see Figure 5A-C).

## T1, T2, and T2*

After being excited, a system of spins will return to thermal equilibrium. In 1946, Felix Bloch modelled the magnetisation signal in NMR with two decay constants, which he labelled T1 and T2 (17) (see Figure 5D-E).

**T1** is the time constant for longitudinal relaxation. This time constant is associated with thermal energy exchanges between the spin system and the surrounding environment, which are commonly called spin-lattice exchanges. If a spin system is driven away from the thermal equilibrium, dictated by the Boltzmann distribution, it will come back to equilibrium following an exponential function, with the empirical time constant T1.

**T2** is the time constant for transverse relaxation. This relaxation is driven by interaction between spin magnetic moments, commonly called spin-spin interaction. Every water proton will be in a slightly different environment dictated by the concentration and location of the neighbouring spins, this will lead to a slightly different precession frequency of every single magnetic moment, causing a different phase shift with time for every different particle. That will result in a net signal loss, following an exponential decay with an empirical constant T2.

In the presence of field inhomogeneities and inhomogeneous materials, the transverse free relaxation is called **T2*.** This time constant is similar in nature to T2, but in this case the main cause of dephasing and a signal loss is not the spin environment, but the main field inhomogeneity and/or a difference in magnetic susceptibility between neighbouring materials.

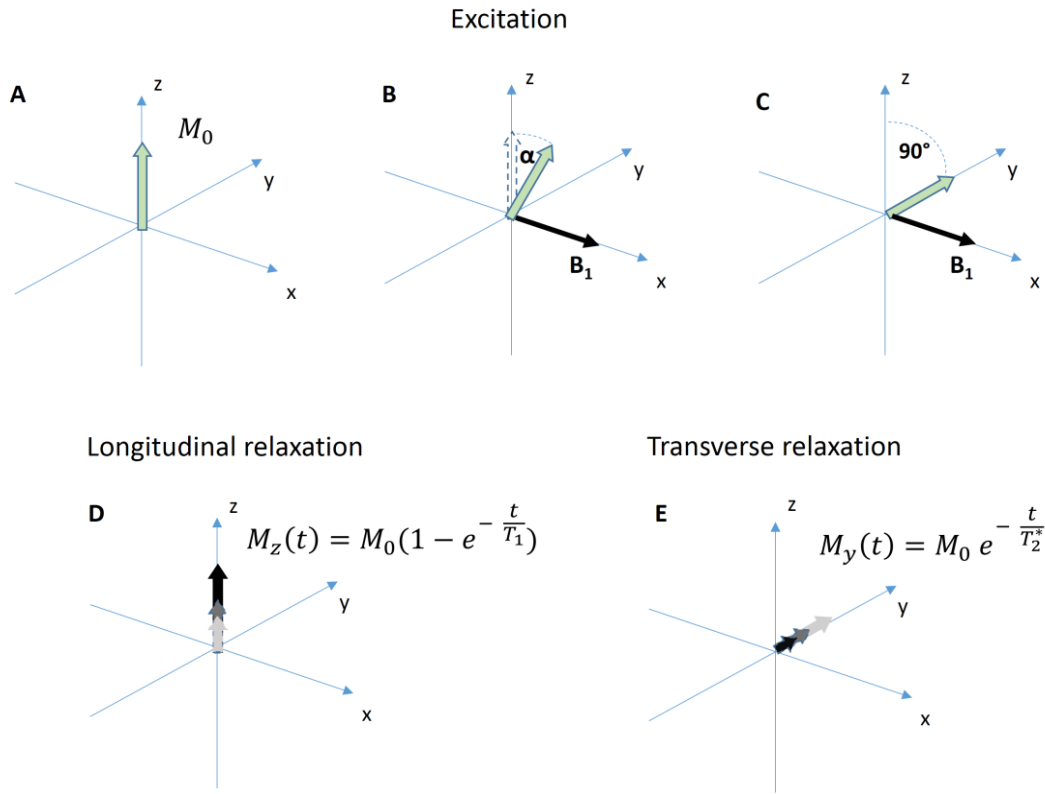

**Figure 5 Diagram of the effect of a radiofrequency pulse on a spin system in the rotating frame of reference. The net magnetization M₀ in A) is rotated using a radiofrequency pulse B₁ B), until the desired angle is obtained C). After the pulse is switched off the magnetization will return to z as shown in D), decaying in the transverse plane as shown in E).**

## *2.2 Acquisition schemes*

After a radiofrequency excitation, the signal decays as displayed in Figure 5E, this is called a Free Induction Decay (FID). This signal, however, is difficult to encode spatially. The way the signal is commonly extracted in MRI consists in the formation of a signal echo at a fixed time after excitation, called echo time (TE).

### 2.2.1 Spin echo

It is possible to refocus the signal loss due to static field inhomogeneities with a radiofrequency pulse.

Even after the signal has decayed with a time constant T2*, it is still possible to measure it by means of a spin echo, introduced by Hahn in 1950 (18).

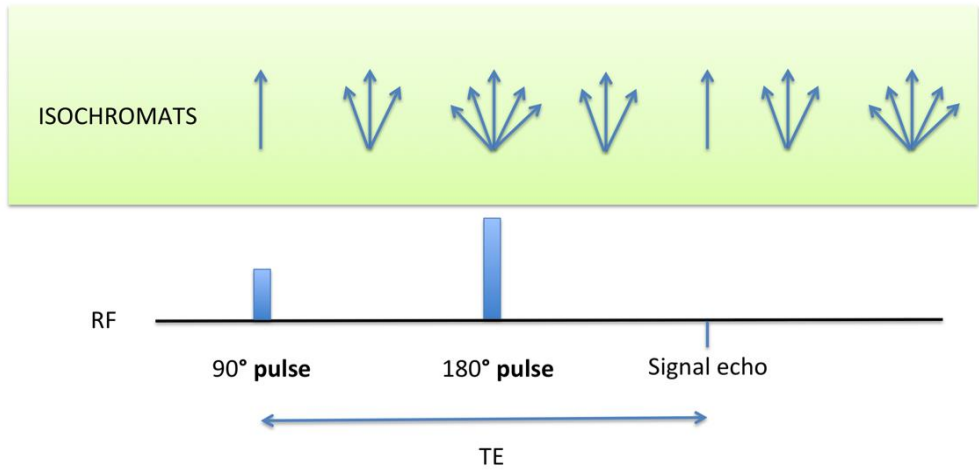

**Figure 6** Spin echo

After the signal has decayed, at a time TE/2 a 180° pulse inverts all the velocities of the dephasing isochromats. All the spins will be coherent again at the time TE, while the signal intensity will have decayed with constants T1 and T2 by that time.

## 2.2.2 Gradient echo

In order to reconstruct an image, instead of a single-voxel spectrum, the position within the sample must be encoded. The position encoding to form an MRI image is achieved using magnetic field gradients. The field gradients are generated by coils adjacent to the internal surface of the magnet bore. When these are switched on, they will apply a linear magnetic field gradient along the x- y- or z- axis. If different static field strength is seen by a spin, this will precess with a different frequency:

$$\omega_0 = \gamma(B_0 + G_x \cdot x + G_y \cdot y + G_z \cdot z)$$

Where $G_x, G_y, G_z$ are respectively $\frac{dB_z}{dx}, \frac{dB_z}{dy}, \frac{dB_z}{dz}$.

In addition, an arbitrary time varying gradient **G**($t$) causes a transverse magnetization vector to change its phase according to the equation $\omega$(r, $t$) = $\gamma$ **G**($t$)·**r**, so that the total phase accumulation at time $t$ is:

$$\theta(r, t) = \gamma \left( \int \boldsymbol{G}(t')dt' \right) \cdot \boldsymbol{r}$$

Utilising this effect it is possible to form a signal echo using only the field gradients.

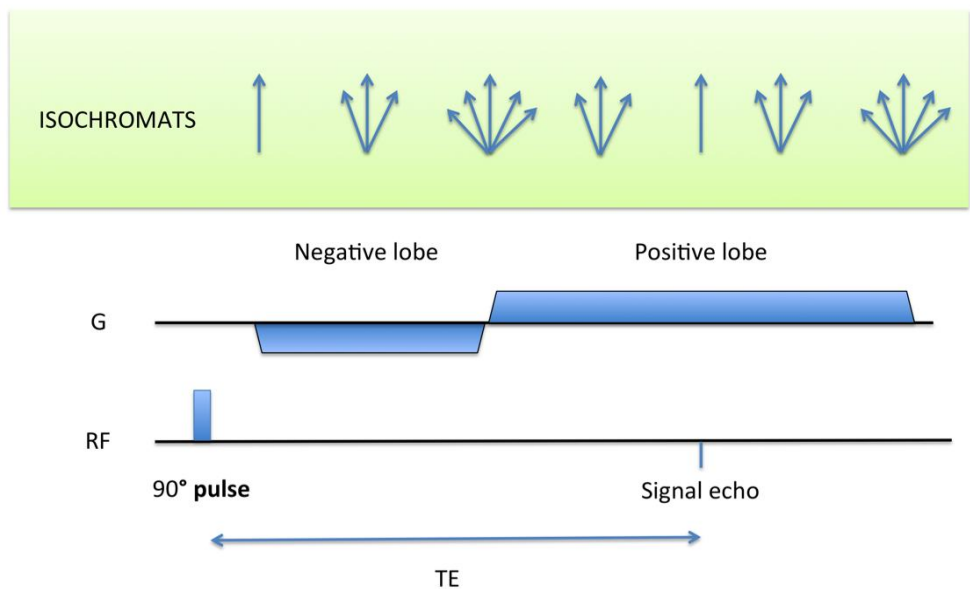

**Figure 7** Gradient echo.

After excitation, all the isochromats are dephased by means of a gradient pulse. Afterwards, another gradient pulse is applied in the opposite direction. At the time TE a signal echo arises, while the signal intensity will have decayed with a time-constant T2* by that time. Imaging pulse sequences employing just a gradient to form an echo are called gradient echo, while imaging pulse sequences employing a refocussing pulse are called spin echo imaging sequences.

### 2.2.3 Slice selection

In MRI, radiofrequency pulses are usually combined with gradient pulses in order to select a slice of the sample. While applying a gradient along an axis, the excitation RF pulse is usually modulated with the Fourier transform of the wanted slice profile. The intensity of the gradient and the bandwidth of the pulse determine the slice thickness, as shown in Figure 8:

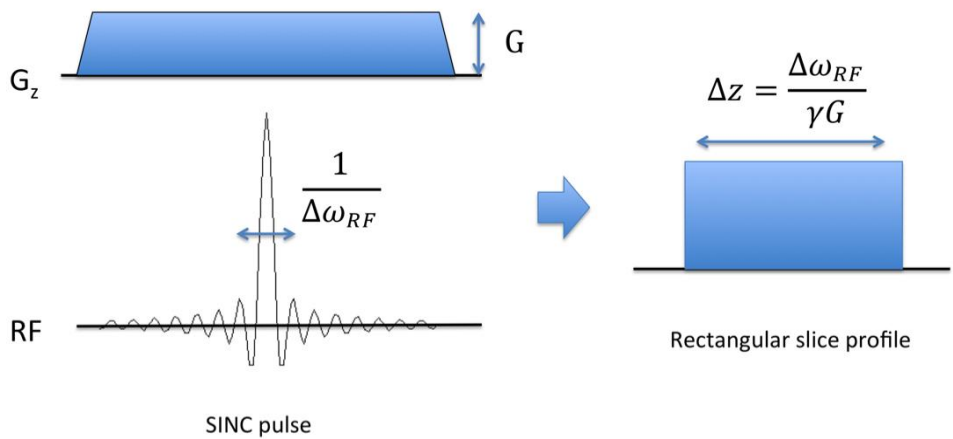

**Figure 8** Slice selection.

### 2.2.4 *k*-space encoding

It is helpful to consider the encoded signal in frequency space, known as k-space. When considering the function of an imaging sequence, the k-space is the space formed by the variables kx, ky and kz, obtained as the product of a field gradient and time:

$$k_i = \frac{\gamma}{2\pi} G_i(t) \cdot dt$$

The k-space is linearly related with the image, which can be obtained through an inverse Fourier Transform:

$$I(x, y, z) = \iiint s(k_x, k_y, k_z) \, e^{i(k_x x + k_y y + k_z z)} \, dk_x \, dk_y \, dk_z$$

Where s indicates the signal in k-space and I indicates the image.

During a signal echo, a k-space line is sampled. The sequence must be repeated several times, obtaining a new k-space line every time. The time between repetitions is called repetition time (TR).

The most common sampling pattern is Cartesian Fourier imaging, illustrated in Figure 9, but the k-space can be sampled with a pattern of choice, like a radial one, illustrated in Figure 10, or a spiral etc.

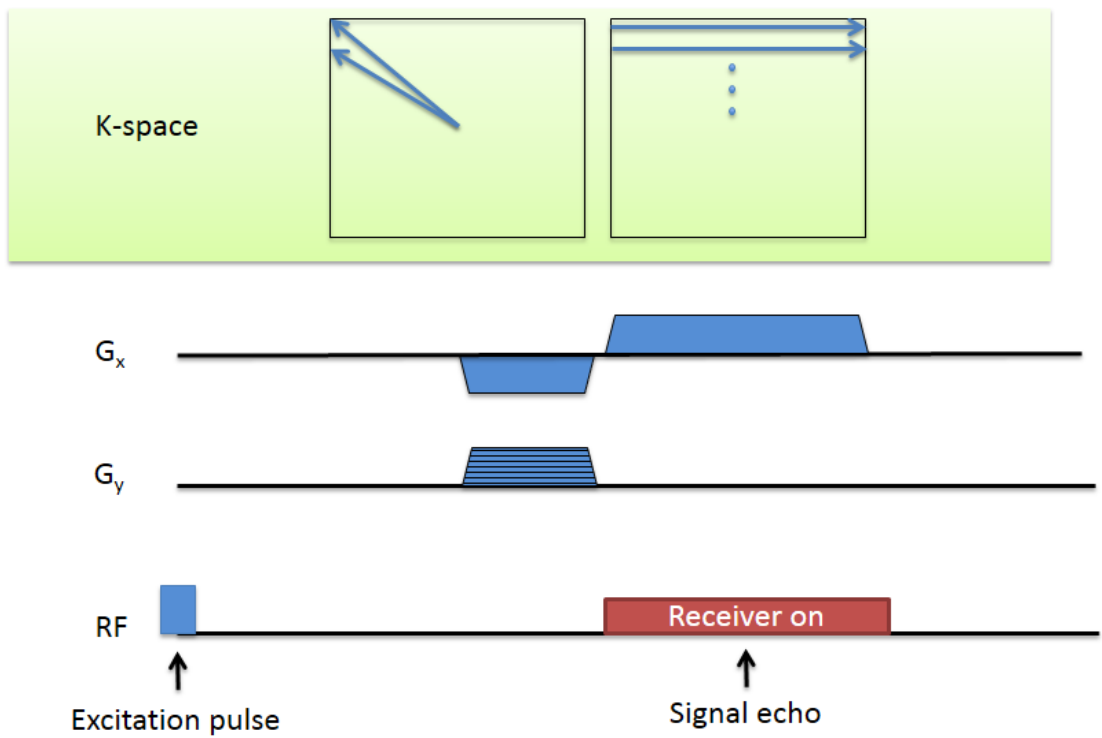

**Figure 9** Acquisition of lines of k-space with a Cartesian sampling scheme

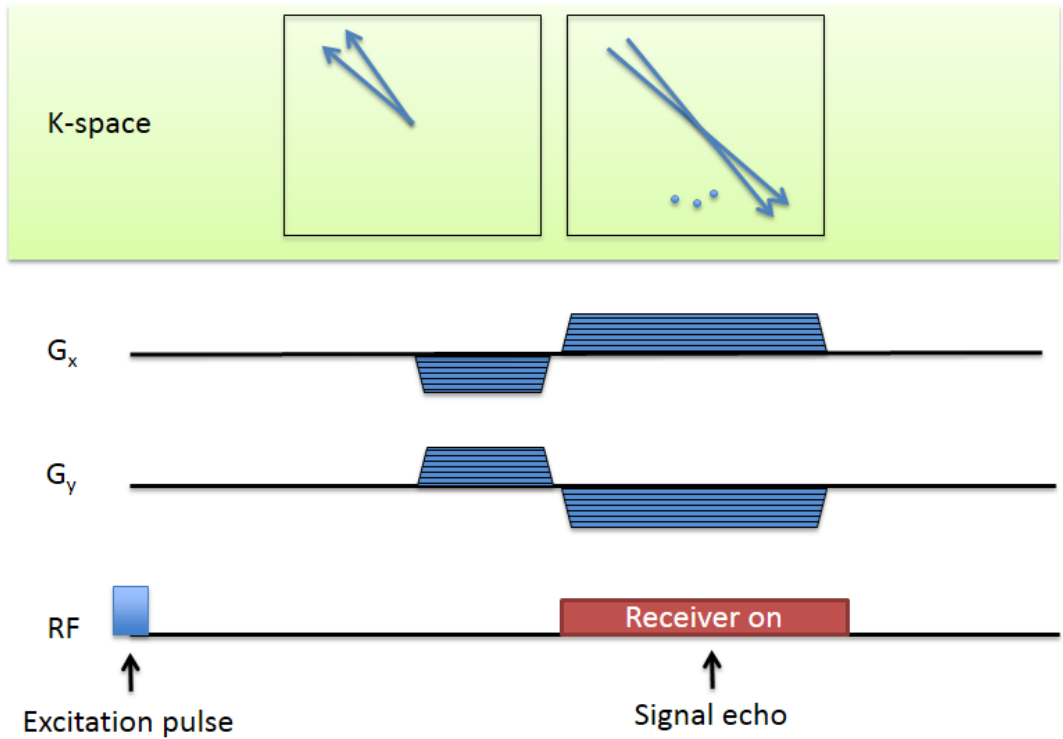

**Figure 10** Acquisition of oblique lines of the k-space with a radial sampling scheme.

## 2.2.5 The contrast of MRI images

Controlling timing between pulses and their intensities, it is possible to create different contrasts between different tissues. Dedicated pulse sequences can be optimised for contrast between T1, T2 or proton density in different tissues. Although the signal in MRI images is rarely quantitative, the tissue contrast between neighbouring tissues and the high spatial resolution produce an anatomical detail of soft tissue which is superior to most imaging technologies.

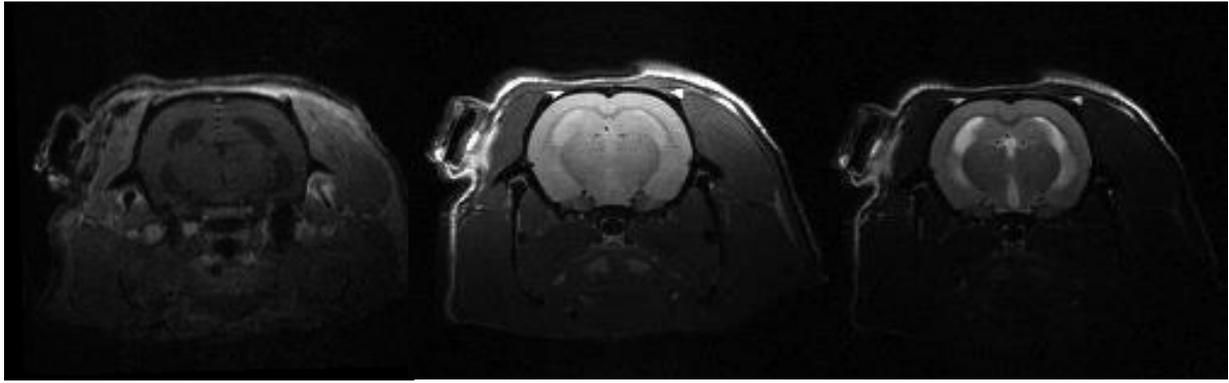

**Figure 11** T1, proton density, and T2 weighted contrast in a rat brain.

Although in MRI endogenous contrast is present between different chemical species, it is possible to manipulate this contrast using contrast agents. Gadolinium is a paramagnetic element that can be used to enhance MR images by affecting relaxation times. The main effect of Gadolinium contrast agents is a reduction of T1. T1-weighted sequences can then be used to visualise the kinetics of contrast agents within organs (see Chapter 7, where this is used to probe tissue perfusion and washout performance, in order to assess viability).

## 2.2.6 Fast gradient echo sequences

When imaging with a time between repetitions (TR) is of the same order or smaller than T1, a saturation of the signal is observed. When an excitation pulse is applied, the magnetization has not completely relaxed from the previous pulse. The number of excited spins will therefore diminish at each successive excitation until the effect of the excitation and the relaxation are balanced: a "steady state". To reach a steady state with sufficient signal intensity, fast sequences usually involve low excitation flip-angles. By changing the flip-angle, the amount of T1 weighting is changed.

This is problematic when TR is of the same order or smaller than T2. In this case, the encoded magnetization of the previous repetition, not completely decayed, will interfere with the freshly excited magnetization, with different encoding.

This problem is addressed using one (or a combination) of two separate strategies:

- Spoiling: the magnetization in the x-y plane is destroyed prior to every new excitation.

- Refocusing: the magnetization in the x-y plane is refocused to be coherent with the new excitation.

The main fast gradient echo strategies are called Fast Low Angle SHot (FLASH) and Fast Imaging with Steady state free Precession (FISP), and these respectively use spoiling and refocusing.

**FLASH**

In FLASH (19, 20), a phase offset is applied to successive excitation pulses to "spoil" the residual magnetization.

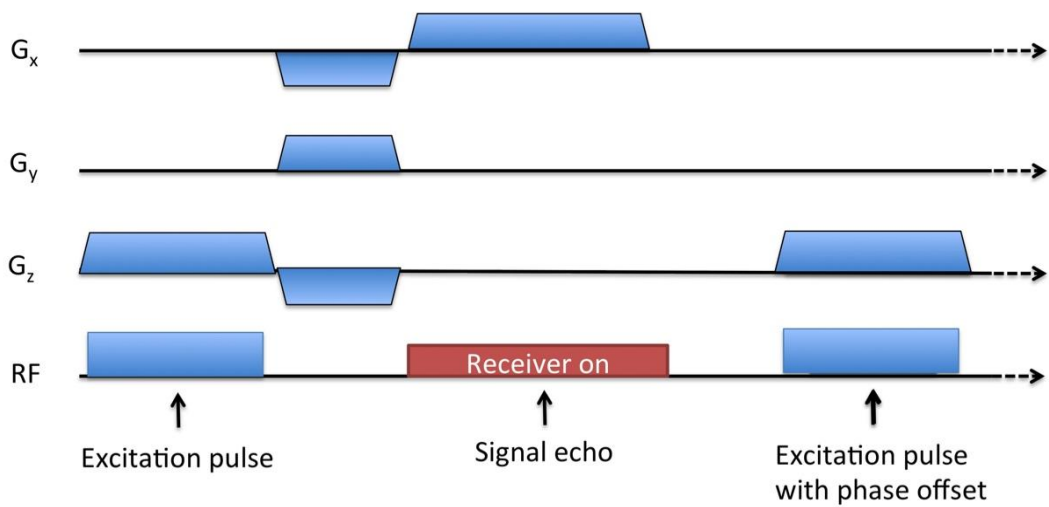

**Figure 12** Acquisition of a k-line with FLASH.

**FISP**

In FISP (21), the magnetization from previous TRs is refocused to be coherent with the current signal. This increases the total amount of signal and introduces additional T2 weighting. Sequences where the phase is refocussed in all three directions are called TrueFISP, while when only in-plane magnetization is refocused we speak of "fid" FISP, or simply FISP.

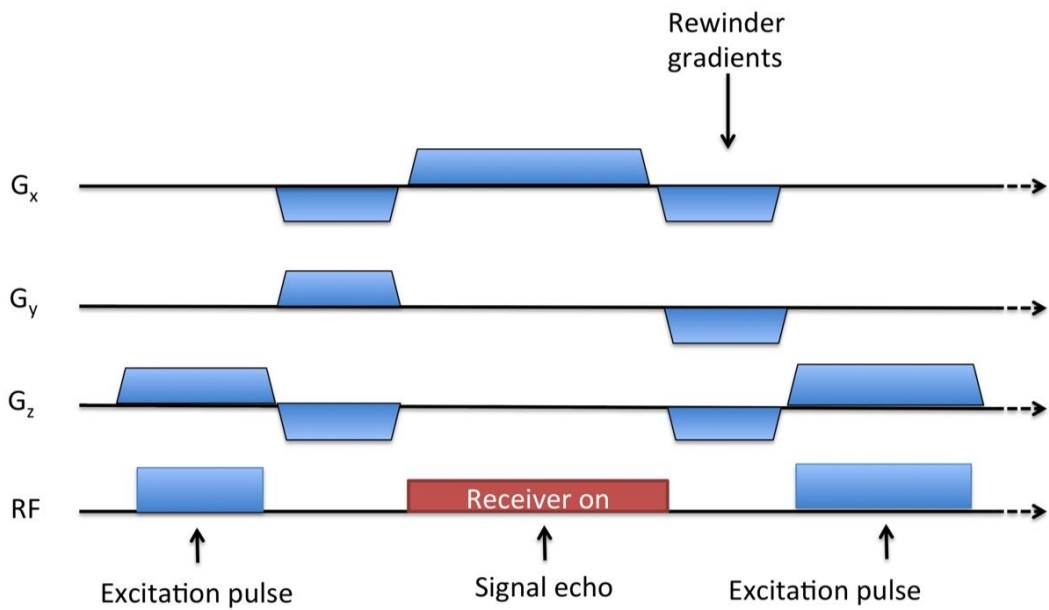

**Figure 13** Acquisition of a k-line with TrueFISP.

## 2.2.7 Inflow enhancement

Blood moves in vessels at high speed and in large amounts. If a slice perpendicular to a vessel is imaged with a short TR (less than T1) and high flip-angle, the signal in the slice becomes saturated. However, as fresh blood flows in the slice at every excitation, the blood will not saturate and will appear much brighter than the rest of the slice. This technique can be used to maximise the contrast between blood and other anatomy in angiography or cine-MRI, where the visualization of the fluid is pivotal.

A          B

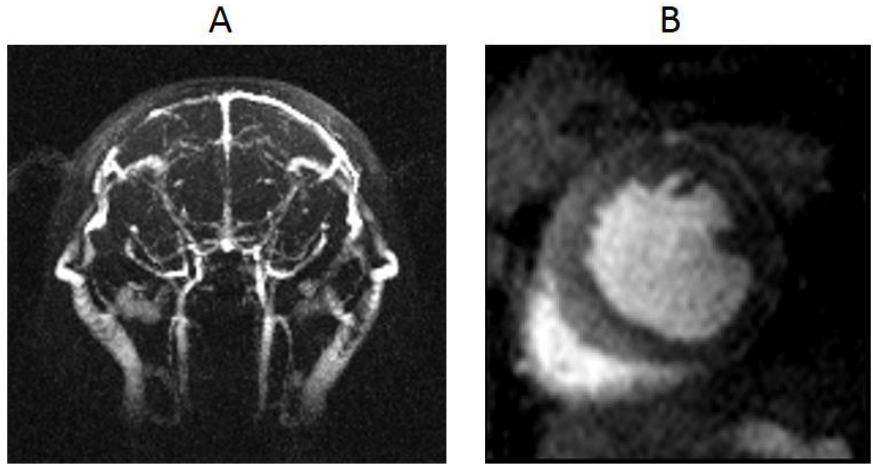

**Figure 14 Examples of cases where inflow enhancement is used for better contrast: A) Maximum intensity projection of a mouse brain. B) Cine MRI of a mouse heart (FISP). In both cases fresh blood is flowing in the slice at each acquisition avoiding saturation, appearing brighter than if it were stationary.**

## 2.2.8 Inversion recovery

In some applications, maximal T1 contrast is required between two or more distinct spin species. In these situations, an inversion recovery sequence can be used to null the signal from one of the species, maximising the contrast with the others. This is achieved by applying an inversion pulse, tipping the magnetization by 180°, at the start of the imaging sequence. At this point, the magnetization will be left to relax with a time-constant T1 along z for a time called the inversion time (TI). Following this time, the residual magnetization will be read out by an imaging sequence such as FLASH.

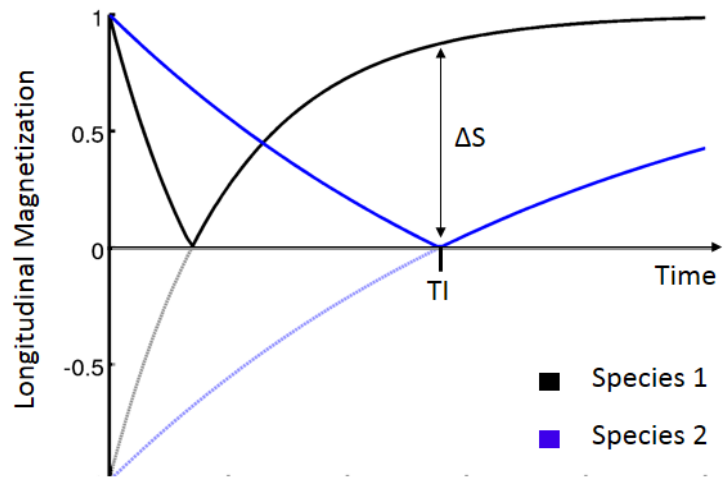

**Figure 15 The longitudinal relaxation of two spin species (shaded lines are the actual magnetization while solid lines represents the signal that can be read from NMR). A value of TI can be chosen to maximise the separation between the two species.**

When applying inversion recovery in practice the time between successive excitations is of the same order of T1. In this case, saturation effects due to successive excitations will be significant as well as the free relaxation (22).

## *2.3 Reconstruction*

If the k-space is acquired on a uniform Cartesian grid, reconstruction is performed by a Fourier transform of the k-space data. The k-space points are complex numbers, and the image voxels also have a real and imaginary part. However, MRI images are usually displayed as magnitude images, discarding the phase information, except for situations where the phase information is important clinically (for instance, see DENSE in 8.1, where motion is encoded in the image phase). The noise of real and imaginary channels of single-channel MRI signals have Gaussian distribution and the noise distribution of the resulting magnitude images is represented by a Rician distribution (23, 24).

## 2.3.1 Image quality metrics

Quantitative metrics are utilised to assess the quality of the images within standardized protocols for checking the status of the scanner, or to compare the effectiveness of different acquisition strategies.

To assess the amount of detected signal within an experiment this is compared to the noise level. The signal to noise ratio (SNR) is defined as the ratio of signal intensity to the standard deviation of the noise. The SNR represents an absolute quantification of the global signal within an object, another useful measure is the capability of discriminating between two different tissues in the image. This is measured with a contrast to noise ratio (CNR). Contrast to noise ratio is defined as the ratio of the difference between the signal in two tissues to the standard deviation of the noise.

## 2.3.2 Radial MRI

Although the majority of MRI experiments sample uniformly spaced Cartesian points, in some applications different sampling schemes that do not sample uniformly can be preferred. One of the most popular acquisition schemes in MRI is radial sampling of k-space (see Chapter 5, where this is used to accelerate the acquisition). To reconstruct images from radial k-space projection algorithms can be used as well as gridding approaches.

Filtered backprojection (FBP )was developed for CT as an approximate solution to the inversion of the radon transform (25). According to the Fourier slice theorem, the radial spokes acquired in k-space represent the Fourier transform of the image projections, like those that would be acquired in CT. If backprojection is applied by simple inversion of the radon transform, blurring arises due to the radon transformation being non unitary. In FBP, the profiles are filtered with an M- or ramp filter prior to backprojection to correct for the blurring. Although FBP is still widely used in nuclear medicine, it has been replaced in MRI by gridding techniques, due to their superior efficiency (26). In gridding, the acquired MRI data are interpolated onto a Cartesian grid after convolving with a convolution kernel. Usually, a Kaiser-Bessel window is used in order to achieve a good image quality at a reasonable window size. Density correction is applied prior to convolution to correct for non-uniform sampling of the k-space. For radial trajectories it is possible to derive an analytic density correction function, yielding the Ram-Lak filter (27).

Means of performing k-space interpolation with high efficiency and speed have been developed for use in iterative algorithms. For instance, one such algorithm used in practice is the non-uniform fast Fourier transform (NUFFT) (28). NUFFT calculates complex kernels based on the given sampling pattern to minimize the interpolation error. Utilising NUFFT the kernel can be pre-computed and stored for use in fast iterative algorithms.

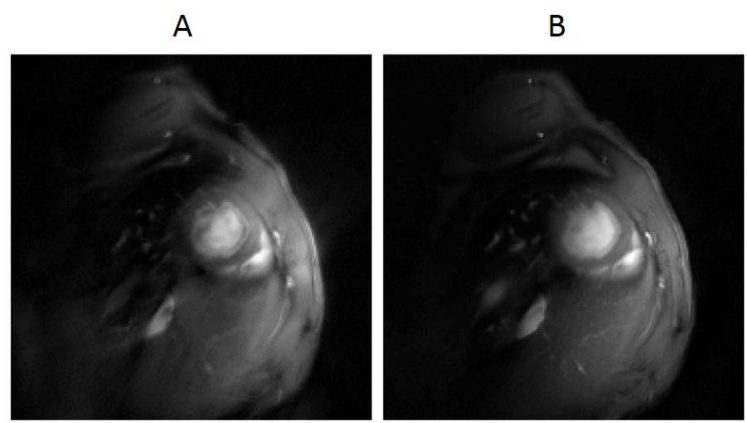

Figure 16 A) Radial MRI in the mouse heart reconstructed with FBP. B) The same slice reconstructed with NUFFT.

### 2.3.3 Trajectory errors

When gradient coils are switched rapidly, the varying magnetic field results in the induction of currents in the conducting surfaces within the rest of the system. These currents, known as eddy currents, generate their own magnetic field as they decay, altering the total magnetic fields thus distorting the desired effect of the gradient waveforms. Artifacts generated by these can be treated as delays of the effective gradients seen (29).

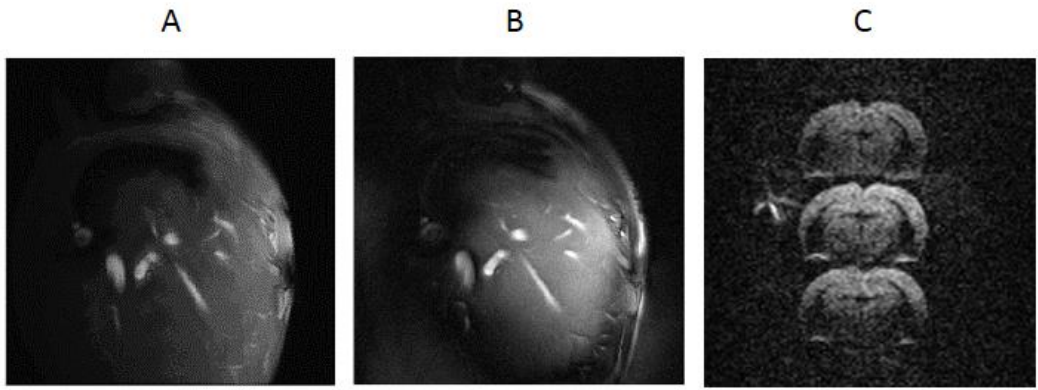

Figure 17 Eddy currents generate artifacts. A) Cartesian image of a mouse liver with Cartesian imaging. B) The same image with radial imaging. Modulation on the intensity of the image is due to eddy currents. C) Four shots echo planar imaging of a rat brain. Multiples copies of the brain are generated by eddy currents.

For standard Cartesian MRI, linear eddy currents produce a global shift of k-space in the readout direction, which cause a global phase-shift of the reconstructed image, with no observable effect on the magnitude image. In contrast, when segmented k-space acquisition is employed, significant artifacts are seen. A very common example where this effect is significant is echo planar imaging (EPI), where each successive k-line is acquired in opposed directions. Here, opposed k-space shifts in successive k-lines result in coherent ghosting artifacts.

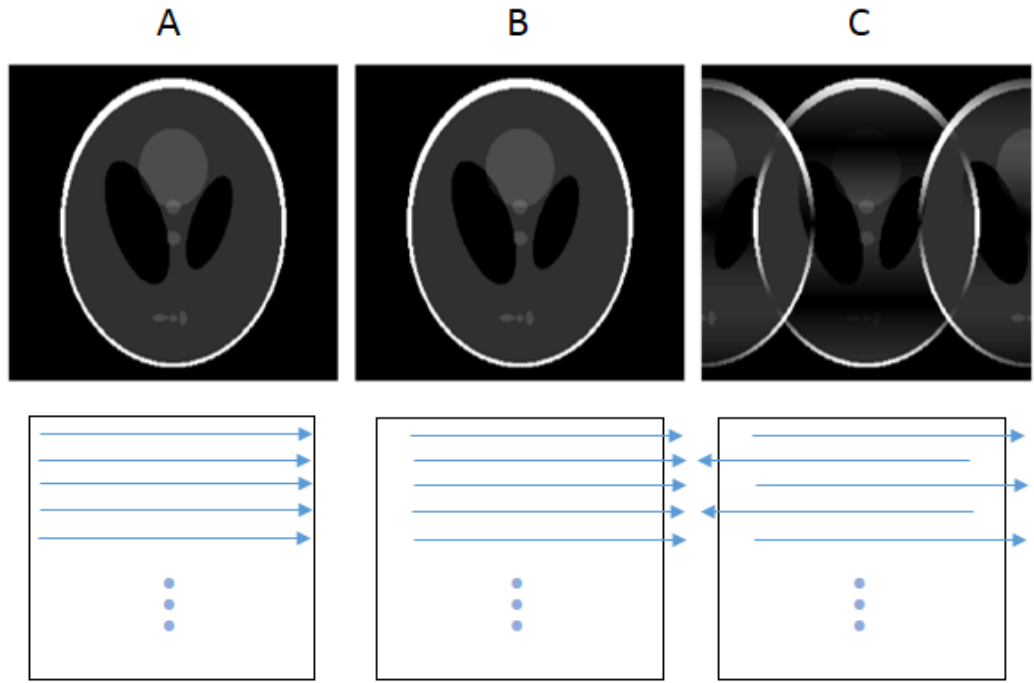

**Figure 18 Simulated effect of linear eddy currents phantom (1 pixel shift in k-space) on a numerical Shepp-Logan phantom: A) The Shepp-Logan phantom reconstructed without eddy currents; B) Linear eddy currents applied to standard Cartesian imaging, no effect is noticeable; C) Effect of Linear eddy currents to EPI acquisition, visible ghosting is present.**

Similarly, in a radial acquisition, shifted lines will result in an erroneous assignment of k-space coordinates to the acquired spokes. However, given the circular trajectory, in this case the effect will result in radial phase interference between projections rather than coherent ghosts. The effect of this trajectory error will be mostly visible as a low-frequency modulation of the image intensity. Some signal will "leak" from the object to areas around it, generating "shading" and "halos".

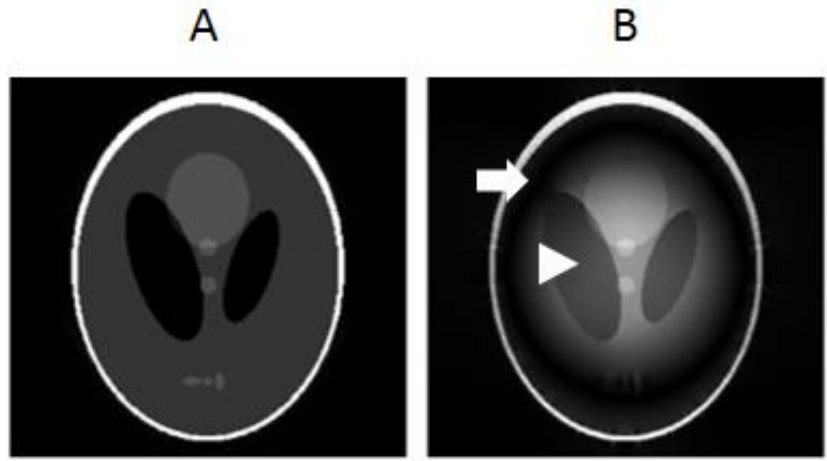

Figure 19 Simulated effect of linear eddy currents (1 pixel shift in k-space) on radial MRI: A) Radially acquired and reconstructed Shepp-Logan Phantom. B) The same phantom in presence of linear eddy currents. Arrow points to shading, arrowhead to halos.

In some situations, especially for smaller bores (29), eddy currents will produce not only gradient terms but field ($B_0$) terms. These terms will produce a constant phase-shift along the readout. When MRI is acquired with standard Cartesian sampling this will produce a constant displacement of the whole image, while in radial MRI or EPI different lines will get different displacements, producing artifacts (see Figure 20).

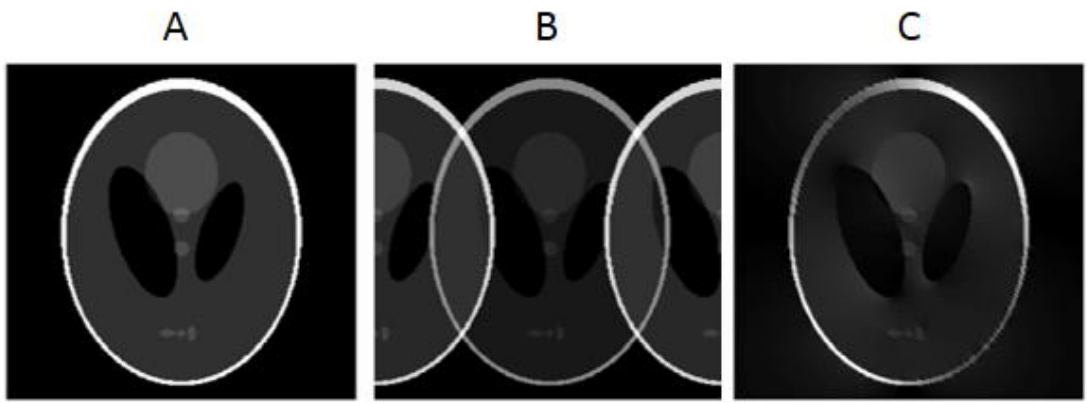

Figure 20 Simulated effects of B0-induced phase errors (1 radians) for A) Cartesian MRI; B) EPI C) Radial MRI.

In Chapter 6 strategies to correct for eddy currents will be demonstrated in the context of free-breathing radial MRI in mouse hearts.

## *2.4 Acceleration*

MRI is time consuming due to the need to fill the k-space one line at a time. In some time-critical applications, however, there is not enough time to acquire the whole k-space. For these acquisitions, methods for reconstructing partially-acquired k-space are required.

The classical reconstruction of MRI images from k-space data can be seen as a linear operation going from coefficients in k-space to image pixels. When undersampled acquisitions are performed for speeding up data acquisitions, the number of coefficients acquired in the k-space is smaller than the number of required image pixels. As the system of equations for linear reconstruction becomes underdetermined, multiple solutions exist for the linear reconstruction problem. As a result, if data is zero-filled and reconstructed linearly, aliasing will be present in the image.

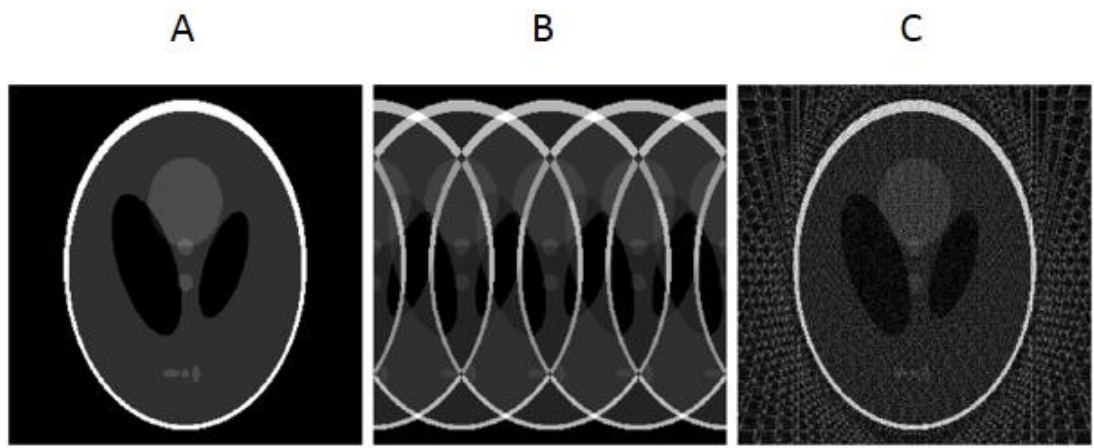

**Figure 21 Simulated effect of undersampling on image reconstruction. Underdetermined systems create aliasing, dependent on acquisition geometry. A) Fully sampled Shepp-Logan phantom. B) Shepp-Logan phantom with ¼ of the k-lines acquired with a cartesian scheme. C) Shepp-Logan phantom with ¼ of the k-lines acquired with a radial scheme.**

To solve this, independent information from different receivers can be used, a technique known as parallel imaging (30). Otherwise, iterative reconstruction can be used instead of linear techniques, making assumptions on the data through regularization. Methods to perform these techniques will be introduced here.

## 2.4.1 Parallel imaging

Parallel imaging techniques used for acceleration of MRI are based on acquisition of the signal from different receivers. If the k-space is fully sampled, different receivers can be combined with a sum of squares.

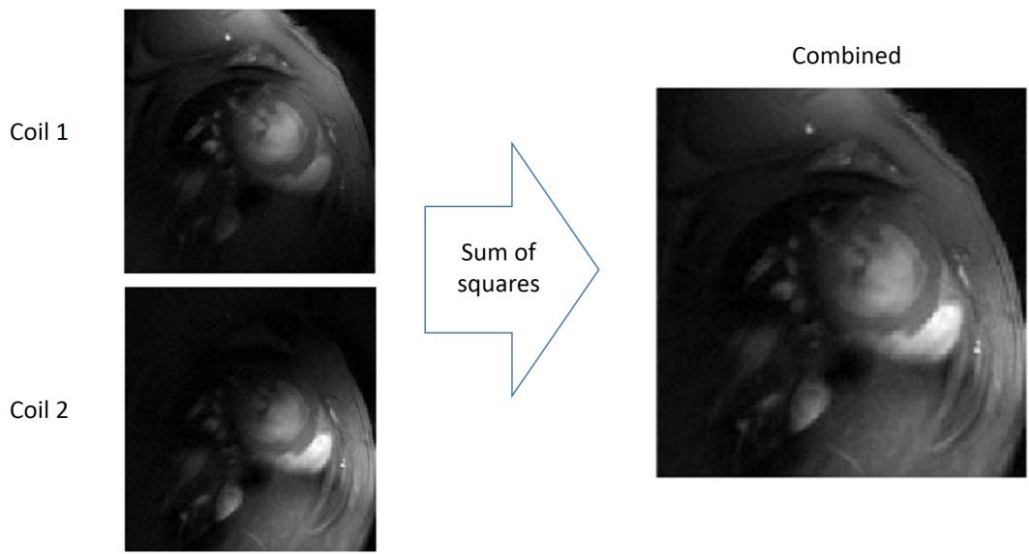

**Figure 22 Different receivers images have different sensitivity profiles. The single coil images are combined using a sum of squares.**

Although the signal is highly correlated between receivers, each of these sees at least partially uncorrelated noise and has a different sensitivity profile. Due to this property, independent receivers can be utilised to resolve aliasing in undersampled acquisitions. The most commonly used algortithms used for parallel imaging are SENSE and GRAPPA.

### Sensitivity encoding (SENSE)

If the full image is reconstructed linearly from undersampled data, this results in an undetermined system. However, the linear system has a unique solution if the image is reconstructed with a smaller FOV. SENSitivity Encoding (SENSE) starts with reconstructing an image on a smaller FOV, which will contain a folded image. Sensitivity encoding can then be seen as an unfolding algorithm:

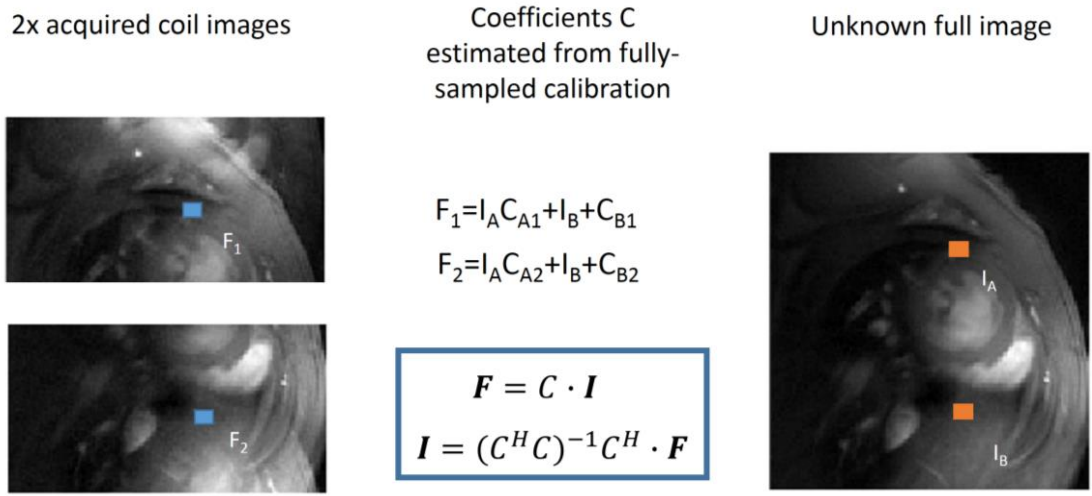

**Figure 23 In SENSE folded coil images are unfolded utilising coil sensitivities derived from fully sampled calibration.**

SENSE strongly relies on accuracy of prior coil sensitivity estimation (30), but works in general for any k-space acquisition scheme, as data are handled in the image space.

## Generalized autocalibrating partially parallel acquisitions reconstruction (GRAPPA)

In GeneRalized Autocalibrating Partially Parallel Acquisitions reconstruction (GRAPPA), full individual coil images are reconstructed from the undersampled data. Combination of different coils is formulated in k-space, synthetizing non-acquired datapoints from acquired datapoints across coils, weighted with coefficients obtained from a calibration.

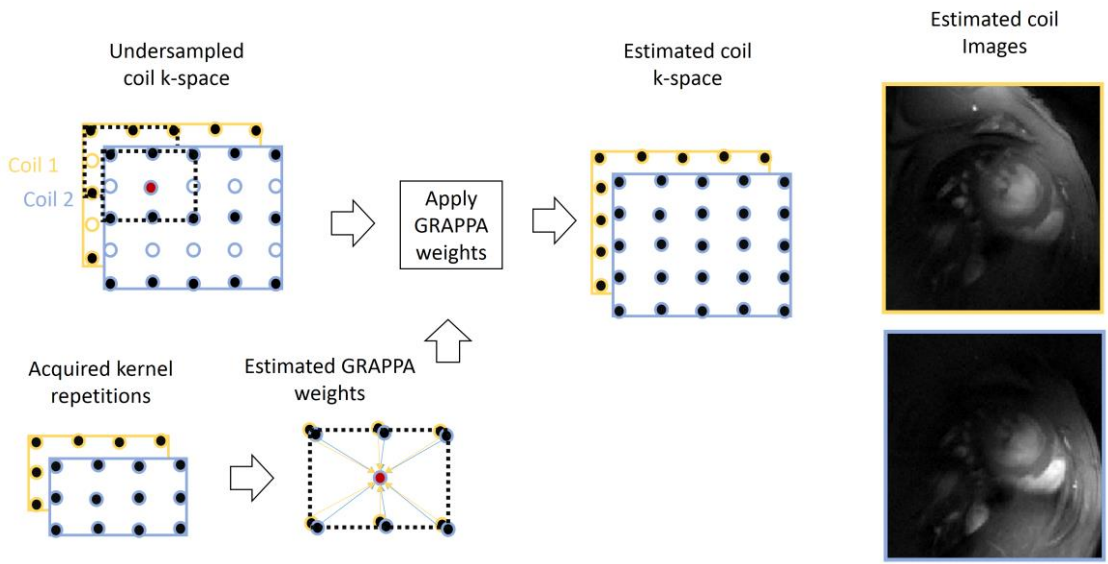

**Figure 24 In GRAPPA, kernel repetitions are acquired to estimates weights, used to estimate non acquired k-space coordinates from acquired k-space coordinates across coils. Individual coil images are reconstructed.**

GRAPPA possesses a robust estimation of sensitivity (as this is done in k-space at its centre, in a condition of high SNR) and aims to restore each coil separately, however it does not handle different acquisition strategies efficiently (30). Different implementations of GRAPPA have aimed to formulate the problem for a number of different k-space trajectories, however these all rely on some approximation of the data, reducing the efficiency of the method when compared to SENSE (30).

## 2.4.2 Compressed sensing

Parallel imaging algorithms utilise multiple coils to increase the number of variables or to reduce the number of equations in a linear system. In contrast, compressed sensing techniques do not combine multiple coils, but they challenge the assumption of independence of image pixels, utilising non-linear iterative reconstruction instead of a linear system of equations.

The assumption that compressed sensing makes is that voxels are not independent and the image is "compressible" in a given space. So, if data are transformed to that space, only a small number of coefficients will be different from zero. A space with this characteristics is called a "sparse" domain. Different domains can be used in different reconstruction problems, based on the properties of the data.

A very common example of a sparse representation is given by the discrete wavelet transform, at the basis of the compression algorithm adopted by JPEG. The sparse property of this transform can be readily demonstrated as JPEG compression can be applied to most medical images, reducing file size without noticing significant alterations (see Figure 25).

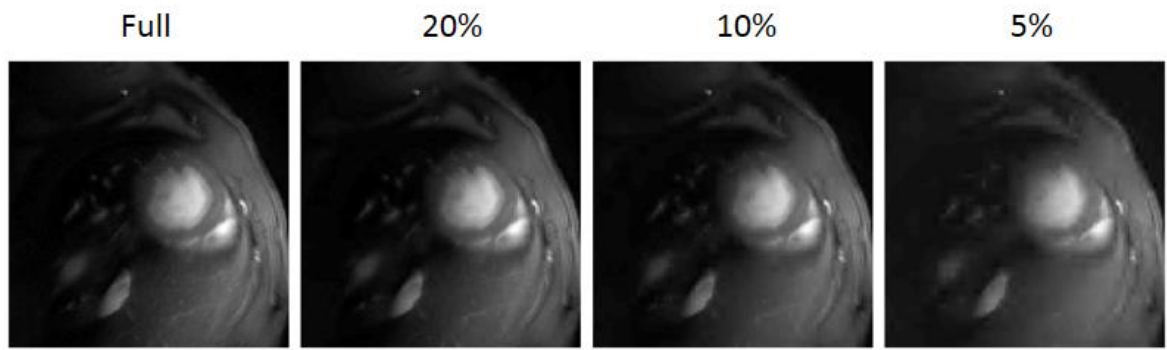

Figure 25 Compression in the wavelet domain. An image from a mouse heart keeping all of the data, only the 20% biggest wavelet coefficients, 10% and 5%. This image can be compressed 5 fold or 10 fold in the Wavelet domain with a small impact on quality, while blurring is observable on the 20 fold compressed image.

The main idea for acquisition in compressed sensing MRI is to minimise coherence in the sampling of k-space, so that aliasing is incoherent in the sparse domain, and the relevant coefficients can be easily distinguished from others containing only aliasing. Once a sparse transform and an acquisition scheme have been established for an imaging problem, an image can be reconstructed utilizing an iterative algorithm. The solution presenting the highest number of null coefficients in the sparse domain is chosen utilising regularized iterative reconstruction.

Although images can be compressed successfully, even higher success can be achieved when compressing a video (31). In a video, several frames share spatial features, and intrinsic correlation between frames can be exploited in the temporal domain. Some MRI techniques aim to measure the temporal evolution of an organ trough time. These acquisition can theoretically benefit from temporal compression (see Figure 26).

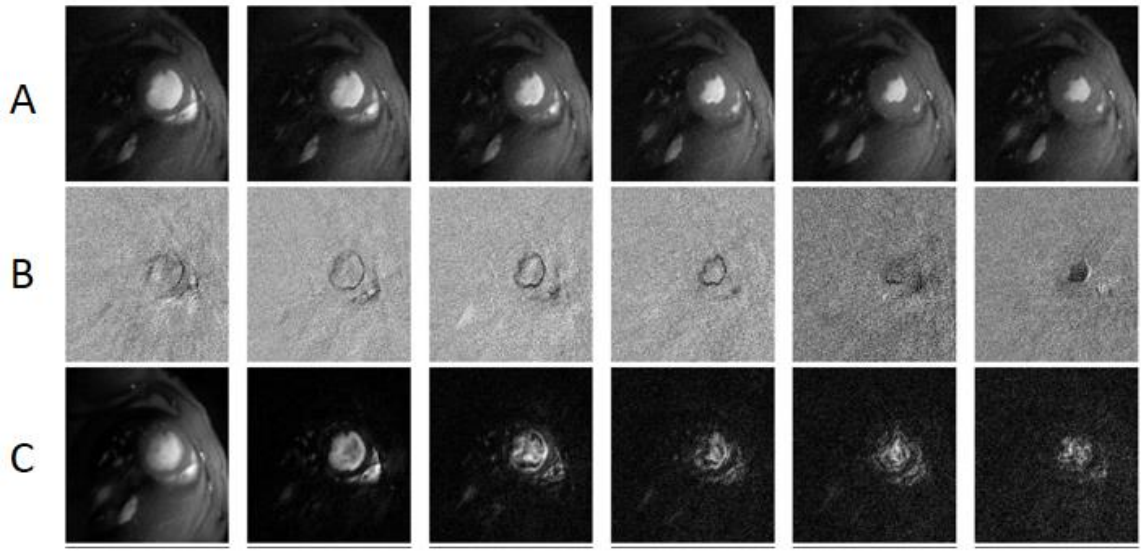

**Figure 26 Sparse representations for a video of a beating mouse heart. A) Frames. B) Frames after applying finite differences in time domain (each frame has been subtracted from the previous one). C) Fourier transform in the time domain. B) and C) are sparse representations for cine MRI in the heart.**

To summarize, compressed sensing needs three main ingredients:

- *Sparsifying transform*: A sparse domain where most coefficients in the data are equal to zero must exist.

- *Incoherent aliasing*: Undersampling must generate aliasing which is equally spread amongst datapoints in the sparse domain. This is most likely obtained when using incoherent sampling.

- *Iterative reconstruction*: Non-linear reconstruction is needed to get rid of the aliasing preserving consistency with the acquired data.

## Mathematical formulation

In the previous paragraphs an overview of compressed sensing has been given. Here a description of an algorithm to implement this in practice will be reviewed briefly.

If $\mathbf{x}$ represents the acquired data in the k-space and $\mathbf{y}$ the resulting image, in a fully sampled case:

$$\mathbf{y} = \mathbf{A}\,\mathbf{x} \qquad \mathbf{x} = \mathbf{A}^*\,\mathbf{y}$$

Where $\mathbf{A}$ represents the linear transform between k-space and image space, while $\mathbf{A}^*$ is its inverse such that the direct application of $\mathbf{A}$ without using an iterative algorithm would produce an aliased image.

The compressed sensing problem can be formulated as a constrained minimization problem. As proposed by the seminal paper in compressed sensing MRI (32), the condition of sparsity, formulated theoretically as the condition of most coefficients being equal to zero in the sparse domain, is practically implemented as a minimization of the l1 norm:

$$\underset{\mathbf{x}}{\mathrm{argmin}} \quad \|\mathbf{A}\mathbf{x} - \mathbf{y}\|_2^2 + \lambda \|\psi\mathbf{x}\|_1$$

Where $\psi$ is the sparsifying transform. The l2 term is called the data consistency condition, while the l1 is called the sparsity condition. The factor $\lambda$ is called the regularization weighting parameter. To solve this we use a projection over convex sets (POCS) that alternates steps that project the solution perpendicularly to the conditions (the convex sets).

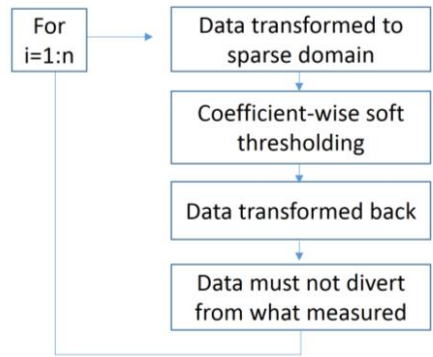

**Figure 27 Scheme of a POCS algorithm. Each iteration alternates soft thresholding and data consistency steps.**

## 2.4.3 Combination of parallel imaging and compressed sensing

As special applications require faster and faster acquisition, there is no reason why parallel imaging and compressed sensing should not be combined to achieve higher speedup. However, this combination requires some practical considerations to be performed effectively. Compressed sensing requires incoherent sampling and an iterative reconstruction where each receiver is handled separately. SENSE is an optimal image reconstruction framework, where a full battery of k-space sampling schemes can be utilized. However, this technology handles all the receivers at once. In contrast, GRAPPA handles each receiver separately, but is optimally applied only to specific sampling schemes.

To overcome these limitations, a novel framework generalising GRAPPA was developed by Lustig and Pauly, called SPIRiT (33). The main idea behind SPIRiT is to formulate the reconstruction problem as an approximate minimization problem. An iterative algorithm is used, where the solution which is most consistent with the calibration and acquisition data is found. A scheme of SPIRiT is reported in Figure 28.

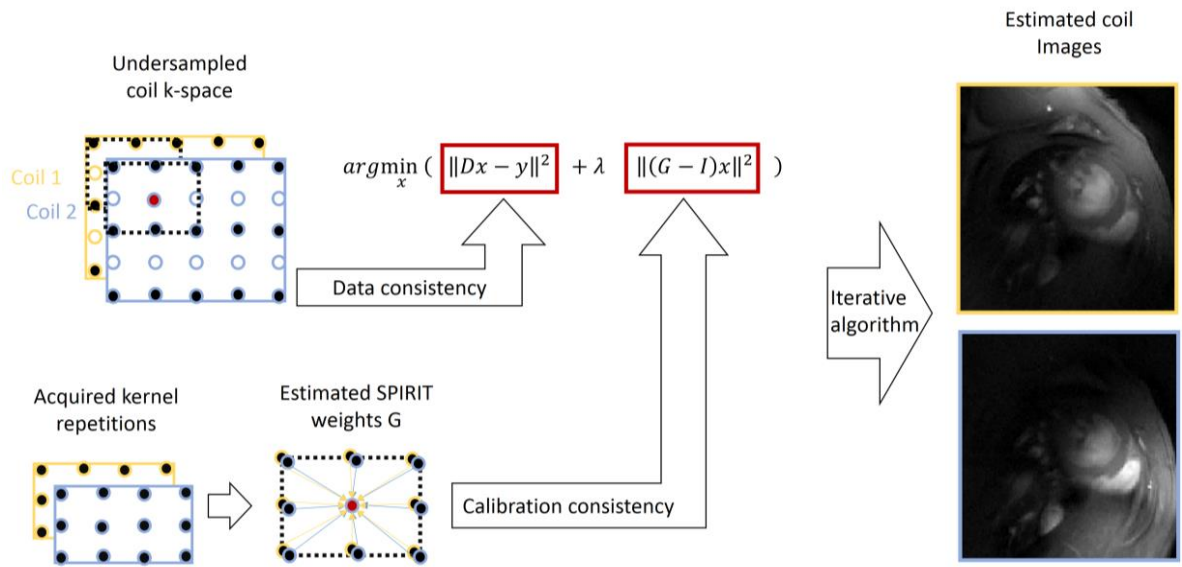

**Figure 28 SPIRiT generalises GRAPPA, here kernel repetitions are acquired to estimates weights, used to estimate non acquired k-space coordinates from the current value of both acquired and non-acquired k-space coordinates across coils. Individual coil images are reconstructed utilising an iterative algorithm where the solution most consistent with both acquired data and calibration is found. (D is the projector over acquired data, y is the acquired data, G is the matrix of SPIRiT weights, I the identity matrix).**

Due to the shift-invariance of $G$, SPIRiT can be formulated either in k-space or in the image space. This represents a great advantage over GRAPPA as any k-space trajectory can be used within the same formulation.

## 2.5 Chapter summary

This chapter introduced basics of MRI required for the subsequent chapters, starting from its physical principles and basic equipment. The generation of the signal, as well as how this can be spatially encoded to form an image was discussed. Imaging sequences and mechanisms to generate contrast were presented, as well as image reconstruction challenges, with a main focus on the techniques used throughout the thesis. Techniques for acceleration of MRI were also introduced, from parallel imaging to compressed sensing.

The following chapters will discuss how these theoretical concepts are put into practice in the context of cardiac imaging in mice. The next chapter will give an overview the techniques for cardiac MRI in mice found in recent literature.

# Chapter 3

# Cardiac MRI in mice: state of the art

Cardiac MRI is capable of assessing global (whole-organ) as well as local function, perfusion, viability, and metabolism within a single imaging session. Due to this, as well as its accuracy and reliability, the use of MRI as a diagnostic tool in cardiology has increased over the last decades (34). A summary of cardiac MRI techniques, including standard and experimental methods is reported in Table 1.

In the preclinical environment it is possible to utilise similar methodologies in order to test new treatments and translate these to patients. However, the implementation of cardiac MRI in small animals has encountered hurdles during its early development (35). When performing cardiac MRI in mice, the small size ($\sim$ 1 cm) of the heart and the high heart rates ($\sim$ 600 bpm) are a challenge. Due to these spatial and temporal constraints, development of cardiac MRI in mice has lagged behind its clinical equivalent. While MRI methods for measuring global functional parameters were developed in the 1980s, the first results of a mouse cardiac MRI study were reported in 1997 by Siri et al. (36). Since then, several techniques to image the heart have been developed or adapted for use in mice, and now the basic technologies for scanning mouse hearts are commercially available. Techniques for experimental MRI in mice will be briefly reviewed in the following chapter.

| Parameter measured | Imaging technique | Application | Target |
|---|---|---|---|
| **Global function** | Cine | Functional cine imaging | Volumes |
| **Local muscle function** | Tagging methods, velocity encoding, DENSE | Local tissue motion | Strain and regional motion |
| **Myocardial perfusion** | First-pass contrast, arterial spin labelling | Myocardial blood flow | Myocardial perfusion reserve, Inducible ischemia. |
| **Myocardial infarction/viability** | Late gadolinium enhancement | Fibrosis/scar | Infarct size, fibrosis. |
| **Acute ischemic injury** | Oedema imaging/ T2w | Myocardial oedema and salvage (when compared with LGE) | Acute Myocardial infarction, reversible vs. irreversible injury |
| **Myocardial oxygenation** | BOLD and T2* | Deoxygenated haemoglobin | Ischemia, haemorrhage, iron content |
| **Myocardial metabolism** | $^{1}$H and $^{31}$P spectroscopy | Quantitative metabolic assessment | Area at risk, genetic diseases. |
| **Tissue characterization** | T1 mapping, T2 mapping | Tissue composition, oedema. | Salvageable myocardium, scar. |
| **Extra cellular volume** | T1 mapping with contrast | Fibrosis/scar | Diffuse fibrosis, infarct quantification, subtle abnormalities. |
| **Microvascular obstruction** | Early gadolinium enhancement | No reflow | Acute Myocardial infarction |

Table 1 Main techniques to image the heart. Abbreviations: DENSE – displacement enconding with stimulated echoes; LGE – late gadolinium enhancement; DTI diffusion tensor imaging.

## 3.1 Global function

Cine MRI can be used to study the global contractile function of the heart in a healthy or diseased state by measuring the volumes of the ventricles in different phases (see Figure 29).

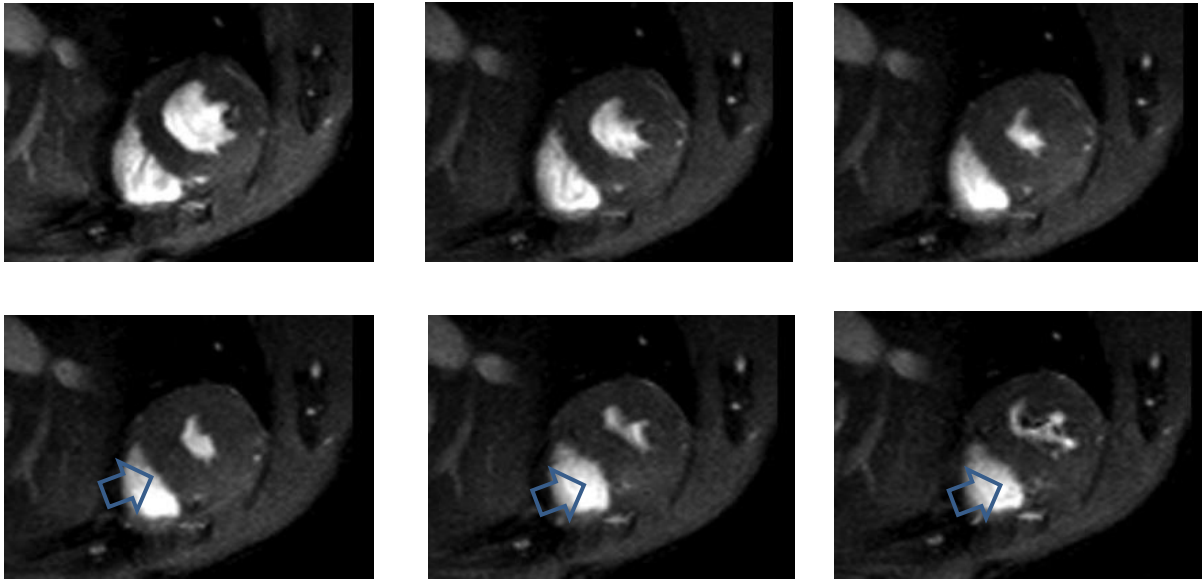

**Figure 29** Movie frames from short axis stacks in a mouse showing a congenital defect: a hole in the heart. While the start of systole looks "O" shaped, at the end of systole the myocardium shows a characteristic "D" shape indicated by the arrow, due to impaired LV pressure.

As shown by the example in Figure 29, impairment in heart function is better seen in some phases of the cardiac cycle with respect to others. Information from different phases can therefore be combined and used for diagnosis of different conditions affecting heart function. Global parameters for systolic function commonly measured with MRI are defined as follows (see chapter 1 for description of heart phases):

- **End Diastolic Volume (EDV) [l]:** Volume of the LV in end diastole.

- **End Systolic Volume (ESV) [l]:** Volume of the LV in end systole.

- **Stroke Volume (SV) [l]:** difference between EDV and ESV.

- **Ejection Fraction (EF) [%]:** ratio between SV and EDV.

- **Cardiac Output [l/min]:** SV times heart rate.

- **Left Ventricular Mass (LVM) [g]:** Volume of the myocardium. As the heart is incompressible this should be the same for all the cardiac phases.

During imaging, mice are usually anaesthetised with gaseous isoflurane, although other anaesthetics have been tested (37). Different anaesthetics have different hemodynamic effects on mice, therefore consistent protocols need to be in place in order to compare data between studies (38). Scouts are first acquired, followed by imaging in the short and long axes (39). Gating of the scans can be performed prospectively using ECG and respiratory sensors, or retrospectively utilising navigators (40) or the centre of the k-space with radial k-space acquisition (41). Although radial scans have advantages for acceleration purposes and are more robust to motion, the Cartesian scheme has achieved better gating quality in comparison (42).

To capture multiple frames quickly, cine MRI protocols utilise fast 2D gradient echo images covering the whole heart. In patients, cine MRI is usually performed with trueFISP, while mice studies commonly use gated FLASH (43) (44, 45) or FISP (46, 47). Once the gated images are acquired these are analysed in post-processing. Contours on the frames can be traced manually or semi-automatically (48) (49) (50) to obtain the relevant volumes. Measures from each slice are then combined to obtain whole-organ volumes.

Early studies compared MRI with invasive measurements of function showing agreement (51) and better reproducibility of MRI with respect to conductance catheters (52). MRI measurements of systolic function have also been compared with echo, showing that the former technique was more precise in assessing ejection fraction (53). Since the early 2000s cardiac MRI has been used to assess function in a number of disease models (see Figure 30 for an example in myocardial infarction). Several studies have utilised this non-invasive tool for longitudinal experiments to investigate ageing or disease progression.

Pharmacological stress tests have also been used to observe the inotropic response of the heart in several disease states (45) (54).

High resolution cine-MRI is also used for the assessment of muscle dimension. Left ventricular mass and wall thickness have been successfully measured in mice (36) (55, 56). Standard methods to measure LV mass have used cardiac and respiratory gating and multiple slices to cover the whole heart with no gap (39). This method is in excellent agreement with ex-vivo measurements (57) and it significantly outperforms echo as it does not require geometrical assumptions (36).

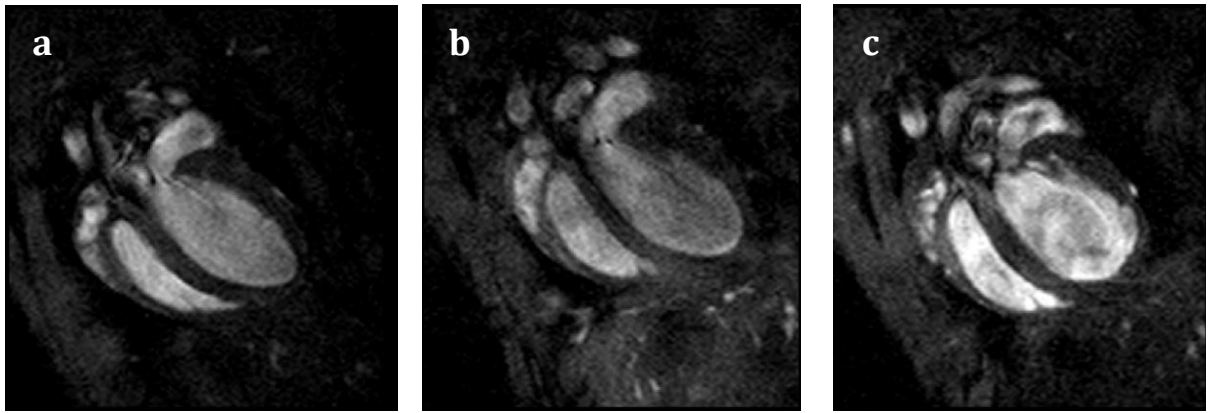

Figure 30 Cine-MRI can be used to assess heart function longitudinally: a) End-diastolic view of a mouse heart at baseline. b) End-diastolic view of the same mouse one day after induced myocardial infarction, by transient LAD occlusion. c) End-diastolic 4ch view of the same mouse four weeks after induced injury. Dilatation of the left ventricle indicates heart failure.

Although cine MRI is a highly accurate method, the long scan duration is a limitation of this technology (58). From the initial studies, where significant averaging was performed, advances in hardware and sequence optimization have made standard protocols for cine MRI much shorter, and typical protocols have a duration of about fifteen minutes. In addition, to further accelerate the acquisition, techniques used in clinical scanners such as parallel imaging and compressed sensing can be adapted in mice. Studies in mice have used either parallel imaging (58, 59) or temporal compressed sensing (60) to accelerate cine imaging.

### 3.1.1 Other techniques for global function *in vivo*

The most commonly used technique to measure heart function is two-dimensional echocardiography, due to availability and cost, as well as efficiency. This technique can be used for real-time acquisition in rodents for fast high throughput screening, but is operator dependent and relies on geometrical assumptions of heart geometry.

To compare MRI and echocardiography techniques in mice within our centre, an experiment was performed utilising *n=5* wild type mice (aged 10 wks) and *n=5* chronic heart failure mice (aged 14 wks, the heart failure was a consequence of acute myocardial infarction, four weeks after ischemia and reperfusion injury as described in 7.2.3). MRI was performed as described in Chapter 4, while echo was performed and analysed by an independent operator. Briefly, echo was performed with 2% isoflurane, using a Vevo 770 ultrasound system (Visual Sonics, Toronto, Ontario, Canada). Hearts were visualised in the two dimensional short-axis plane and analysis performed in M-mode in the consistent plane of the papillary muscles. Ejection fraction (EF) was calculated from end-systole and end-diastole measurements in at least three repeated cardiac cycles.

A Bland-Altman plot was used to assess agreement between echo and MRI (61). Results, reported in Figure 31, show the degree of agreement between the two techniques, with echocardiogram overestimating ejection fraction, especially in chronic heart failure mice where the geometric assumptions of echo lose validity due to remodelling.

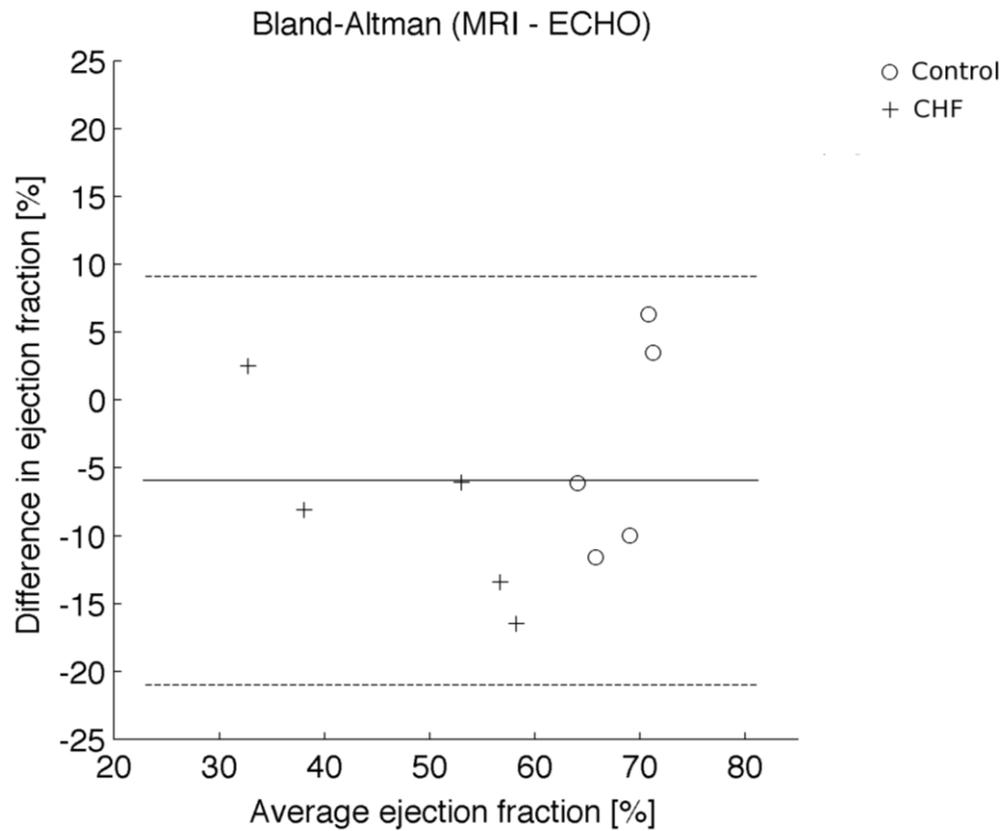

**Figure 31 Bland-Altman plot comparing echocardiogram and cine MRI in assessing ejection fraction. The graph includes control mice and chronic heart failure (CHF) mice.**

Micro CT is another technique used to assess global heart function in mice, which is less user-dependent compared to echo, and can still achieve high throughput acquisition for heart function in three dimensions in less than 80 seconds (62). However, despite its promise, CT has not seen widespread adoption in cardiac studies in rodents, perhaps due to concerns over the radiation dose involved.

MRI is considered the gold standard measurement of systolic function (53), however it suffers from long scan durations when compared to echo or CT.

## 3.1.2 Meta-analysis: aging of the C57BL6 mouse heart

Since the early 2000s the majority of the studies have used C57bl6 mice with similar regimens of gaseous isoflurane (1-2 % in 1l/min $O_2$). Here left-ventricle data from control groups of studies from different centres are grouped together in order to derive a time evolution of the typical C57bl6 mouse heart. Unpublished data from 27 mice acquired during this doctoral training in Cambridge are also included (the method used to obtain the measurements is described in Chapter 4). Different study protocols and analysis methods generate different parameters, and these differences are visualised in the data reported here. This emphasizes the necessity of fixed rules and protocols to compare data within a centre. Although there is some centre-induced variability, typical parameters can be extracted from these data that can be used as a reference when planning studies or evaluating results. The individual volumes change significantly with age, but ejection fraction remains relatively stable between 8 weeks and 26 weeks of age for a mouse. As the range for healthy patients is similar to the range in healthy mice, this parameter can be used for translational comparisons.

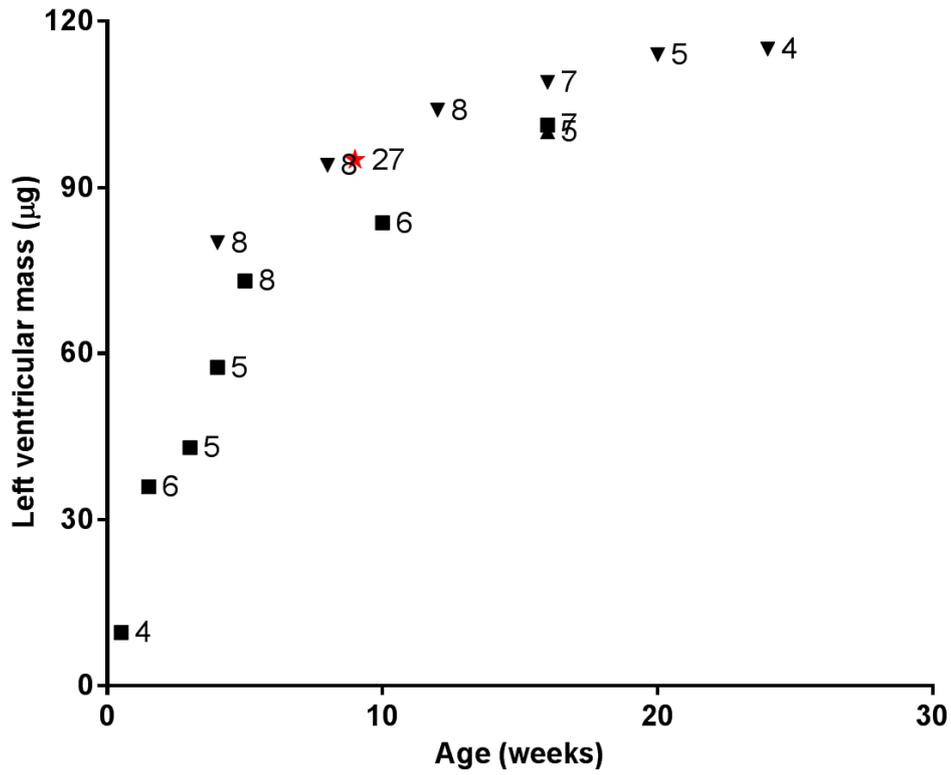

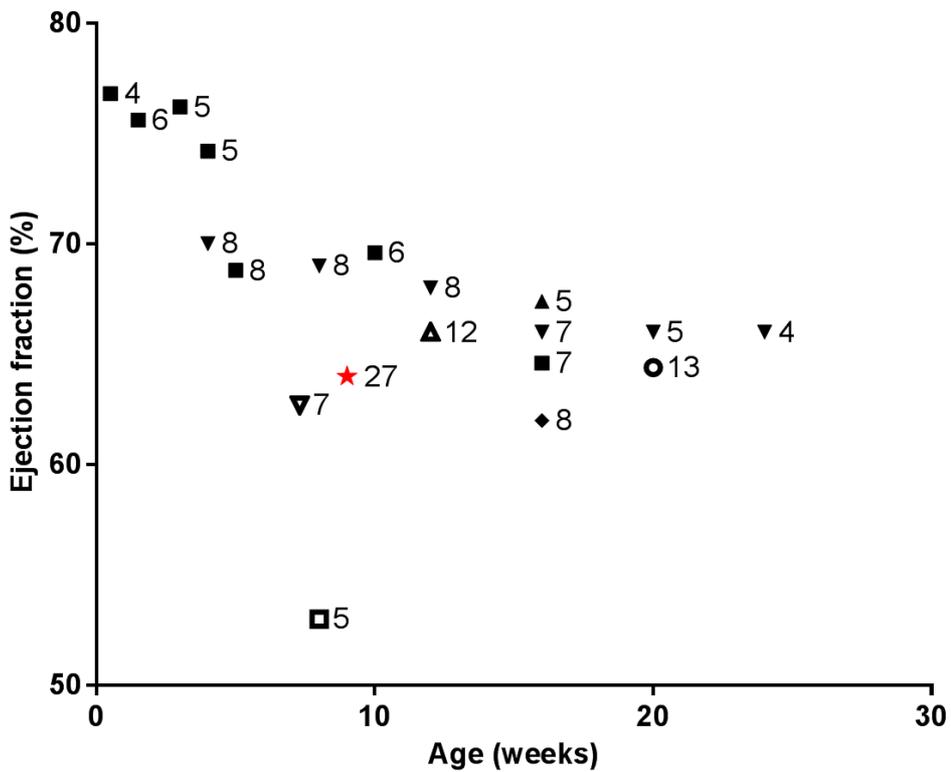

The figure continues on the next page.

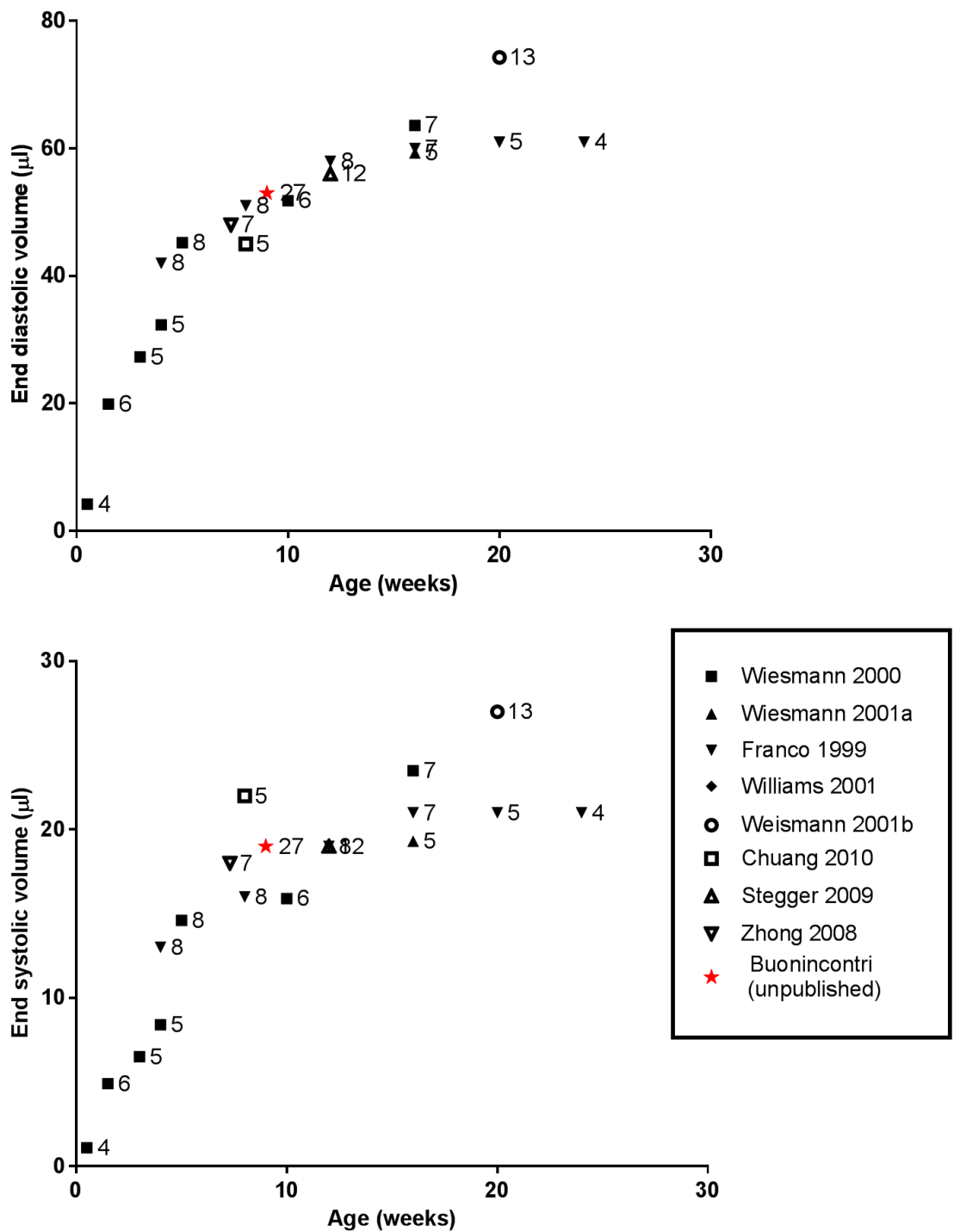

**Figure 32 Functional parameters from the ageing C56BL6 mouse. Next to each symbol the number of C57BL6 mice utilised in the experiment is reported. The legend reports reference of the relevant papers.**

### 3.1.3 The right ventricle

In the past, research on the cardiopulmonary system has lagged behind compared to work on systemic circulation. Procedures such as Fontaine surgery and cauterization of the RV lateral wall in dog models have led clinicians to underestimate the role of the RV (63). Only recently have technical developments in echocardiography and MRI led to renewed interest in the RV, seen in the broader context of the cardiopulmonary system (63). Although there is limited information on the role of RV function, impairment plays a significant part in many diseases, where the extent of RV dysfunction is a predictor of outcome. In 2006, the National Heart, Lung, and Blood Institute identified RV physiology as a priority area in cardiovascular research (64).

Imaging the RV remains a challenging task because of fibre orientation, complex geometry and contraction pattern (65). MRI is an accurate tool for measuring RV function in humans (66), as well as in mice (67), since it is capable of high spatial, temporal resolution and of arranging oblique slices to form 3D stacks. For this reason, studies investigating right ventricle pathology non-invasively in mice have utilised MRI (68) (69).

## 3.2 *Motion and strain estimation*

Most forms of heart disease involve some extent of muscle function deficit, and changes in local contractility can lead to global organ failure in the longer term. MRI can be used to measure tissue displacements and this information can be converted to stress and strain measurements to assess contraction and relaxation of the muscle.

The first methods to measure tissue displacement in the myocardium were based on tissue tagging (70). This method encodes a modulation (tags) on the image at a particular phase of the heart cycle. With heart motion, the tags move with the muscle, showing the spatial deformation. Variations of this technique include spatial modulation of magnetization (SPAMM) (71) or delay alternating with nutations for tailored excitation (DANTE) (72). Several studies have successfully adapted tagging techniques in mice.

One limitation of tagging techniques is that the data analysis is lengthy and requires significant user interaction. To overcome this limitation, harmonic phase (HARP) was developed (73) and adapted for mice (74). Another limitation of tagging techniques is spatial resolution, as the distance between tag lines must be larger than image resolution. In contrast, phase encoded techniques utilise phase information to derive voxel-wise velocity or displacement maps.

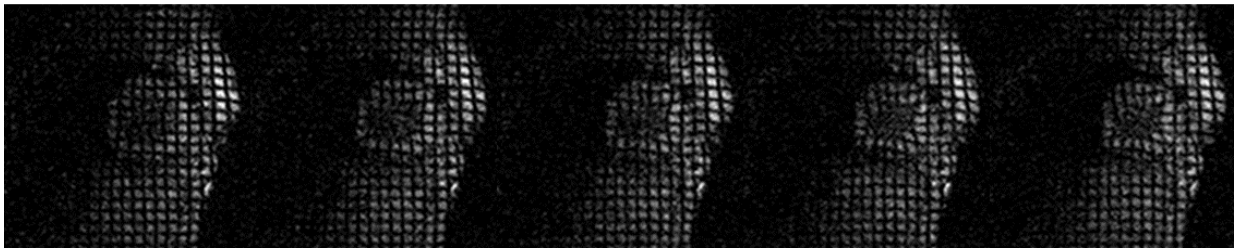

**Figure 33 Tagging in the left ventricle of a mouse through systole.**

Velocity-encoded phase-contrast methods have been successfully applied to mice, demonstrating pixel-wise measurements of LV myocardial velocities (75). However, this technique is less efficient in assessing strain than techniques that use displacement measurements (38).

Displacement encoding with stimulated echoes (DENSE) MRI encodes spatial displacements in the image phase, by applying two pulsed gradients (76). This technique has been successfully adapted in mice both in 2D (77) and 3D (78) (79). In addition, highly-automated analysis has been developed for temporally-resolved DENSE imaging (80).

All of the tissue deformation methods described above can be utilised to extract strain values, which can be used for local or global metrics of tissue strain. Global values from tissue deformation imaging have recently shown high prognostic value for remodelling in acute myocardial infarction (AMI) patients, and it has been suggested that these might more closely reflect myocardial contractility than traditional measures of systolic function (81).

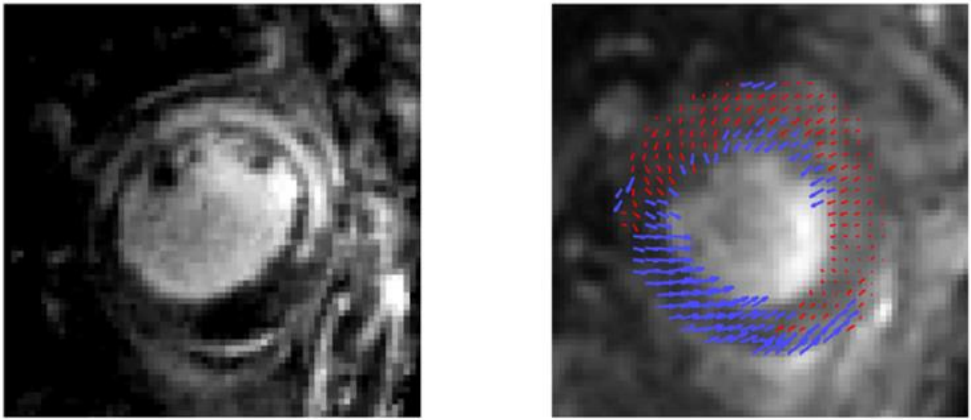

**Figure 34 LGE (left) and DENSE (right) on the same slice 24 hours after I/R injury. Hypokinetic areas marked in red (frozen myocites), extend beyond the infarct.**

### *3.3 Perfusion*

Perfusion measurements in the myocardium play an important role in the early detection of occlusive coronary disease. First-pass perfusion techniques observe the transit of a gadolinium-based contrast agent within the myocardium. Here, T1-weighted fast gradient-echo images are acquired after injection, measuring different amounts of tissue enhancement dependent on gadolinium concentration, which can be used to extract quantitative perfusion parameters (82, 83). Clinically, if this exam is performed in the presence of a vasodilator, the response of the coronary arteries can be probed and the area of inducible ischemia can be delineated in order to plan intervention.

Although this technique presents some extra complications in mice due to faster heart rate, lower SNR and lower bolus injection reproducibility, quantitative perfusion studies have been performed recently (84).

As an alternative to first pass perfusion, arterial spin labelling (ASL) techniques can be used to measure myocardial blood flow. Here, water within blood is used as an endogenous contrast agent comparing two T1 maps, obtained respectively with a global and a local inversion pulse. The difference between the two maps represents the blood that perfuse the slice between the inversion and the excitation. Techniques to perform ASL in the myocardium of rodents have been used in a number of studies (85-87).

## 3.4 Viability

Myocardial viability is another important index in the assessment of tissue following ischemia. After ischemia, a complex cascade of events leads to the development of a scar. As the extent of nonviable myocardium is a predictor for the severity of heart failure, an accurate measurement of the non-viable area is required. Techniques to assess myocardial viability are usually performed using late gadolinium enhancement (LGE). Here, a Gadolinium contrast agent is injected in the peripheral circulation. After the first pass of contrast agent perfusing the tissue with arterial blood, healthy areas will have normal washout characteristics while the contrast agent will remain longer in the infarct area, and be washed out only later. T1-weighted MRI performed in the late phase of contrast kinetics shows bright infarcted areas (see Figure 35). Viability measures obtained with MRI have shown a good correlation with single photon emission computed tomography SPECT and histology (88).

 In patients, LGE is commonly performed with TI-optimized segmented inversion recovery (IR) within a breath-hold, where images are typically acquired 10-30 minutes after intravenous injection of contrast (0.1-0.3 mmol/kg). Techniques to perform this in mice have used different approaches due to different heart rates and kinetics of contrast. Some methods have used segmented inversion recovery measurements (89-91), while others have utilised cine MRI with high flip-angles to generate heavy T1-weighting (92). Recent accounts, including our method (discussed in Chapter 7), have shown that multi-slice inversion recovery is highly contrast-efficient (13, 93).

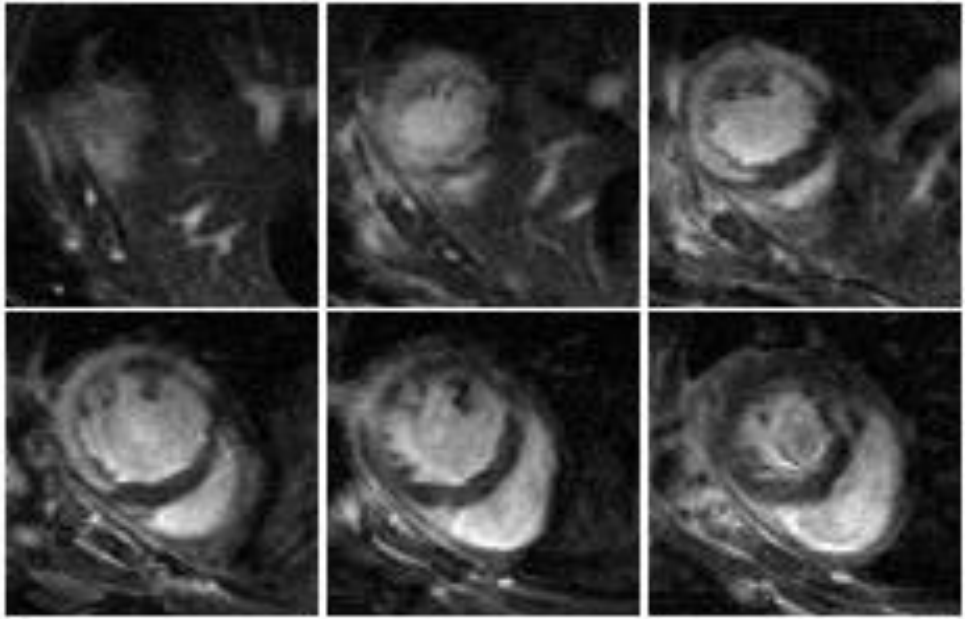

**Figure 35** Late gadolinium enhancement in short-axis slices of the mouse heart (from a myocardial infarction study).

## *3.5 Oedema and haemorrhage*

In ischemia and reperfusion injury, only part of the area at risk supplied by the blocked artery becomes necrotic. In MRI, an elevated T2 due to tissue oedema is characteristic of the area at risk and can be detected with MRI. The difference between this area and the LGE is the salvageable myocardium. In mice, techniques for T2 imaging of the mouse heart have been developed based on spin echo (94) or T2 preparation (95). Techniques for T2 mapping have also been developed for imaging fibrosis (96).

Another important MRI parameter correlated with disease is T2*. An elevation in T2* values in the myocardium is an indication of haemorrhage in myocardial infarction (97). In addition, this parameter correlates to iron content in thalassemia (98). Due to the technical difficulties associated with inhomogeneities, a T2-prepared SSFP BOLD sequence was proposed to overcome the limitations of traditional BOLD, and demonstrated in a dog model (99). In mice, studies have attempted to perform blood oxygen level dependent (BOLD) MRI in order to obtain T2* weighted images (100).

### 3.6 Angiography, tissue structure and metabolism

Magnetic resonance angiography in small animals is usually performed with time of flight methods (see 2.2.7). Vascular anatomy can be visualised in the images and rendered in 3D. In mice this technique has been rarely used to look at coronary anatomy due to the small size, however reports have been produced attempting this technique (101).

Fibre direction in the heart can be probed with diffusion tensor imaging (DTI), in order to assess structure development and disruption due to disease. Although this method has been mostly experimented on ex-vivo samples, it holds promise for the assessment of fibre integrity in response to injury (102). In addition, the development of muscle fibres in the healthy or diseased foetal heart could be assessed with this method.

Tissue metabolism can be probed utilising magnetic resonance spectroscopy (MRS) in the heart. The relative abundance of metabolites can be used as a marker for tissue energetics in genetic diseases as well as in the area at risk of infarction models. Although *in vivo* hydrogen spectroscopy studies have been performed in the mouse heart to obtain information on metabolites, phosphorus has been the most investigated nucleus in cardiac MRS (103). By measuring $^{31}$P spectra, it is possible to derive phosphocreatine to adenosine triphosphate ratios, which can be used as a biomarker of cell metabolism. To guide the prescription of the spectroscopy in vivo, $^{31}$P spectroscopy usually employs dual-tuned probes (104).

### *3.7 PET/MRI*

Magnetic resonance imaging (MRI) gives excellent views of the heart non-invasively with clear anatomical detail, which can be used for accurate global and local functional assessment. Contrast agents can provide basic measures of tissue viability but these are non-specific. Positron emission tomography (PET) is a complementary technique that tracks radioactively-labelled molecules to assess cell metabolism. PET is highly specific for molecular imaging, but lacks the anatomical detail of MRI. Used together, these techniques offer a sensitive, specific and quantitative tool for the assessment of the heart in disease and recovery following treatment. Experimental studies combining these techniques have been performed in mice. Stegger *et al* utilised sequential acquisition to validate a method for assessing function with PET (105), while Buscher *et al* demonstrated the first simultaneous measurements of PET/MRI in the mouse heart utilising a PET insert (106). Lee *et al* combined sequentially acquired LGE MRI with FDG PET to assess tissue inflammation in follow ups after myocardial infarction, validating this as a novel clinical tool (107). As shown by this study, PET/MRI technologies in mice offer promise for the validation of new diagnostic markers. Although most of the first PET/MRI studies utilised FDG (the method developed within this thesis is reported in Chapter 8), this field opens to the validation of new tracers with specific molecular bindings for diagnosis and follow-up of disease.

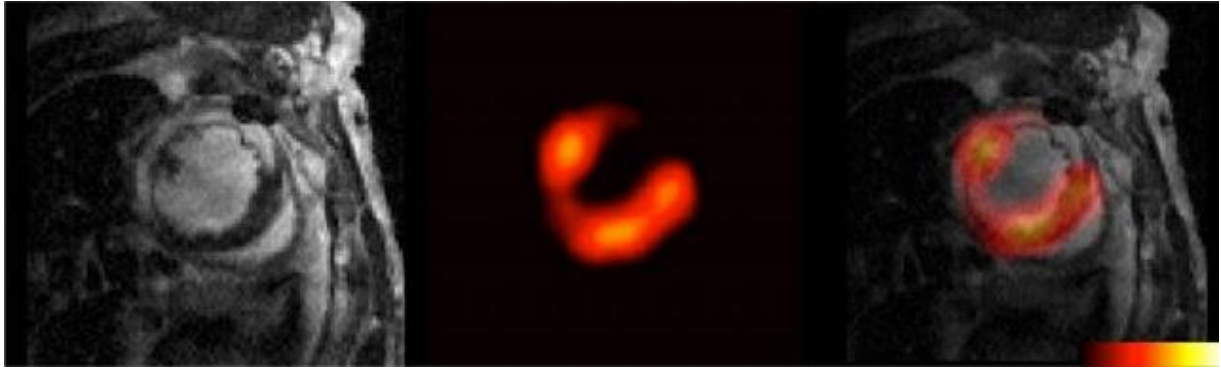

**Figure 36 Overlaid LGE-MRI and of an infarcted mouse heart obtained during a drug study. FDG-PET hypointense areas match LGE-MRI hyperintense regions.**

### 3.8 Chapter summary

In this chapter methods to image the mouse heart at different levels were introduced in the context of recent literature. The review covered global function, local contractility, perfusion, viability, metabolism as well as structure. This concludes the introductory section of this thesis. The next section will focus on standard protocols and innovations for measuring global heart function in mouse models of disease.

# Section 2: Heart function

# Chapter 4

# Current techniques for cardiac MRI in mice

As seen in previous chapters, performing cardiac MRI in mice is a challenging task given the small dimension and high heart rates. To obtain a good outcome, particular attention is paid to some critical steps of the imaging and data processing protocol. First, a reliable monitoring signal is obtained, as incorrect gating produces errors in the volumes estimates. Secondly, slice planning is performed. Sub-optimal geometry produces inconsistent data that cannot be corrected in post-processing. Last, segmentation of MRI data, which is performed in post-processing, requires rigorous and fixed rules in order to avoid inconsistency in the volume estimates.

This chapter focuses on practical aspects of MRI in the mouse heart. Inter and intra-observer agreements are measured for the MRI method, and this is applied to current problems in experimental cardiology.

## 4.1 Animal anaesthesia, positioning and monitoring

For the experiments performed here, volatile gas anaesthetics are used. Before starting the procedure, the mouse is anaesthetised in an induction chamber utilising 3% isoflurane in 1l/min O2. Within the first five minutes of induction the respiration rate of the mouse drops significantly, then the mouse can be moved and carefully placed on the MRI bed, delivering 1-2 % isoflurane in 1l/min O2 with a nose cone for anaesthesia maintenance as required. Mice are breathing freely. During the scan, respiration rates are kept between 20 and 70 cycles per minute.

The animal is positioned prone above the receiver coil. The heart is roughly 0.5 cm below the forepaws. Accurate placement is crucial in order to maximise the signal intensity. Initially, an approach with the animal supine and the coil on top was used, which despite the ease of placing the coil exactly centred on the heart without moving the animal, was less practical and more time consuming, as care had to be taken not to apply extra pressure on the chest when securing the coil, especially when mice had undergone a chest surgery.

For reliable monitoring signals, sensors are carefully placed and their correct functioning is verified prior to starting the imaging session. To monitor respiration, a small pillow is placed under the chest slightly below the diaphragm. For ECG monitoring, three electrodes are located on the anterior paws and on the left rear paw, making sure that the palm of the toe is completely open. The pair of electrodes achieving the best signal quality is used for ECG. Cables are twisted together avoiding corruption the ECG signal. In diseased mice ECG signals can be weak or abnormal, which makes reliable gating difficult. In addition, the signal quality can degrade during the exam. For these reasons, ensuring high quality for the ECG signal before the examination is pivotal.

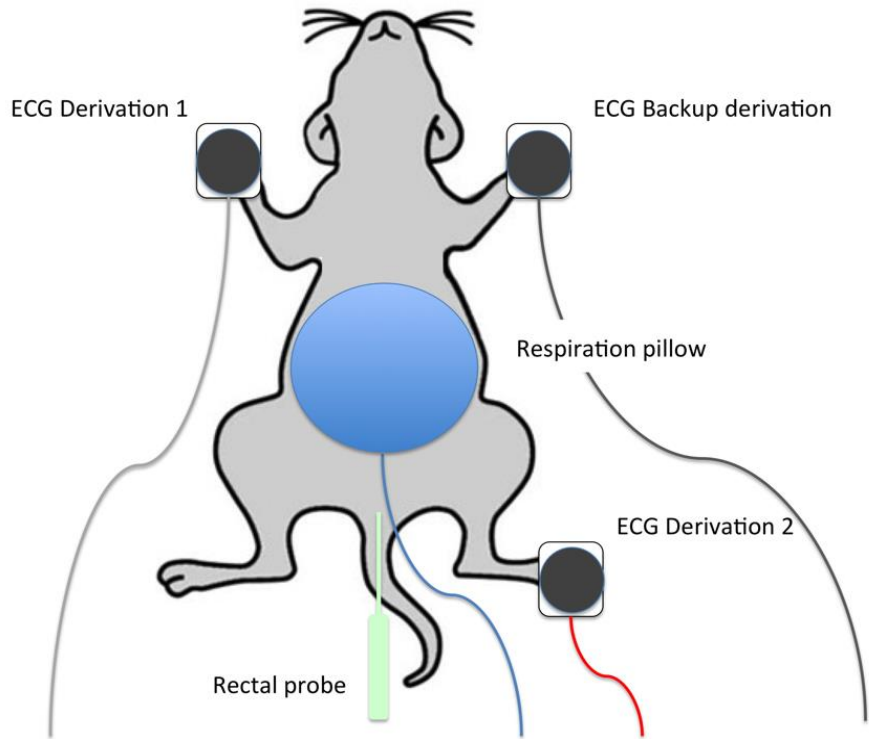

**Figure 37** Monitoring setup.

To avoid changes in the position during the scans, the electrodes are firmly attached to the bed using tape. The electrodes and the nose cone will hold the animal in place during the whole procedure. Temperature is measured with a rectal probe thermometer, placed using a lubricated cover. A water-heated blanket is positioned over the mouse, encapsulating the monitoring and coil leads to maintain body temperature. If the temperature is not kept constant during the experiment, the heart rate changes as a result, leading to incoherent timing of the frames over the course of the experiment. Once good signals are detected for monitoring, the mouse is positioned in the magnet using an automatic bed. A laser is used to position the heart in the centre of the bore, where there is maximal field homogeneity. To identify the heart, the forepaws can be used as a landmark.

For maximum SNR, fine tuning and matching of both the transmitter and the receiver is obtained adjusting the capacitance of two trimmer capacitors included in the coils circuit.

Prior to the acquisition of any imaging sequence, the following standard adjustments are performed, including:

- <u>Central frequency adjustment</u>: the spectrometer is set to acquire at the present resonant frequency of the system.

- <u>Shim adjustment</u>: the line width of the spectrum is minimized by acting on the currents of shim coils. That makes the main static field as homogeneous as possible.

- <u>Transmit gain adjustment</u>: The transmit gain is set to obtain a 90° pulse, used to calibrate the power needed to obtain a given flip-angles later in the experiment.

## 4.2 Cine MRI acquisition

Accurate positioning of the slices for acquisition is vital for quantitative analysis. This section describes how this can be achieved.

### Scout images

The first challenge for cardiac imaging of small animals is to reliably identify the heart and plan a fast scout with enough resolution and SNR to identify the relevant landmarks in order to plan the cine imaging. The following paragraph will briefly describe how the relevant views are obtained starting from the localizers.

#### Tripilot, 6 cm fov

The aim of this first quick scout is to check the positioning of the coil and localise the heart among the other organs. In this image, the flow artefacts arising from blood flow are helpful to identify the heart. If the heart is correctly placed in the centre of the receiver coil, it will be roughly in the centre of the RF field profile, as shown in Figure 38.

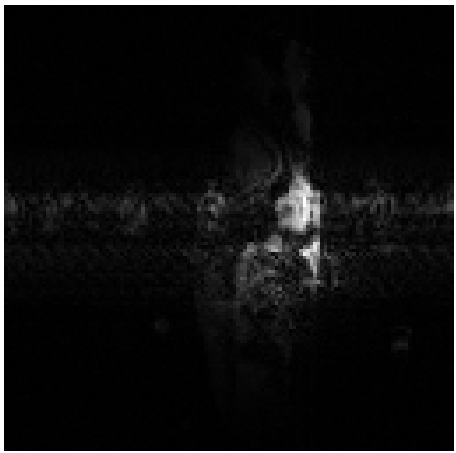

**Figure 38 Image from a Tripilot including the characteristic flow artifacts** .

If the animal is not correctly positioned in one of the views, the bed is taken out of the magnet and positioned again, with major adjustments repeated.

*Gated multipilot, 3cm FOV*

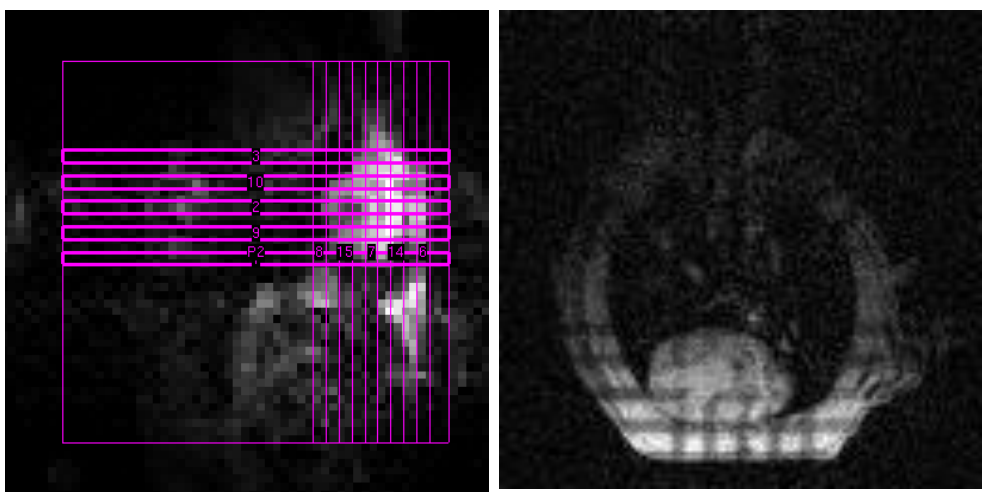

**Figure 39** Multipilot planning on a tripilot and one example of the result in an axial plane.

This scan is acquired in the scanner frame of reference, i.e. axial sagittal and coronal, as opposed to the heart frame of reference, i.e. long axis, short axis. In the multipilot, five slices per orientation are acquired with ECG gating. The aim of this scan is to get the relevant landmarks to plan the long and short axis views.

## Four chamber view (4ch)

A four chamber view scan is planned using the multipilot, and should cut through the apex and the tricuspid and mitral valves, showing the whole 4 chambers during the heart cycle.

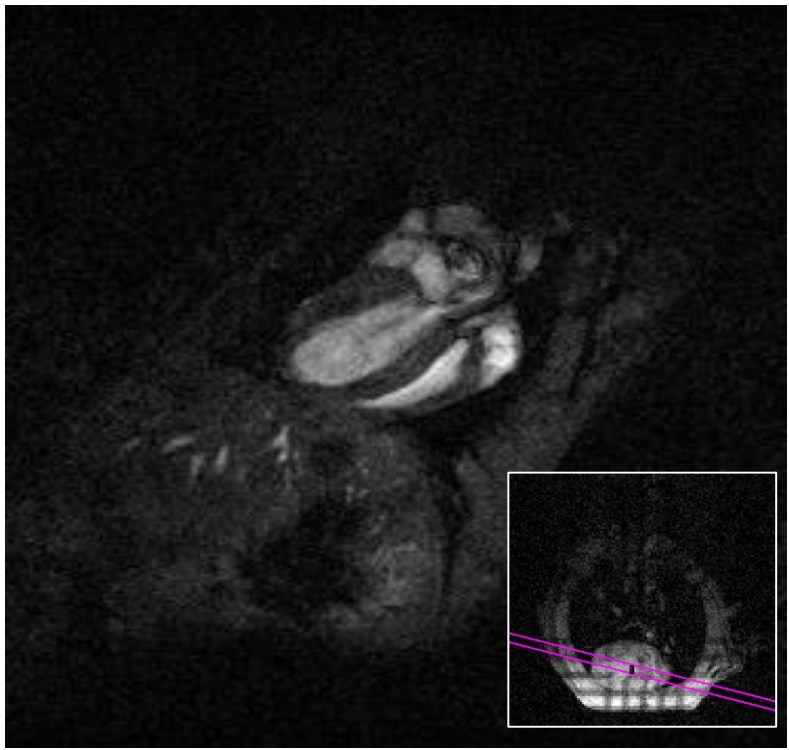

**Figure 40** A frame from a 4-chamber view with an example of the planning on the multipilot

## Two chamber view

A two chamber view is planned using the multipilot and the 4ch view, it should cut through the apex and the tricuspid valve, showing the left atrium and ventricle during the heart cycle. At this point, the geometry of the two long-axis views can be cross-checked. If the slice planning is poor, the scans can be repeated.

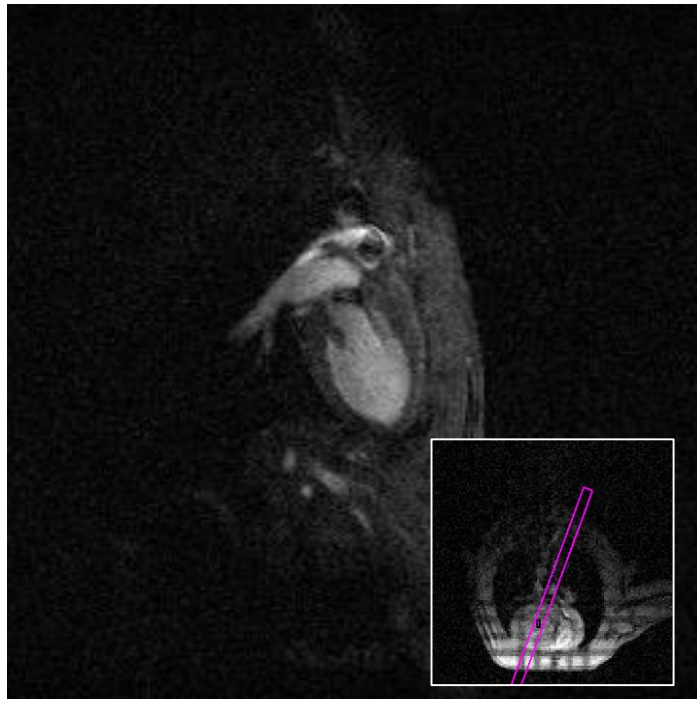

**Figure 41** A frame from a 2-chamber view with an example of the planning on the multipilot.

## Short axis view

A stack of short-axis slices is planned orthogonal to both the four-chamber and to the two-chamber view. To assess LV and RV function the slices should start from the first apical slice without blood pool and be repeated with a fixed distance until the first basal slice without any right ventricle. Slices should be equally spaced with no gaps.

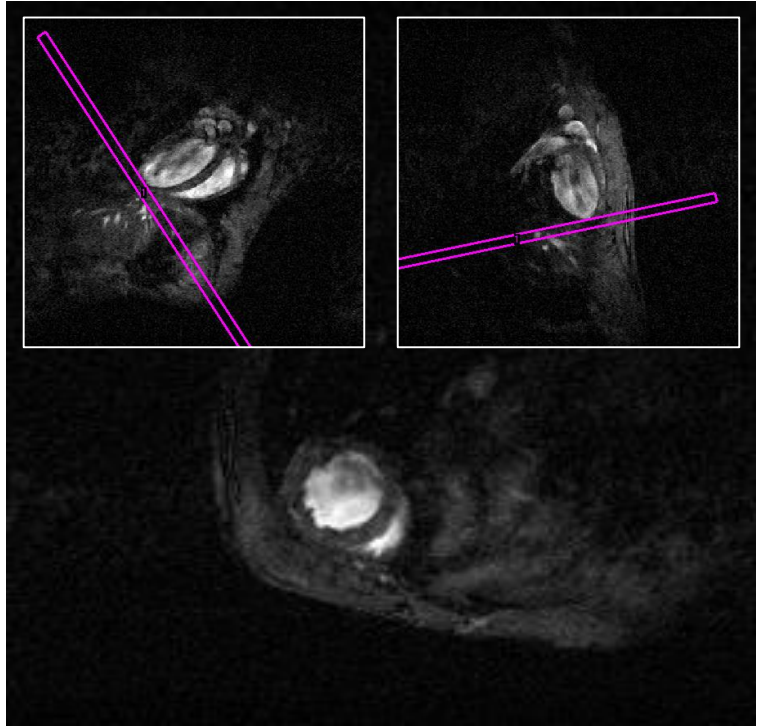

**Figure 42** A frame from short-axis stacks, including the planning of the first slice on 2-ch and 4-ch views.

## 4.3 Cine MRI segmentation

To achieve a semiautomatic segmentation of the cine images, these are loaded in Segment v1.9 (48). First, the automatic tools for segmentation are ran. Starting from a point inside the ventricle, these will inflate the estimated ventricle walls until the best match is found. Then, LV and the RV volumes are manually delineated at end systole (ES) and end diastole (ED), defined respectively as the frames with the maximal and with the minimal global LV volume. Papillary muscles and trabeculations are excluded throughout. Image contrast is not adjusted between subjects, as inconsistency in the contrast leads to a different interpretation of partial volume effects. The epicardium is delineated at ES and ED for LV mass calculations, as shown in Figure 43. The cardiac muscle is incompressible, so the LV mass must be consistent in ES and ED. If the difference is more than 5% operator error is likely and these should be reviewed.

At the base of the ventricles, a straight line is used to discriminate ventricles from atria as shown in Figure 43, identifying the angle of the slice with the help of the long-axis views. After the segmentation has been performed on the short-axis slices, it can be cross-checked on the long-axis views.

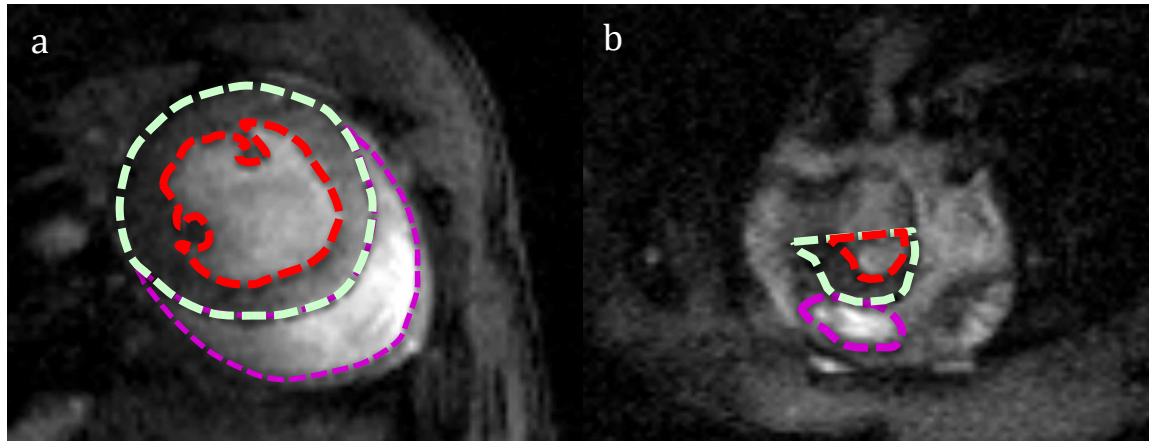

**Figure 43 a) Segmentation of a mid-ventricular short-axis slice. b) Segmentation of a basal short-axis slice. Endocardial and epicardial borders are identified and segmented for each short-axis slice. Fixed are applied to achieve reproducible volume estimates.**

## *4.4 Intra-observer and inter-observer variability*

To measure inter-observer and intra-observer variability, C57 male mice (n=7, age 10 weeks) were scanned with the cine-MRI protocol described above. Segmentation was performed twice to assess intra-observer variability. For inter-observer variability segmentation was performed once by an independent operator.

### 4.4.1 Methods

MRI was performed at 4.7 T with a Bruker BioSpec 47/40 system (Bruker Inc., Ettlingen, Germany). A quadrature birdcage coil of 12cm was used for signal excitation and a four-channel cardiac receiver coil for signal reception. Animals were positioned prone. After initial localization images, 4-chamber and 2-chamber views were acquired. Using these scans as a reference, short axis slices were arranged perpendicularly to both the long-axis views to cover the LV (FISP, TR/TE 6 ms/2.1 ms, 13-20 frames, 3.5 cm FOV, 256x256 matrix, 1 mm slice thickness, bandwidth 78 kHz, flip angle 20°, NEX 2). Full LV coverage was achieved with no slice gap with 8-10 slices.

Data were anonymised and analysed twice by the author (operator 1). To assess inter-observer agreement data was analysed by an independent assessor (operator 2). Inter and intra observer bias were reported as well as inter and intra observer 95% confidence intervals (CI), obtained by multiplying the standard deviation of the difference between values by the factor 1.96 (61).

## 4.4.2 Results and discussion

Inter and intra-observer comparisons are reported in Table 2. Estimates of LV volumes were relatively unbiased between observers, while RV volumes and LV mass were more subjective. Rules utilised to segment the hearts achieved high consistency both inter and intra observers. Values are in line with similar accounts in the literature (108).

|  | Group mean ± s.d. | Inter-observer bias | Inter-observer 95% CI | Intra-observer bias | Intra-observer 95% CI |
|---|---|---|---|---|---|
| LVM (µl) | 101 ± 16 | 13 | 3.6 | 0 | 6.0 |
| LVEDV (µl) | 60 ± 13 | -0.7 | 3.0 | 0 | 3.2 |
| LVESV (µl) | 16 ± 4 | -1.6 | 2.6 | -0.9 | 1.4 |
| LVSV (µl) | 43 ± 9 | 0.6 | 2.6 | 0.5 | 3.8 |
| LVEF (%) | 72 ± 3 | 2.6 | 1.4 | 1.1 | 2.2 |
| RVEDV (µl) | 49 ± 9 | -3.4 | 3.7 | -1.3 | 3.3 |
| RVESV (µl) | 11 ± 3 | -2.6 | 4.5 | -1.4 | 2.5 |
| RVSV (µl) | 38 ± 7 | -0.9 | 3.0 | -0.14 | 5.5 |
| RVEF (%) | 77 ± 5 | 4.7 | 4.7 | 0.9 | 5.5 |

**Table 2 Intra and inter observer comparisons. Group mean and standard deviation values are reported as obtained by operator 1 (n=7).**

### 4.4.3 Conclusion

Mean values obtained from different observers should be compared with care due to significant biases originating by different interpretation of anatomy. The rules utilised to segment MRI of mouse hearts achieve high consistency and reproducibility both inter and intra-observers.

To demonstrate practical applications of the technique described in this chapter, three studies are reported in the following.

## 4.5 Application 1: cine MRI in a cardiomyocite-specific KO

In this section results are presented measuring heart function in a cardiomyocite-specific Complex I knock-out (KO) model. The data shown here represent an example of how the techniques described in this chapter can be used to detect hypertrophic cardiomyopathy in a transgenic mouse model *in vivo* with high sensitivity.

### 4.5.1 Background

The Ndufs4 gene encodes a 18 kDa subunit of complex I that is not directly involved in electron transport but plays a role in assembly or stability of the entire complex. Ndufs4-null mice have been established as a model for Leigh syndrome and manifest severe developmental and motor problems, and typically die by post-natal day 55 (109, 110). In contrast, a recently developed heart-specific Ndufs4-null mouse strain shows no visible pathology to one year of age despite significant inhibition of complex I activity (111). Recent studies have demonstrated that a transient inhibition of complex I is protective in acute myocardial infarction (112) (see also 7.5). The present experiment employed the heart specific Ndufs4-null mouse as a model for persistent inhibition of mitochondrial complex I activity to explore the effects of chronic inhibition of this key node of myocardial function.

## 4.5.2 Methods

MRI was performed at 4.7 T with a Bruker BioSpec 47/40 system (Bruker Inc., Ettlingen, Germany). A quadrature birdcage coil of 12cm was used for signal excitation and a four-channel cardiac receiver coil for signal reception. Animals were positioned prone. After initial localization images, 4-chamber and 2-chamber views were acquired. Using these scans as a reference, short axis slices were arranged perpendicularly to both the long-axis views to cover the LV (FISP, TR/TE 6 ms/2.1 ms, 13-20 frames, 3.5 cm FOV, 256x256 matrix, 1 mm slice thickness, bandwidth 78 kHz, flip angle 20°, NEX 1). Full LV coverage was achieved with no slice gap with 8-10 slices. Both male (n=5 controls, n=5 KO) and female mice (n=3 controls, n=2 KO) at 14 $\pm$ 3 weeks of age were used in this experiment.

## 4.5.3 Results

Although the genotype did not impact on behaviour and external appearance, severe heart disease was visible from the MRI scans. The appearance of a typical transgenic and a typical control are shown in Figure 44, showing an enlarged heart for the transgenic animal, with severely impaired contraction capabilities. Quantitative measurements confirmed this, showing a reduction in function driven by an increase in systolic volumes and LV mass.

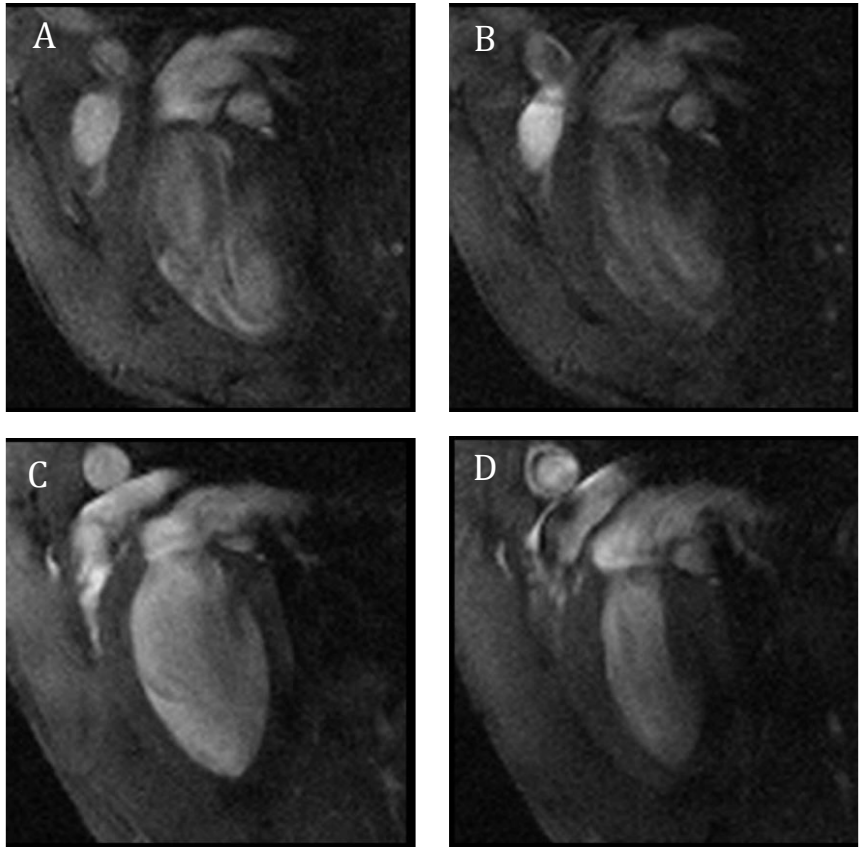

**Figure 44 Long-axis views: A) End diastole in C1KO mouse, B) end systole in the same mouse. C) End diastole in wildtype control, D) end systole in the same mouse. The C1KO mouse heart fails to contract, furthermore, the turbulent flow of blood within the ventricle results in a visible reduction of the contrast between the muscle and the blood.**

|  | **Controls (n=8)** | **C1KO (n=7)** |
|---|---|---|
| LVM(µl) | $96 \pm 3$ | $115 \pm 7$ * |
| LVEDV(µl) | $56 \pm 2$ | $63 \pm 4$ |
| LVESV (µl) | $19 \pm 2$ | $44 \pm 5$ *** |
| LVSV(µl) | $37 \pm 1$ | $19 \pm 3$ *** |
| LVEF(%) | $67 \pm 3$ | $31 \pm 5$ *** |

**Table 3 MRI-derived left ventricular volumes, mean ± SEM. * p <0.05, *** p<0.001. (*t*-test)**

### 4.5.4 Conclusions

Cine MRI was successfully used to detect hypertrophic cardiomyopathy in a cardiomyocite-specific knockout. The data shows that in contrast to the protection observed by transient inhibition of mitochondrial complex I (112), persistent disruption by NDUFS4 ablation causes hypertrophic cardiomyopathy.

## 4.6 Application 2: Cardiac phenotyping of R6/2 mice

In this section cardiac MRI is used to characterise the cardiac phenotype of transgenic mice obtained by modification of the gene causing Huntington's disease in humans.

### 4.6.1 Background

Huntington's disease (HD) is a hereditary disorder of the central nervous system characterised by psychological, neurological and physiological symptoms. The disease is carried by an autosomal dominant gene (IT15, chromosome 4). The defect is caused by a trinucleotide CAG repeat for a protein called huntingtin.

Although primarily considered as a disease of the CNS, the mutant protein responsible for disease is expressed throughout the body (113). There is growing awareness of cardiac dysfunction in HD. Heart failure (HF) is a major cause of death in HD patients (114-116), and HD patients are 15 times more likely to suffer from heart disease than age-matched controls (117). It has been suggested that these symptoms are related to loss of dopamine receptors, microglia activation and neural inclusions in the hypothalamus of PHD individuals (118), although it is possible that the cardiovascular system is directly involved.

The R6/2 mouse has proved to be a useful model of HD, displaying neurodegeneration (119, 120), behavioural abnormalities (121-126), and altered daily activity patterns (127-129).

## 4.6.2 Materials and methods

### Animal model

Mice were taken from a colony of R6/2 transgenic mice (130) established at the University of Cambridge, and maintained by backcrossing onto CBA × C57BL6N F1 female mice. Genotyping and CAG repeat length measurement were carried out by Laragen (Los Angeles, CA, USA) as described previously (131). The transgenic mice used in this study had a mean CAG repeat length of 242 ± 1 (range 237-251).

### Experimental design

Two groups of mice (10 WT, 10 R6/2 mice) were imaged at 13 and at 16 weeks. One of the transgenic died after the first MRI session.

### MRI acquisition and analysis methods

After initial localization images, a 4-chamber view was acquired (FLASH, TR/TE 8ms/2.8ms, 20-25 frames, 2.56cm FOV, 256 matrix, 1.5mm slice thickness, bandwidth 70kHz, flip angle 30°, NEX 8). Using this scan as a reference, short axis slices were arranged perpendicular to the septum to cover the left ventricle (LV).

### The black-blood cine-FLASH sequence

Saturation slices were placed over the pulmonary veins/left atrium for a black-blood sequence (FLASH, TR/TE 10.9ms/2.8ms, 15-18 frames, 2.56cm FOV, 256 matrix, 1.1mm slice thickness, bandwidth 70kHz, flip angle 20°, NEX 6). Full LV coverage was achieved with no slice gap with 6-8 slices.

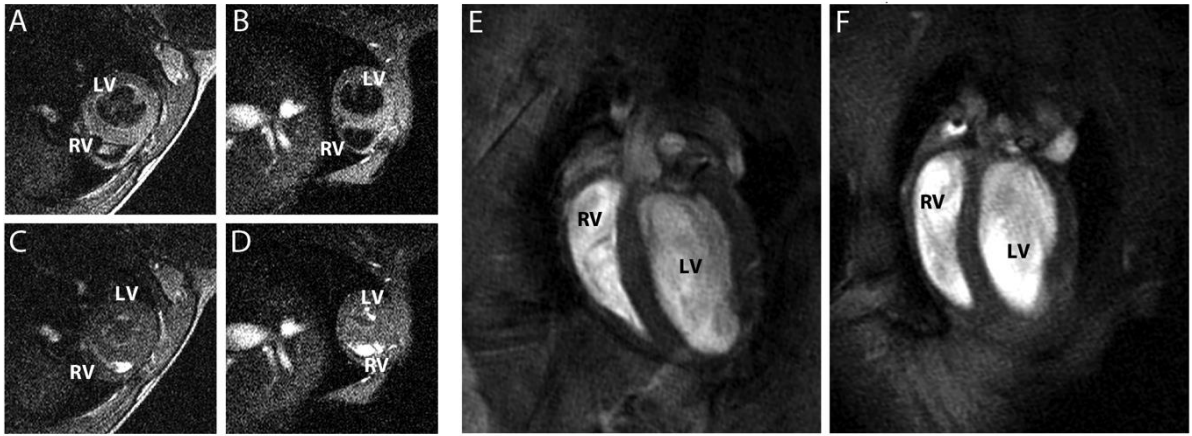

**Figure 45** A-B) Images from a short-axis view black-blood CINE stack in a wildtype, C-D) in a transgenic mouse. E) four chamber FLASH in a wildtype. F) Four chamber FLASH in a transgenic mouse.

Even if it was possible to segment the blood and the myocardium, this implementation of a black-blood sequence has shown problems of partial saturation of the blood, especially in end-systole, that complicated the segmentation of the blood pool, this is discussed in 4.8.2.

**MRI assessment of septal bowing**

In viewing the MRI cine frame, it was observed that the R6/2 hearts appeared to have a different shape during contraction, in particular septal bowing seemed to be present.

To assess this, a simple measure of 'bending' of the left ventricle was conceived. Three lines were fitted to the centroids of delineated LV blood pool regions. The first line was the best fit to the centroids of each slice to determine the 'midpoint' of the LV. Two further lines were found as the best fits to the slices above and below this midpoint respectively. The metric considered was the angle made between these lines.

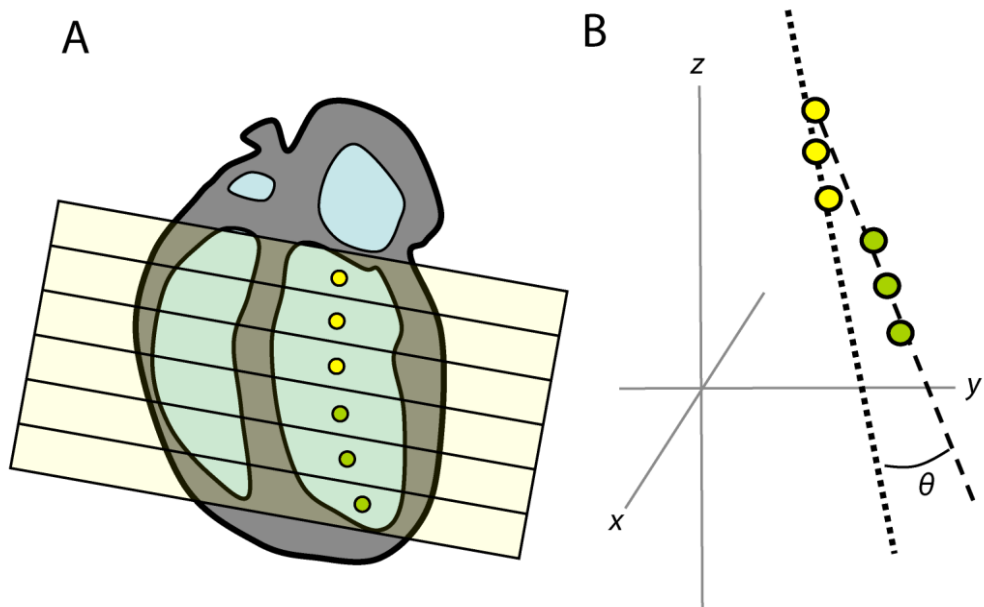

**Figure 46** A) The barycentre of each LV slice is calculated. B) The angle θ between the higher and the lower portion of the heart is used as a measure of distortion.

## 4.6.3 Results

At 13 weeks, end diastolic volume (EDV) and end systolic volume (ESV) were reduced by 17 ± 4% (p < 0.002) and 28 ± 9% (p < 0.02) respectively, in R6/2 compared with WT mice (Table 2). By 16 weeks, abnormalities increased to 37 ± 7% (p =< 0.00001) and 53 ± 11% (p <0.00001), respectively (Table 2). became significantly reduced in R6/2 compared with WT mice (EF, 18 ± 5%, p < 0.005; SV, 26 ± 6%, p < 0.0001; CO, 31 ± 7%, p < 0.0005; Table 2).

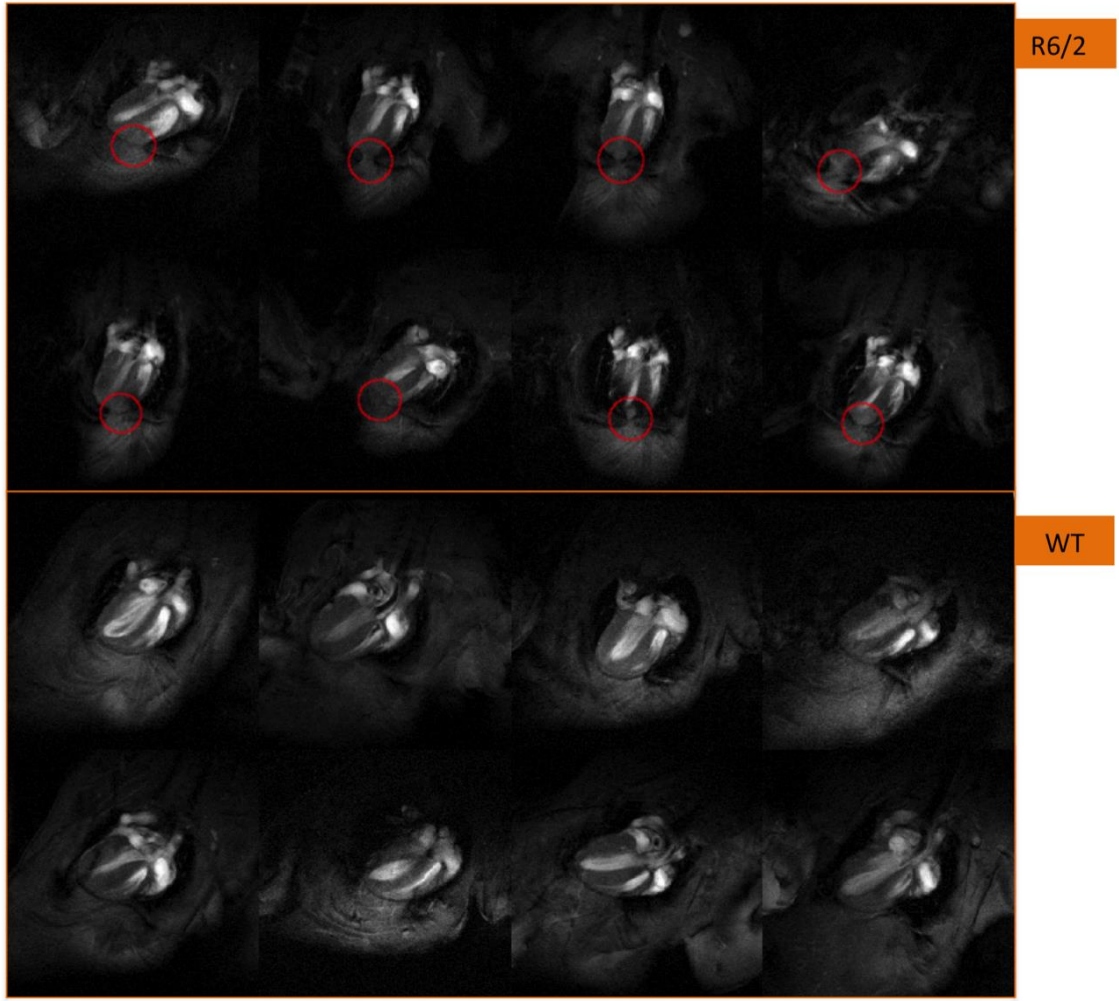

**Figure 47** MR images in the long axis at end-systole from individual R6/2 mice show apparent adhesions of the apex of the heart to the pericardium (red circles). These apparent adhesions were not seen in the hearts of WT mice

Visual inspection of the MRI images revealed that R6/2 hearts were smaller than WT hearts (Figure 47), and had marked visual differences in the cardiac cycle. Video recordings of the beating hearts suggest that hearts from R6/2 mice have a distorted shape during contraction, relative to WT mice. Measurement of bending revealed no difference between genotypes at 13 weeks, but by 16 weeks the bending angle in R6/2 mice was significantly greater than in WT mice (p<0.05, Table 4). An additional finding is the appearance of "adhesions" of the apex of the heart to the pericardium in the images from 16 week old R6/2 mice (Figure 47).

These phenomena were not seen in 13 week old R6/2 or WT mice at either age.

| | Scan 1 (13 weeks) | | | Scan 2 (16 weeks) | | |
|---|---|---|---|---|---|---|
| | WT (n=10) | R6/2 (n=9) | | WT (n=10) | R6/2 (n=9) | |
| **Age (weeks)** | 13.1 ± 0.2 | 13.3 ± 0.1 | | 16.7 ± 0.1 | 16.4 ± 0.1 | |
| **Body weight (g)** | 27.7 ± 1.0 | 21.1 ± 1.4 | ** | 29.3 ± 1.3 | 19.0 ± 1.1 | *** |
| **EDV (µl)** | 45.3 ± 1.5 | 37.3 ± 1.6 | ** | 46.0 ± 1.4 | 29.1 ± 1.5 | *** |
| **ESV (µl)** | 20.9 ± 1.7 | 15.0 ± 1.4 | * | 18.2 ± 1.4 | 8.6 ±1.1 | *** |
| **SV (µl)** | 24.4 ± 1.0 | 22.3 ± 1.4 | | 27.8 ± 1.2 | 20.6 ± 0.7 | *** |
| **CO (ml/min)** | 6.9 ± 0.5 | 6.1 ± 0.4 | | 8.1 ± 0.5 | 5.5 ± 0.5 | *** |
| **EF (%)** | 54.0 ± 3.0 | 60.0 ± 3.0 | | 61.0 ± 2.0 | 71.0 ± 2.0 | ** |
| **LV bending angle (°)** | 13.3 ± 1.5 | 14.3 ± 1.5 | | 11.0 ± 1.2 | 15.0 ± 1.4 | * |

**Table 4** Cardiac parameters derived from mice used for MRI scans. EDV, end diastolic volume; ESV, end systolic volume; SV, stroke volume; CO, cardiac output; EF, ejection fraction; LV, left ventricle. Data are means ± SEM. Significant differences are R6/2 compared to WT mice. * p<0.05, ** p<0.01, *** p<0.001.

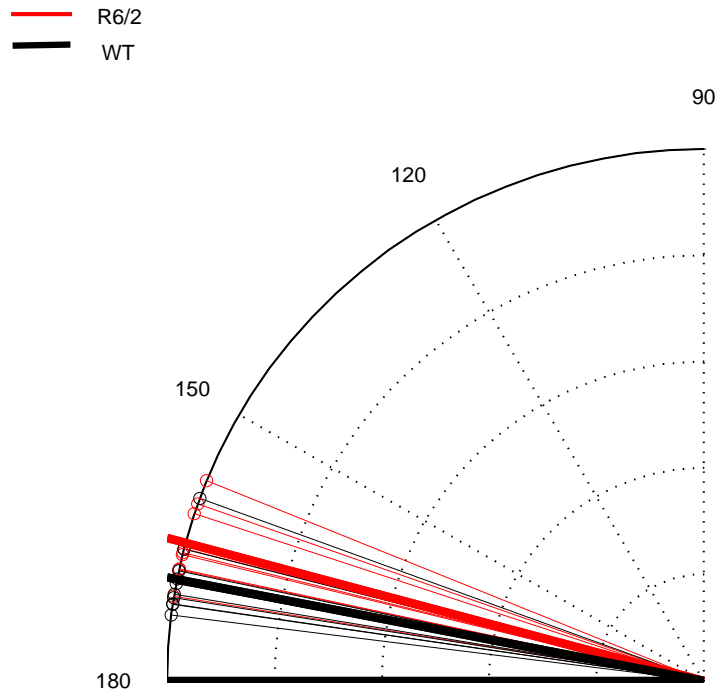

Figure 48 Bending angle at 16 weeks, solid lines represent group averages: The R6/2 (250 cag) hearts display significant bowing with respect to WT at 16 weeks (p<0.05).

### 4.6.4 Conclusion

This study has demonstrated in vivo abnormality of heart function in the R6/2 mouse model of HD, consisting of a reduced cardiac output and septal bowing, but with preserved systolic function. These data add to the body of work demonstrating the usefulness of the R6/2 mouse as a model of HD.

## 4.7 Application 3: Right ventricular dysfunction in R6/2 mice unmasked by dobutamine

During the preliminary observations in the R6/2 mouse the mice had a reduced CO, significant septal bowing and an altered anatomy of the diaphragm, though with a preserved LV systolic function. In this follow-up study we were interested in the role of the right ventricle (RV) in the cardiac phenotype seen in R6/2 mice. It was hypothesized that impaired RV function, and therefore reduced pulmonary circulation, might be an important component of the circulatory problems seen in HD mice, contributing to a reduction of LV cardiac output through mechanisms of ventricular independence. In this study we used in vivo MRI to determine whether the RV was involved in R6/2 cardiac pathology. We characterised this over time, and also used a pharmacological stress test to determine whether or not these effects could be unmasked earlier in the disease. Stress was pharmacologically-induced in vivo with dobutamine, a drug that is commonly used to increase sensitivity and specificity in the assessment of patients with suspected coronary artery disease (132).

### 4.7.1 Materials and methods

**Animals**

Mice were taken from a colony of R6/2 transgenic mice (130) established at the University of Cambridge, and maintained by backcrossing onto CBA × C57BL6N F1 female mice. Genotyping and CAG repeat length measurement were carried out by Laragen (Los Angeles, CA, USA) as described previously (131). The R6/2 mice used in this study had a mean CAG repeat length of 226 ± 7 (mean ± s.d.).

**Magnetic Resonance Imaging**

Two experiments were performed: the first to evaluate the functional parameters longitudinally in transgenic mice (n=5) that were not exposed to dobutamine, the second to evaluate the difference in the effect of dobutamine on C57BL6 WT (n=6) and R6/2 (n=6) mice at a presymptomatic timepoint.

*Longitudinal characterization*

Five male R6/2 mice were scanned at three time points, designed to capture an initial period before severe symptoms developed (t1=7±1 weeks of age; mean ± s.d.) and two later scans reflecting different stages of pathology. In R6/2 mice, body weight increases with development and, after the onset of the symptoms, it falls as a consequence of disease progression. Disease-induced weight loss was therefore used for pathology staging. To evaluate changes in heart size independently of bodyweight, a second scan (t2=14±1 weeks of age) was performed when the bodyweight had decreased to the same point as it was at the initial scan. A third scan was performed at a later stage of the disease, when the bodyweight was 75% of the initial weight (t3=16±1 weeks of age).

*Stress test*

Six male R6/2 mice and six male WT controls (C57Bl6N/CBA) were scanned at 10 weeks of age. This corresponded to the last time point before the disease phenotype starts to show and body weight starts to decrease. In addition to the protocol described below, pharmacological stress was induced with an in situ intraperitoneal (i.p.) bolus injection of dobutamine (12 μg/g) to induce chronotropic, as well as inotropic and lusitropic response (54). For 30 minutes following injection three short axis mid-ventricular slices with no gap were acquired in order to observe the changes in LV and RV volumes caused by beta-adrenergic stimulation.

 *Imaging protocol*

Anaesthesia was induced with 3% isoflurane in 1l/min O2 and maintained with 1-2% isoflurane in 1l/min O2. A pressure sensor for respiration rate was used to monitor anaesthesia depth and a rectal sensor was utilised to monitor the core temperature, which was maintained at 36-37° with the use of a flowing-water heating blanket. Gating of the MRI sequences was achieved prospectively with electrocardiogram (ECG). The ECG trace was visually inspected between MRI sequences to identify severe arrhythmias if present. MRI was performed at 4.7T with a Bruker BioSpec 47/40 system (Bruker Inc., Ettlingen, Germany). A birdcage coil of 12cm was used for signal excitation and a 2cm surface coil for signal reception with the animals positioned prone. After initial localization images, 4-chamber and 2-chamber views were acquired (FISP, TR/TE 7ms/2.4ms, 13-20 frames, 3.5 cm FOV, 256x256 matrix, 1 mm slice thickness, bandwidth 64kHz, flip angle 20°, NEX 2). Using these scans as a reference, short axis slices were arranged perpendicularly to both the long-axis views (FISP, TR/TE 7ms/2.4ms, 13-20 frames, 3.5 cm FOV, 256x256 matrix, 1 mm slice thickness, bandwidth 64.1kHz, flip angle 20°, NEX 2). Full LV and RV coverage was achieved with no slice gap with 8-10 slices following Weissman et al (67).

*Data analysis*

To analyse the effects of inotropic stimulation, ejection fraction from the slices acquired 30 minutes after the injection of dobutamine was compared with the same slices acquired at baseline performing a paired Student's t-test.

In the longitudinal analysis parameters were tested for significant differences using a paired Student's t-test. Differences between WT and R6/2 mice were tested with a two samples different variance t-test.

## 4.7.2 Results

During MRI examinations, mean heart rates were in the range of 350-450 beats/minute, and respiratory rates were between 25-65 breaths/minute. Visual inspection of the ECG signals during the MRI studies did not reveal arrhythmia in any subject.

| Age (wk) | *7± 1* | *14± 1* | *16± 1* |
|---|---|---|---|
| | *(presymptomatic)* | *(symptomatic)* | *(late stage)* |
| | mean ± s.d. | mean ± s.d. | mean ± s.d. |
| Body weight (g) | 23 ± 1 | 23 ± 1 | **16 ± 1\*\*** |
| Heart rate (bpm) | 425 ± 40 | 400 ± 80 | 375 ± 40 |
| LVM (µl) | 76 ± 10 | **65 ± 7\*\*** | **58 ± 9\*** |
| LVEDV (µl) | 42 ± 5 | 40 ± 5 | **30 ± 5\*** |
| LVESV (µl) | 15 ± 1 | 11 ± 2 | **8 ± 4\*** |
| LVEF (%) | 65 ± 4 | 72 ± 6 | 73 ± 7 |
| LVSV (µl) | 28 ± 3 | 29 ± 5 | **21 ± 2\*** |
| LVCO (ml/min) | 12 ± 1 | 11 ± 3 | **8.0 ± 0.7\*** |
| RVEDV (µl) | 40 ± 4 | 42 ± 9 | 44 ± 6 |
| RVESV (µl) | 12 ± 2 | 14 ± 6 | **22 ± 6\*** |
| RVEF (%) | 71 ± 5 | 68 ± 10 | **51 ± 9\*** |
| RVSV (µl) | 28 ± 4 | 29 ± 6 | **22 ± 2\*** |
| RVCO (ml/min) | 12 ± 1 | 11 ± 3 | **8.2 ± 0.7\*** |

**Table 5** Cardiac parameters for R6/2 mice at different disease stages (n=5).

**\*: p<0.05, \*\*: p<0.01.**

Cardiac functional parameters and body weight at different stages of the disease in R6/2 mice are shown in Table 1. R6/2 mice develop normally, and body weight increased as expected in the first 8-10 weeks. Body weight then falls as a result of disease progression. For this reason, weight was the same at 7 weeks (t1) and 14 weeks (t2), and had declined further by 16 weeks (t3). No significant change in heart rate was observed due to disease progression. Nevertheless, a significant decrease in LV mass starting from middle stage of the disease (t2) was observed. At a later stage (t3), a substantial decrease in cardiac output was observed. The LV volume measurements (Table 1) show a progressive reduction of EDV and ESV, consistent with the reduced cardiac output, but LVEF was unaffected.

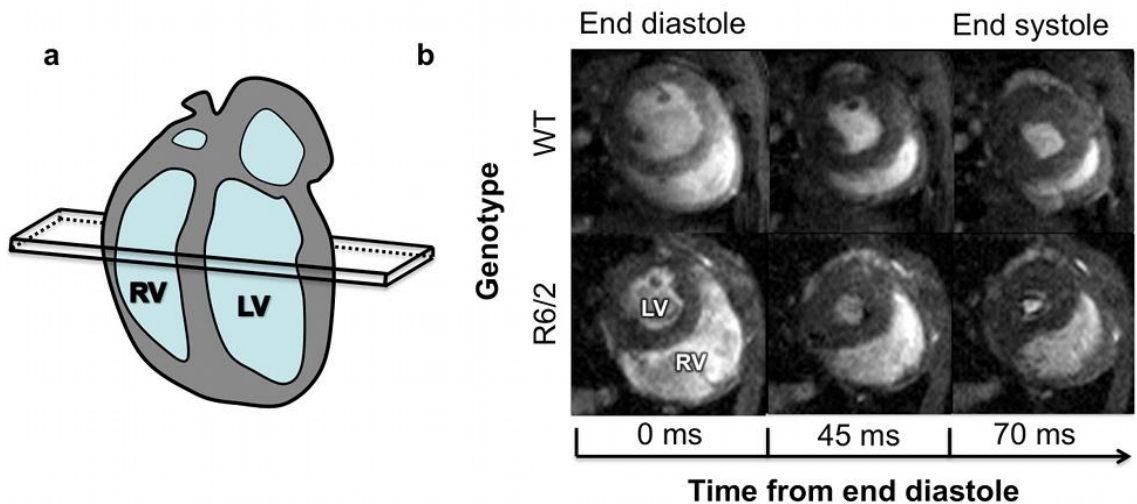

**Figure 49 Right ventricular function during systole in WT and R6/2 mice. (a) Long axis view of the mouse heart, showing the position of the mid-ventricular short axis slice seen in (b). (b) Images from mid-ventricular short-axis slices from WT (upper panels) and R6/2 (lower panels) mice, showing an enlarged RV throughout systole in 16 weeks old R6/2 mice.**

Visual inspection of the short-axis images (Figure 1) showed a substantial impairment of RV systolic function. Remodelling of the RV was represented by a substantial increase in ESV, as confirmed by the volumetric measurements (Table 1). RVEF measurements showed significant deterioration from 71% to 51%.

|  | Baseline values (mean ± **s.d.**) | |
| --- | --- | --- |
| **Genotype** | **WT (n=6)** | **R6/2 (n=6)** |
| **Age (wk)** | 10 ± 2 | 10 ± 2 |
| **Body weight (g)** | 25 ± 5 | 25 ± 4 |
| **LVM (µl)** | 84 ± 15 | 79 ± 14 |
| **LVEDV (µl)** | 46 ± 11 | 43 ± 10 |
| **LVESV (µl)** | 13 ± 5 | 14 ± 5 |
| **LVEF (%)** | 73 ± 4 | 68 ± 5 |
| **RVEDV (µl)** | 42 ± 8 | 38 ± 10 |
| **RVESV (µl)** | 11 ± 3 | 10 ± 4 |
| **RVEF (%)** | 74 ± 5 | 75 ± 4 |

Table 6: Cardiac parameters in WT and R6/2 mice at 10 weeks, prior to dobutamine administration.

At 10 weeks of age, when R6/2 mice show no overt symptoms, LV mass and ventricle volumes were similar in WT and R6/2 mice (Table 2). Furthermore, no significant differences in LVEF or RVEF were present between WT and R6/2 before inotropic stimulation. After dobutamine administration, a significant increase in the heart rate was observed for both groups (p<0.001), although the increase was not significantly different between the two groups (20 ± 15 % in WT, 30 ± 15 % in R6/2 mice, p>0.10). Nevertheless, under dobutamine-induced stress, RV systolic function dropped dramatically in the R6/2 mice, but not in the WT mice (Figure 2).

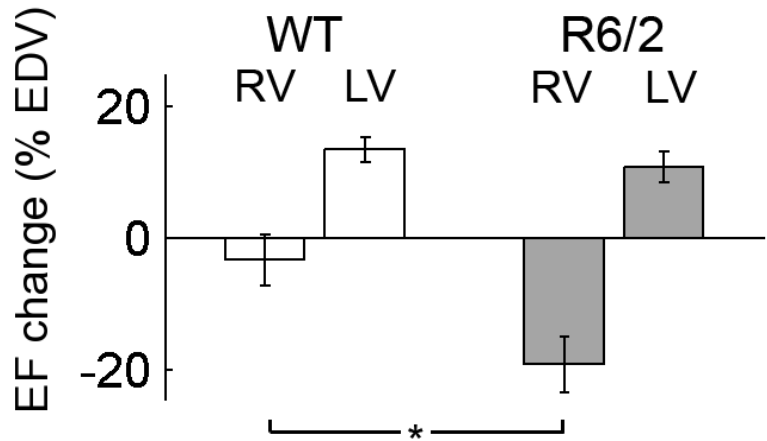

**Figure 50: Changes in ejection fraction during dobutamine stress test.**

**RV = right ventricle, LV = left ventricle.**

**Data are mean ± SEM.  *=p<0.05**

Decrease in RV function of R6/2 mice was driven by a significant increase in both RVEDV and RVESV (Table 3). LV ejection fraction significantly increased in both groups (p<0.001), but there was no significant difference in the effect of dobutamine on LVEF on mice of different genotypes.

| | % increase (mean ± **s.d.**) | |
|---|---|---|
| **Genotype** | **WT (n=6)** | **R6/2 (n=6)** |
| **LVEDV (µl)** | -32 ± 10 | -35 ± 10 |
| **LVESV (µl)** | -71 ± 30 | -50 ± 15 |
| **RVEDV (µl)** | -5 ± 15 | 15 ± 40 |
| **RVESV (µl)** | 30 ± 15 | 115 ± 50 |

**Table 3: Percentage changes in ventricular volumes caused by dobutamine.**

### 4.7.3 Conclusions

MRI was successfully used to investigate the RV in the R6/2 mouse model of Huntington's disease. These data add to previous findings of LV pathology in the R6/2 mice by showing that RV function is also impaired. Importantly, these changes can be unmasked in young, otherwise asymptomatic animals with the beta-adrenergic agonist dobutamine. Using a serial and non-invasive method, this work successfully identified a specific dysfunction of the cardiovascular system in this model of human disease.

## *4.8  Discussion*

### 4.8.1 FLASH vs. FISP

As seen in 2.2.6 fast gradient echoes use spoiling or refocussing. To compare the approaches in the system utilised throughout the thesis (Bruker Biospec 47/40), a long-axis view was acquired with FLASH and FISP, keeping all the other parameters constant (TE/TR= 2.8/8 ms, 20° flip angle and 2 NEX). The SNR of the blood was measured in the first frame and found to be higher in FISP with respect to FLASH with a factor of 3/2, while the SNR of the myocardium was higher in FISP by a factor of 1.14. Higher SNR in FISP over FLASH was consistent over movie frames. This is not surprising, as FISP rescues signal from precedent repetitions instead of spoiling. The reason behind the wider adoption of FLASH in mouse studies has been the robustness to artifacts (103).

Cine FISP                    Cine FLASH

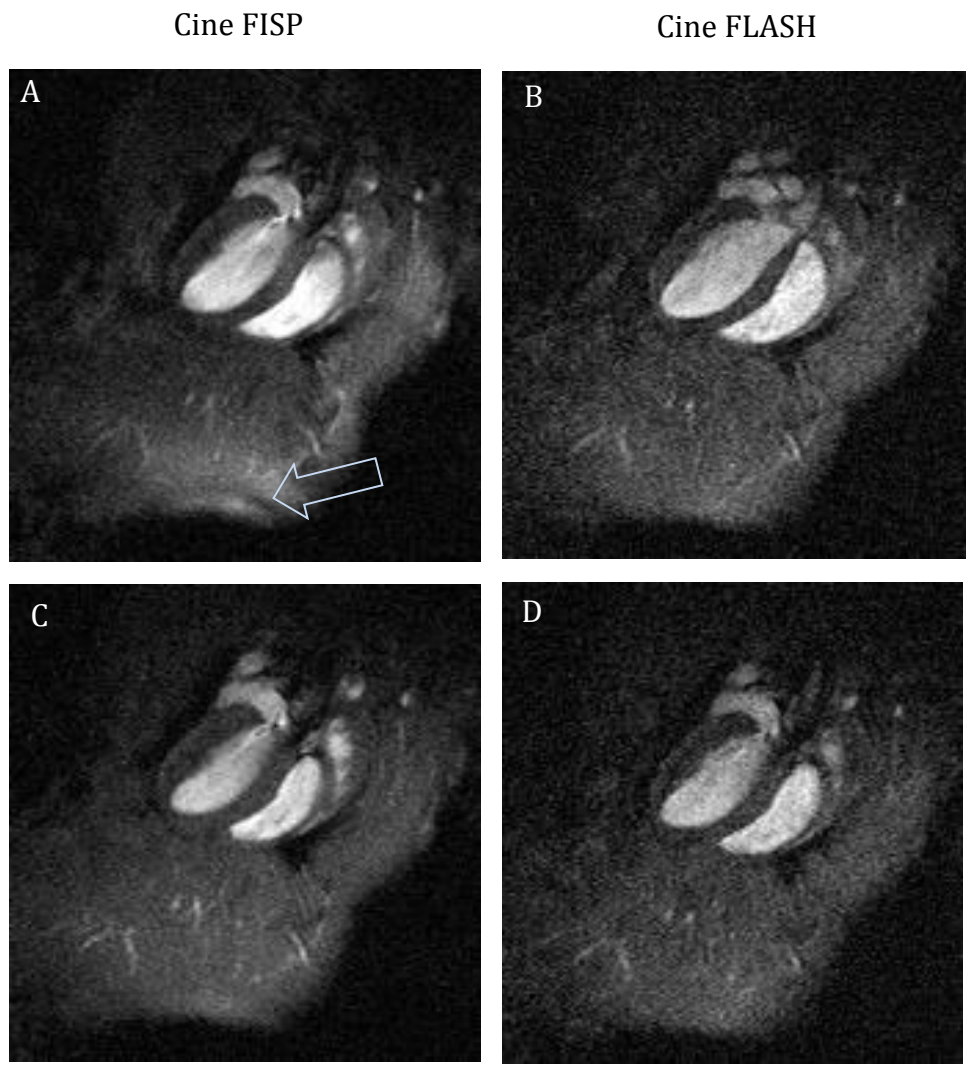

**Figure 51** Visual comparison between FISP and FLASH. A-B) First frame acquired after the ECG trigger. C-D) Third frame acquired after the trigger. The arrow points to a banding artifact, typical of FISP images, not present in FLASH.

Typical banding artefacts were observable in FISP images, but these were always outside of the heart. Banding artifacts have not caused problems during any of the experiments performed throughout this thesis. Other mouse studies that have used FLASH due to banding artifacts in the heart were often performed at higher field and with smaller bore systems. The fact that these artifacts are less prominent in a 47/40 system may be due to lower field and larger bore (40 cm), producing a larger homogeneous area.

### 4.8.2 Bright blood vs. Black blood

The study in 4.6 employed a black blood sequence achieved by saturation of the blood from incoming vessels. This achieved unreliable saturation of the blood, which caused difficulties in segmenting the blood from the myocardium. More effective methods have been reported in the literature, utilising double inversion preparation in order to achieve a better suppression of the inflowing blood (133). Black-blood methods rely on the null of the blood signal in order to achieve high CNR, and have shown to achieve high reproducibility especially in diastole, where fast inflow of blood produces artifacts in bright-blood images.

In contrast, bright blood techniques can achieve high CNR due to an effect known as inflow enhancement (see 2.2.7). As blood moving into the slice has not received any previous excitation, the full population can be magnetized and much higher signal can be extracted from the blood. Due to the flow enhancement effect, in bright blood cine MRI the myocardial wall can be visualised even in an environment where myocardial SNR is low.

### 4.8.3 Speed

Using the protocol described in this chapter, functional parameters can be extracted with an acquisition time of around 12 minutes. Although this scan time is accepted for state-of-the art cine MRI experiments, in some applications this reduces the feasibility of experimental protocols. For instance in 4.7, when imaging the inotropic response of the heart to dobutamine, the coverage had to be limited to three slices only, due to the change of the heart-rate during these examinations. On one hand, better control of the stress utilising an i.v infusion pump would have achieved higher reproducibility of the dose and more stability for performing the scans. However, lengthy stress times are problematic for diseased animals as the stress agent has counter indications (134).

Faster acquisition would allow freezing of the inotropic state of the heart during stress tests with multiple steps, as performed in the clinic for pharmacological stress functional examinations (132).

## 4.9 Chapter summary

Here standard methods to perform cine MRI efficiently and accurately were presented and demonstrated in two genetic models. The main technical challenges and drawbacks were discussed. Although cine-MRI in mice is now feasible in many centres and can be used for useful translational studies, a constraint of this technique is the length of the examination. In this chapter, this limitation was encountered when studying the inotropic response to dobutamine of the R6/2 mouse model of Huntington's disease. The next chapter will explore methods to accelerate cine MRI in mice.

**Chapter 5**

# Accelerated cine MRI: assessment of systolic function with a one-minute acquisition

As seen in the previous chapter, in-vivo assessment of heart function is useful for phenotyping of transgenic models of heart disease. However, lengthy times under anaesthesia reduce throughput and effectiveness of stress test experiments. In this chapter a method to perform a full functional assessment in one minute is presented and validated.

## 5.1 Introduction

As discussed in previous chapters, Cine MRI is considered the gold standard measurement of systolic and diastolic function (53), producing images with anatomical detail superior to any other in vivo technology. MRI can be used for assessment of the left as well as the right heart, and does not require geometrical assumptions on heart shape and contraction (67). In addition, as discussed in Chapter 3, MRI is a highly versatile technique, capable of assessing the heart at multiple levels.

Despite its high accuracy, the application of cardiac MRI in mice has been limited in the past by several challenges. Fast heart rates and small dimensions, together with the difficulties in gating meant that this technology has lagged behind clinical MRI (35). Most of these problems have now been resolved with the adoption of the latest generation of small-bore MRI systems and techniques such as retrospective gating (40). Nevertheless, the procedure to assess function is still time consuming when compared to echo or CT, imposing lengthy anaesthesia and limiting the number of animals that can be done in each session. Long scan times are prohibitive to techniques such as pharmacological stress testing, which can be used to boost sensitivity and specificity during functional examinations, and are important clinically (132). Speeding up the MRI experiment for mice would therefore make the cardiac examination more translatable to the clinic, allowing a full battery of testing to be performed.

In recent years, acceleration techniques such as parallel imaging and compressed sensing have been successfully applied to cardiac MRI in clinical settings. Parallel imaging uses multiple radiofrequency receivers to reduce the number of samples required (see 2.4.1). Compressed sensing achieves the same goal exploiting redundancy of the image in space or time by means of regularized iterative reconstruction (see 2.4.2). Although these techniques have been used in clinical research studies, only a few reports so far have attempted to accelerate cine MRI in rodents. These studies have used either parallel imaging (58, 59) or temporal compressed sensing (60), achieving 2-4 fold acceleration.

Spatiotemporal compressed sensing has previously been implemented in the mouse heart using retrospective gating, in order to increase temporal resolution (135) or accelerate the acquisition in the rat (136) obtaining scan times of 20-30 seconds per slice reconstructing a limited number of heart phases (8-16 per heart cycle).

To date, there have been no reports of the combination of compressed sensing and parallel imaging in the mouse heart. Here these are used in concert for the first time to accelerate cine-MRI the mouse heart, comparing rectilinear and radial acquisition protocols. The effects of acceleration are evaluated by comparing measured parameters of function.

## 5.2 Materials and methods

### 5.2.1 Animals

Five male C57/bl6 mice (20 weeks) were used. Animals were anaesthetised with gaseous isoflurane both for induction (3% in 1l/min O2) and maintenance (1-1.5% in 1l/min O2). A pressure sensor for respiration was used to monitor anaesthesia depth with breathing rates maintained in the range 30-60 breaths per minute. ECG signals were monitored with graphite electrodes. Body temperature was monitored using a rectal thermometer and a flowing-water heating blanket was used to maintain animal temperature at 37°C throughout the experiment.

### 5.2.2 Imaging

Imaging was performed at 4.7T with a Bruker BioSpec 47/40 system equipped with 400 mT/m gradients (Bruker Inc., Ettlingen, Germany). A birdcage coil of 12cm was used for signal excitation and a four-channel receiver array for signal reception (Bruker Inc., Ettlingen, Germany). Animals were positioned prone. After the initial localizers, standard prospectively-gated cine MRI was acquired (see Chapter 4).

For the rectilinear scheme, the sequence was modified to include a navigator after slice excitation (40). In the radial scheme, spokes were acquired spanning 360°, and the forward and reverse spokes were used to estimate and correct gradient inaccuracies as discussed in Chapter 6. The k-space centre signal was used for navigation.

| Sequence | FISP | FLASH | FLASH |
|----------|------|-------|-------|
| Scheme | Cartesian | Cartesian | radial |
| Gating | prospective | retrospective | retrospective |
| TR/TE | 6.0 / 2.1 ms | 4.8 / 1.7 ms | 5.6 / 1.1 ms |
| FOV | $3.5 \times 3.5$ cm$^2$ | $3.5 \times 3.5$ cm$^2$ | $3.84 \times 3.84$ cm$^2$ |
| Echo position | 50 % | 15 % | 5 % |
| Matrix size | $256 \times 256$ | $192 \times 192$ | $256 \times 1440$ |
| Rec BW | 78 kHz | 100 kHz | 100 kHz |
| Pulse | flip angle 20° | flip angle 20° | flip angle 20° |
| | BW 5.4 kHz | BW 18 kHz | BW 14 kHz |
| Repetitions | 15-20 frames, | 100 repetitions | 12 repetitions |
| | 1 NEX | | |

**Table 7 Sequence parameters.**

## 5.2.3 Retrospective gating

Retrospective gating was achieved by performing principal component analysis (PCA) on the navigator signals, using all of the receivers. For the rectilinear case, data in the navigator signals were concatenated across coils. For the radial case, the whole k-lines were concatenated across coils. The first component of the PCA was high-pass filtered to extract cardiac gates. Prediction error on a band-pass Fourier filter centred on the heart rate was used to identify and exclude respiration events. Cine-MRI images were reconstructed with 20 cardiac phases. Typical self-gating signals are shown in Figure 52.

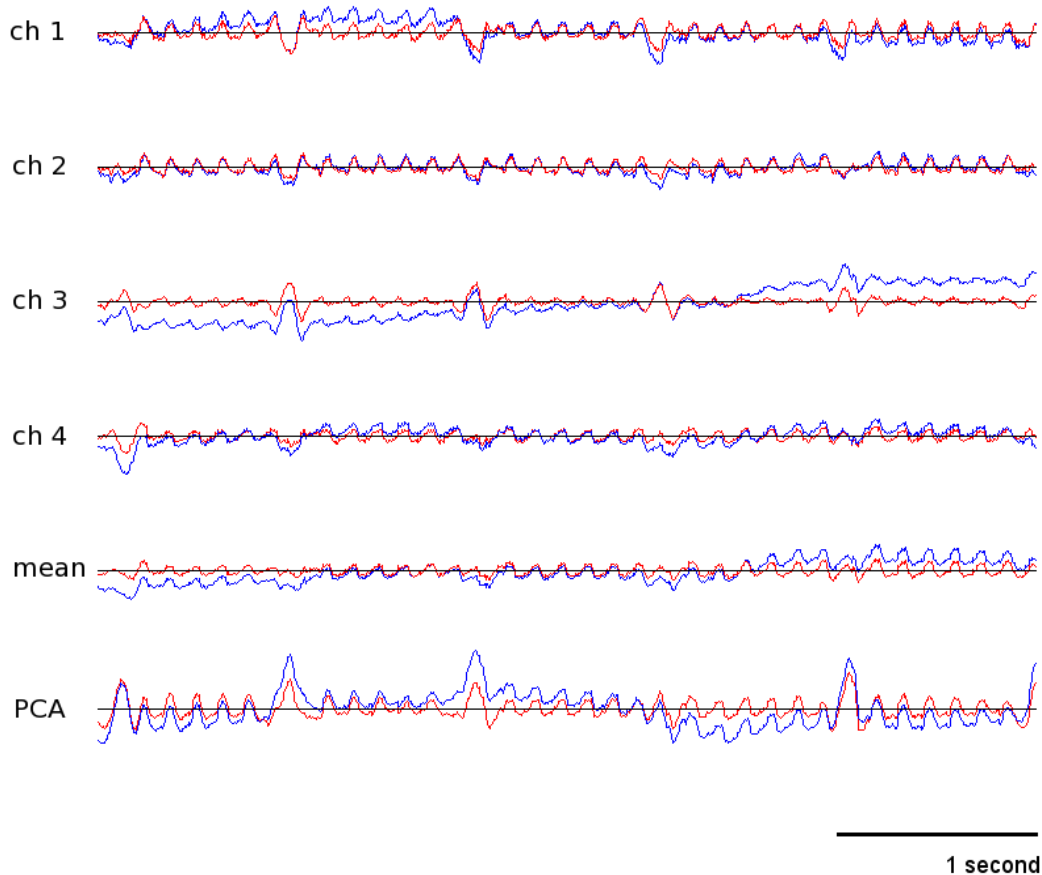

**Figure 52** Comparison between typical gating signals in the radial case. Blue signals are raw data, Red signals are high-pass filtered. Features with the frequency of the order of 1 Hz are respiration events, while higher frequency oscillations represent heart beats.

## 5.2.4 Iterative reconstruction

Compressed sensing was applied to both radially- and rectilinearly-acquired data, using a projection over convex sets (POCS) algorithm (137) (see 3.1). Each iteration of the algorithm included two data consistency steps (acquired k-space consistency and SPIRiT consistency across receivers in the image domain (33)) and two soft thresholding steps in sparse domains (wavelet transform to take advantage of spatial compressibility; temporal Fourier transform to exploit the periodicity of the heart cycle). In the radial case, the algorithm employed implicit gridding with non-uniform fast Fourier transform (NuFFT (28)) to avoid feedback of numerical errors. The code was implemented using

Matlab (Mathworks, USA). Open-access, freely available libraries were used for SPIRiT, wavelet and NuFFT.

Undersampled datasets were obtained taking only the first part of the acquisition and discarding the rest. For each acquisition strategy we reconstructed 6-fold undersampled images (15 seconds per slice) and 12-fold undersampled images (7.5 seconds per slice), to compare with fully sampled images (90 seconds per slice). Standard functional parameters for the LV were derived using Segment v1.9 (48). Precision in the volume estimates between acquisition schemes and undersampling factors was assessed by means of Bland-Altman plots (61). To validate the self-gating strategy, parameters from fully sampled Cartesian and radial acquisitions were compared with standard prospectively gated scans using a t-test.

## 5.3 Results

The heart rate was 454 ± 30 bpm (mean ± s.d.) during the examination. Using PCA, our self-gating routine extracted the relevant information automatically for both radial and rectilinear MRI. The gating procedure required minimal manual interaction to define the threshold for rejection of respiratory events, which was selected on an individual basis by visual inspection. No significant differences were found in measurements using prospective gating and the two retrospectively-gated schemes.

|  | FISP | FISP - Rectilinear FLASH | FISP - Radial FLASH |
|---|---|---|---|
| LVEDV (µl) | 79 ± 7 | -0.8 ± 4.9 | 3.0 ± 5.3 |
| LVESV (µl) | 27 ± 3 | -0.4 ± 3.0 | 1.6 ± 2.5 |
| LVSV (µl) | 51 ± 6 | -0.4 ± 3.3 | 1.4 ± 3.8 |
| LVEF (%) | 65 ± 4 | -0.2 ± 2.8 | -0.4 ± 2.3 |

Table 8: Differences between functional parameters obtained from prospectively-gated Cartesian FISP (reported in the first column) and retrospectively gated scans, respectively rectilinear and radial FLASH (mean ± s.d.). No significant differences were found between prospective and retrospective gating.

The compressed sensing algorithm successfully anti-aliased the cine images, which could be used to produce accelerated whole-heart stacks. Image quality of 6-fold undersampled rectilinear MRI was sufficient to segment the ventricles, while residual aliasing corrupted the measurement on the 12-fold rectilinear undersampling (Figure 53). Residual aliasing was mostly present in the slices with higher diastolic inflow at the base of the heart, while mid-ventricular and apical slices were still artefact-free. On the other hand, 12-fold accelerated radial images were alias-free throughout the heart (Figure 54).

Cartesian                          Radial

zero-filled        iterative        zero-filled        iterative

Full

6x

12x

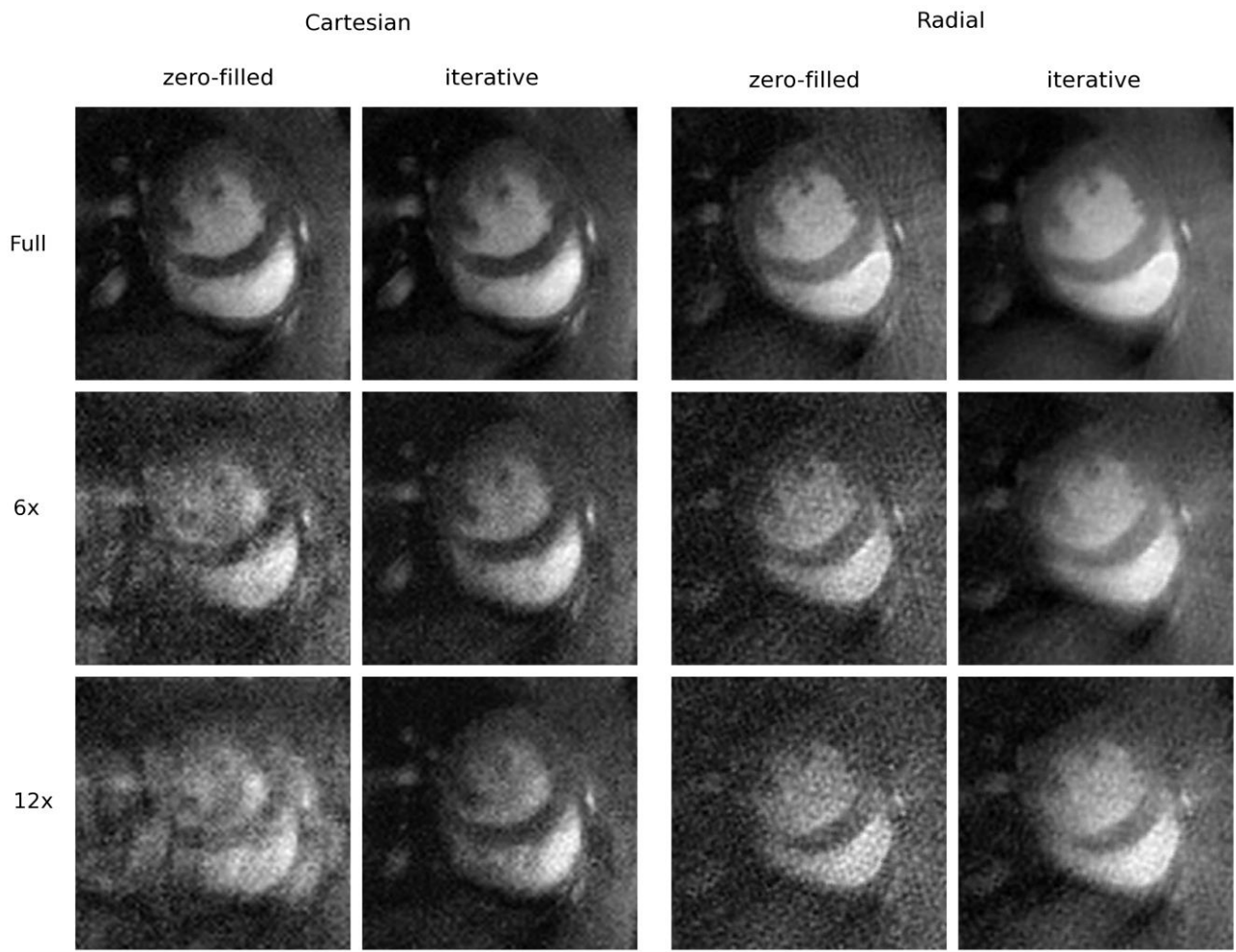

**Figure 53: Mid-ventricular short-axis slice comparing full sampling, 6x and 12x acceleration for Cartesian and radial sampling. Each image is reconstructed both by applying straightforward transformation to zero-filled data and iterative reconstruction including compressed sensing and parallel imaging.**

End diastole    End systole

full    12x    full    12x

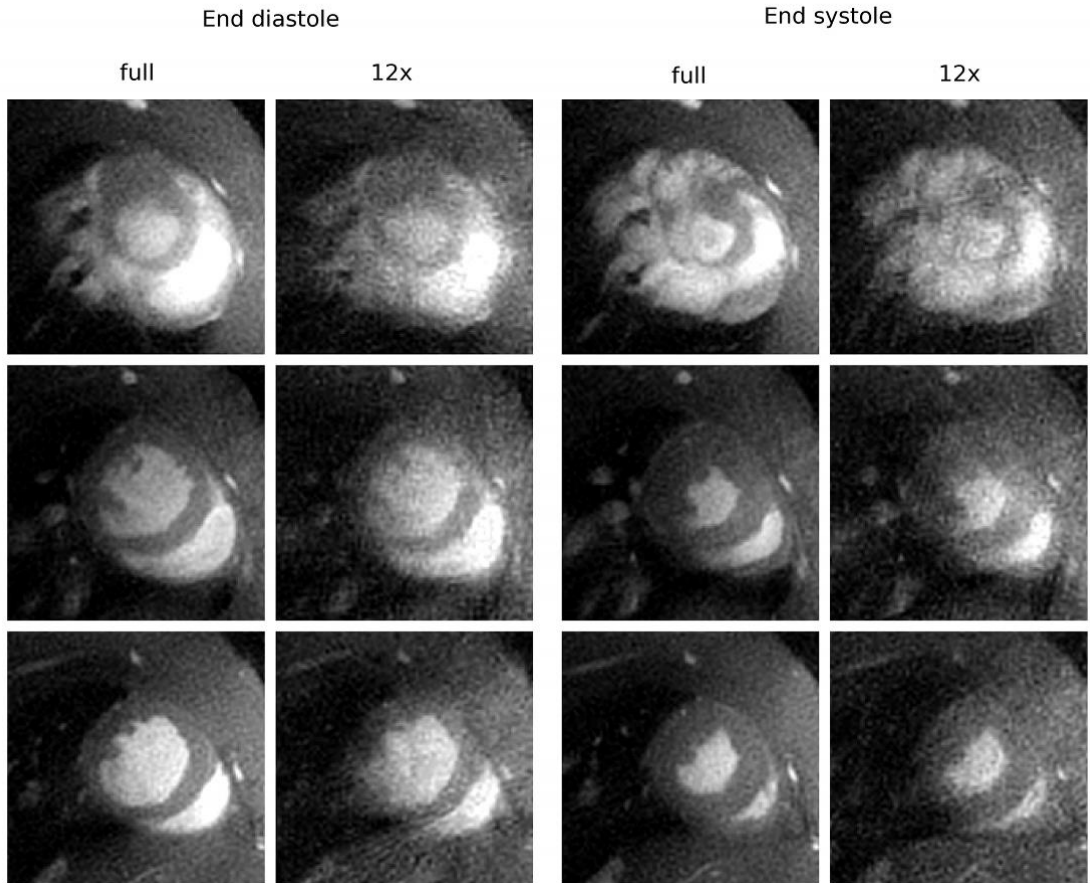

*Figure 54 Comparison between fully sampled and 12x accelerated radial MRI. Although signal to noise ratio is lower in the 12x images due to fewer datapoints acquired, the anatomical detail is preserved for functional examination as aliasing has been successfully removed.*

Despite a reduction in signal to noise ratio due to the fewer datapoints acquired, the anatomical detail was preserved and it was possible to segment all the images. Bland-Altman plots for functional parameters are shown in Figure 55 for Cartesian sampling and in Figure 56 for the radial case.

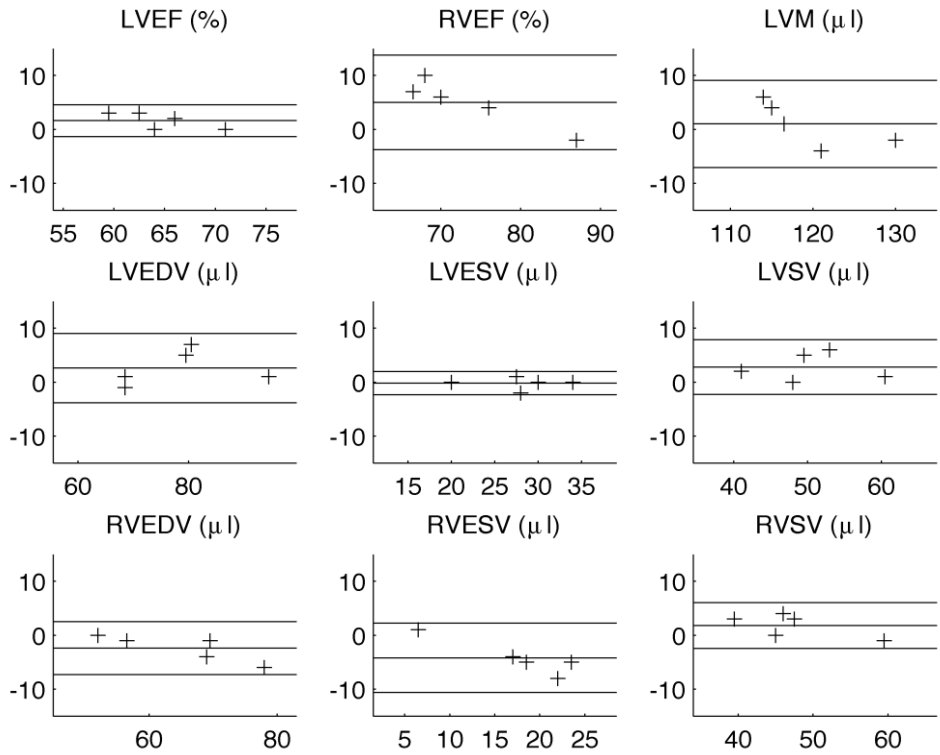

*Figure 55: Bland-Altman plot comparing functional parameters derived from full Cartesian sampling and 6x accelerated Cartesian cine MRI (full-undersampled vs average). Lines are mean with 95% confidence intervals.*

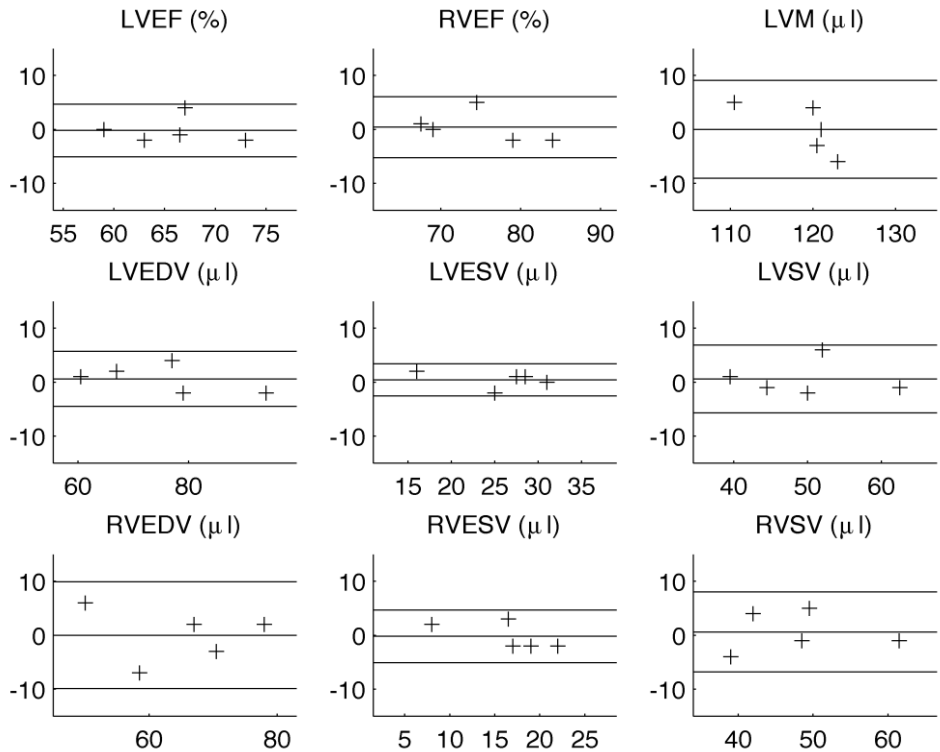

*Figure 56: Bland-Altman plot comparing functional parameters derived from full radial sampling and 12x accelerated radial cine MRI (full-undersampled vs average). Lines are mean with 95% confidence intervals.*

Good agreement was observed for the undersampled two-minute rectilinear and one-minute radial set compared with full sampling. Precision was comparable between acquisition schemes, but undersampling with the radial scheme had lower bias than using rectilinear sampling.

## 5.4 Discussion

The results show that imaging times for complete functional assessment of heart function in mice can be accelerated 12-fold to obtain an acquisition time of one minute.

Importantly this experiment considered not only the visual quality of the images, but also the quantification of the functional parameters. There was no change in the volumes obtained in any of the tested mice and the group variability was not increased. To assess an optimal regime to achieve the required acceleration, the performance of rectilinear and radial acquisition of k-space were compared. Unlike the radial case, rectilinear sampling led to a bias in measures caused by coherent aliasing. This is unsurprising as the radial data achieve a less coherent sampling of k-space, producing noise-like aliasing, which is vital for the application of compressed sensing.

Compared to prospective ECG gating, using navigators for retrospective gating results in higher resistance to arrhythmias (40). Furthermore, the gating procedure can be performed in post-processing and is less prone to human error. Through PCA, the method described here made efficient use of all physiological information at once, minimising operator interaction. Although the procedure for obtaining the cine images is relatively straightforward, preparatory steps such as positioning and imaging scouts are required for careful planning of the short-axis scan. Nevertheless, landmarks such as the apex, the aorta and the LV wall can be easily and quickly identified to guide the prescription of oblique short axis slices (see 4.2).

Parallel imaging and compressed sensing have been used to reconstruct images from partial data, therefore accelerating the acquisition. Importantly, the undersampled data used here were not synthetized, but obtained from the first part of the acquisition, therefore demonstrating directly the acceleration capabilities of this method. In order to reconstruct images from partial data, iterative reconstruction must be employed. As this comes with high computational burden, especially when re-gridding steps are needed, at present reconstruction is performed off-line. However, with multi-core computers and graphical processing units being employed in scientific programming, strategies to perform these algorithms faster have been proposed that may permit online reconstruction in the future (138).

Fast in-vivo techniques for assessing function of both ventricles are required, for longitudinal phenotyping and meaningful translation of new treatments. In echocardiography, volumes are derived from two-dimensional datasets making assumptions on ventricular shape based on a geometrical model. This can potentially produce some discrepancy in diseased hearts with abnormal anatomy and function (139) (see 3.1.1). Furthermore, it is difficult to apply such models accurately to the right ventricle, due to its complex shape and contraction pattern. MRI can derive those volumes directly from three-dimensional data. MRI is considered the non-invasive reference standard (139) when accurate assessment of left or right ventricular volumes is needed, and the method is particularly suited to longitudinal studies as smaller sample sizes are needed compared to other techniques (140).

## 5.5 Chapter summary

In this chapter a protocol to accelerate self-gated retrospective cine MRI in mice using spatiotemporal compressed sensing and parallel imaging was demonstrated. This can be performed with either Cartesian or radial MRI, utilising an optimum self-gating strategy. MRI assessment of left and right ventricular systolic function in mice can be significantly accelerated maintaining high accuracy performing a one-minute acquisition. Fast protocols such as these will enable high throughput phenotyping or translational studies investigating left or right heart, including pharmacological stress experiments, favouring reduction and refinement of preclinical research. The next chapter addresses the issue of trajectory correction in radial MRI.

# Chapter 6

# Trajectory correction for radial MRI in mice

Radial sampling has inherent advantages over Cartesian k-space sampling, such as lower sensitivity to motion, flow, and incoherent aliasing (41). Repeated sampling of the k-space centre yields a signal that can be used as a navigator for retrospective gating of k-lines into frames covering the cardiac or respiratory cycles. Since the coherence between sampled points is reduced compared to Cartesian imaging, the method is popular with compressed sensing for fast MRI techniques employed in real-time imaging applications (141).

As demonstrated in Chapter 5, these advantages extend also to the imaging of mouse hearts. However, images from radial acquisitions are prone to eddy current artifacts (as simulated numerically in paragraph 2.3.3). In this chapter eddy-current induced artifacts are characterized in the mouse heart, and novel methods to correct for retrospective correction are tested.

## 6.1 Introduction

Fast MRI techniques require switching of high magnetic field gradients within the magnet bore, which induces eddy currents in surrounding conductive structures, each of which decays exponentially (142). These eddy currents alter the effective field gradient seen by the imaged object. Linear terms will result in shifts of the k-space, while B0 terms will result in phase accumulation (143). In standard Cartesian schemes, both of these terms produce a uniform shift and phase error throughout and therefore do not have an effect on image quality. On the other hand, using non-Cartesian or segmented acquisition, a different term will corrupt each readout, leading to visible artifacts in the images. While B0 terms are significant only for ultra-high fields or small bores (29), artifacts due to linear terms are also present in standard wide-bore systems. In radial MRI, imperfect k-space trajectories caused by these effects lead to artifactual hypointense regions within images ('shading') and hyperintense regions outside of them ('halos').

Some efforts have been made to address these by pre-scan calibration shots measuring the phase offset between forward and reverse acquisition of the same line in two perpendicular directions (144) and B0 correction (29). In the present study, these correction methods were compared with some improvements specifically for cine MRI of the mouse heart.

Experiments reported in this chapter aimed to test different means of correcting k-space trajectories retrospectively using the imaging data and to evaluate their performance during free breathing and cardiac motion.

## 6.2 Materials and Methods

### 6.2.1 Image acquisition

Radial scan data were acquired using ten male C57/Bl6 mice (16-20 weeks old) at 4.7T using a Bruker BioSpec 47/40 system (Bruker Inc., Ettlingen, Germany). A birdcage coil of 12cm was used for signal excitation and a four-channel cardiac array (Bruker) for signal reception, animals were in a prone position. After initial localizers, 2D cine-MRI slices were acquired in short axis views covering the left ventricle. To obtain radial images, 1,440 spokes were acquired covering 360° with 256 points per spoke with the echo occurring 5% along the spoke length. Other scan parameters were: TR/TE 5.6/1.1 ms, FOV 3.84×3.84cm$^2$ with a receiver bandwidth of 100 kHz and a flip angle of 20°. Cartesian data prescribed over the same geometry were acquired for comparison with the radial data. The scan parameters were TR/TE 4.8/1.7ms with matrix size 192x192, FOV 3.5×3.5cm$^2$ with the echo position at 15%, receiver bandwidth 100kHz and flip angle 20°.

### 6.2.2 Animals

Animals were anaesthetised with gaseous isoflurane both for induction (3% in 1l/min O2) and maintenance (1-1.5% in 1l/min O2). A pressure sensor for respiration was used to monitor anaesthesia depth, rate was maintained in the range 30-60 breaths per minute. Body temperature was monitored using a rectal thermometer and a flowing-water heating blanket was used to maintain animal temperature at 37° C throughout the experiment.

### 6.2.3 Trajectory correction

Trajectory correction was first considered using two perpendicular k-space lines (acquired along the x- and y- directions, respectively) compared to using all antiparallel spokes covering 360°. Each method is based on the projection data from each radial spoke, but the use of magnitude or phase data were considered separately (full details given below). In each case, the additional impact of correcting for B0 effects was considered.

The first correction strategy used spoke magnitudes. This method was based on the assumption that the spoke point with the largest amplitude corresponds to the k-space centre. The correction to apply in each case was derived using quadratic interpolation of the k-line for each projection to find the maximum point and translating it so it would coincide with the k-space centre.

The second strategy was based on the phase of the projections. Previous work to correct radial trajectories measured the phase offset between projections of forward and reverse acquisitions of the same k-space line in two perpendicular directions (144). In the ideal case this offset should be zero. If it is not, the value can be used to adjust the k-space lines to correct the artifacts.

B0-phase induced errors were corrected independently for each receiver following (29). Briefly, a constant shift in phase is estimated for each k-line using the central point in each line.

Images were reconstructed with a non-uniform fast Fourier transform (NuFFT (28)) and reconstructed with 20 heart phases. Radial artifacts caused by trajectory errors tend to manifest as signal loss (shading) inside the object and hyperintense regions (halos) outside the object. As these artifacts were similar in all heart phases, metrics to measure artifact presence were applied to the average of the frames in order to obtain a high SNR environment. The success of the artifact correction was determined with two metrics. The first metric used was the mean signal inside the object, while the second metric was the mean signal in air. The object mask used was derived from the Cartesian images automatically using a threshold.

The correction schemes were also compared in terms of their effect on the navigator signal used for retrospective gating: the central k-space point. Trajectory errors lead to an extra modulation of this signal depending on the angle of each spoke, and can be seen by finding peaks in the power spectrum of the signal at frequencies corresponding to complete cycles of angular variation (the 'sweep frequency'). As a further metric of trajectory correction, the first two harmonics of this frequency were summed under each scheme tested and compared to uncorrected data.

Comparisons between schemes were performed statistically using a two-tailed paired t-test on all slices acquired. Results where $p < 0.01$ were considered to be statistically significant.

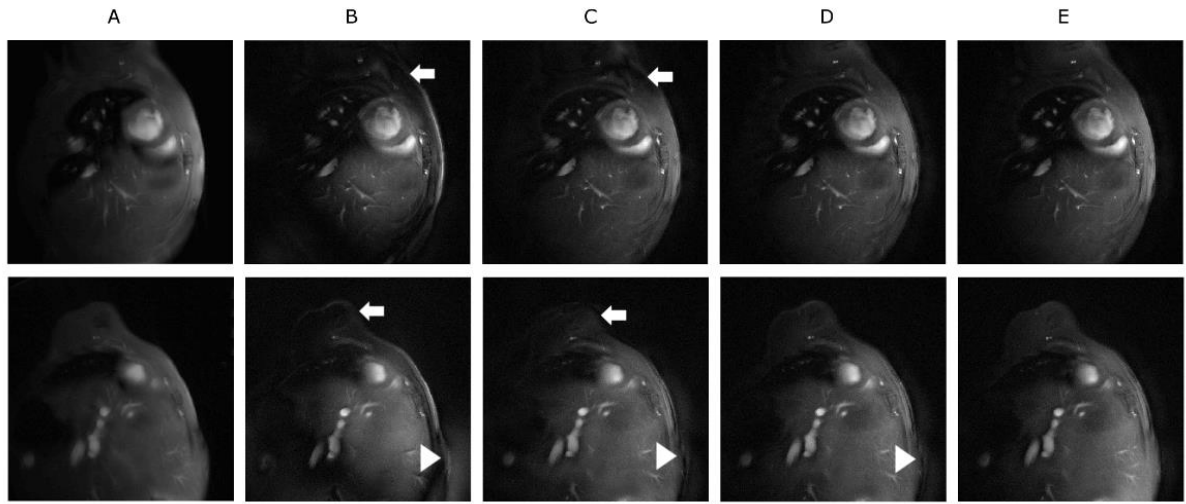

**Figure 57 Effects of trajectory correction using the phase of the projections on a short-axis slice before applying gating. Arrows point to shading artifacts, arrowhead to halos. A) Cartesian image. B) Radial image before any correction is applied. Shading is visible in several parts of the body. C) Correction of linear eddy current terms is applied using the Cartesian axes. Artifacts are still visible especially at the edge between body and air. D) Correction of linear terms using the whole sweep. The image is artifact-free. E) $B_0$ correction is applied, edges are sharper.**

## 6.3 Results

With no correction applied, artefacts were present in all images (Figure 57B). As expected, these artifacts consisted of shading (Figure 57B, arrow) and/or halos (Figure 57B, arrowhead). The figure shows the corrected images using the phase information with two perpendicular directions (Figure 57C) and with every direction (Figure 57D). The additional effect of B0 correction is shown in Figure 57E, little further improvement was seen in our images. Similar results in terms of image quality were obtained with the magnitude images.

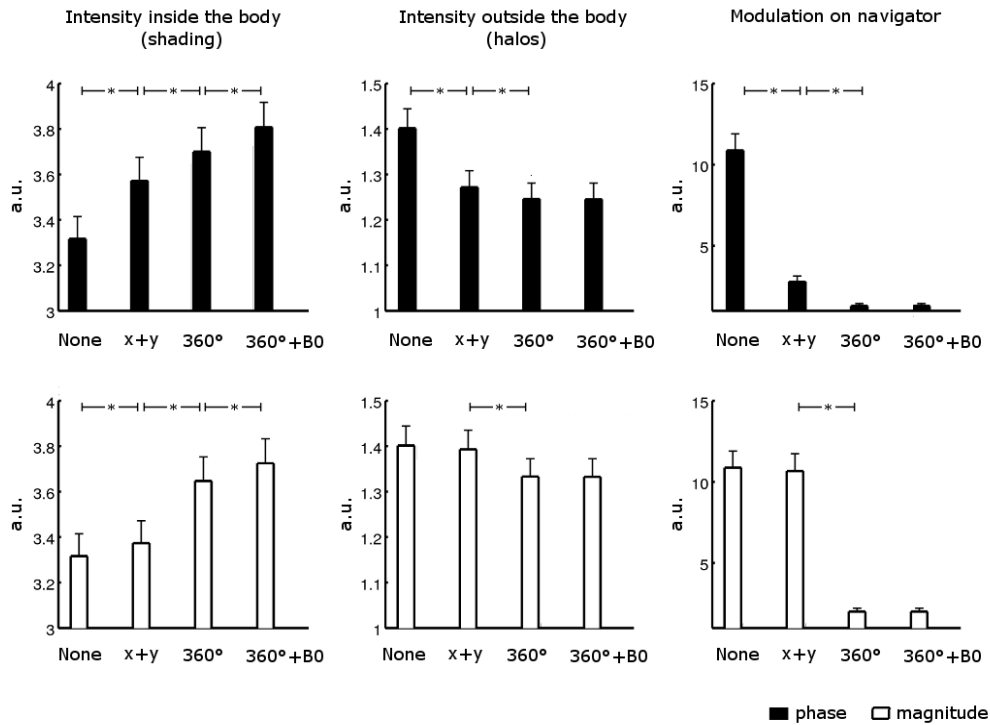

**Figure 58 Performance of phase method (black) and magnitude method (white) comparing correction schemes.**

Quantification of the improvement is shown in Figure 58. An increase in signal inside the object is seen with all correction methods and this improves when more spokes are used, and furthermore with the B0 correction even though the improvement in the image appearance was minor. Furthermore, reductions in signal in air were seen with all improvement schemes and these were significantly lower when more spokes were used. In cases where motion corrupted the acquisition along x and y Cartesian axes, the correction schemes based on these directions failed. Effects of this are shown in Figure 59. In our dataset this was observed on 4% of the acquired slices, mostly at the apex of the heart.

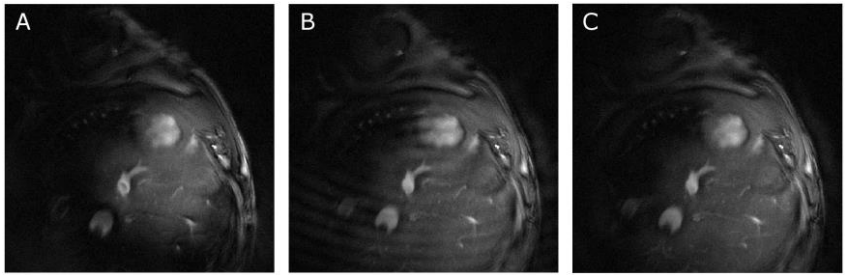

**Figure 59 When a respiration event occurs during the acquisition of two perpendicular directions in the x/y scheme the correction can fail: A) Uncorrected image. B) Linear terms are corrected using x and y axes (phase method). C) Linear terms are corrected using the whole sweep (phase method). Similar results are obtained with the magnitude method.**

The phase correction method gives consistently better results than the magnitude correction method, with more signal inside the image and less signal outside of the image in each case.

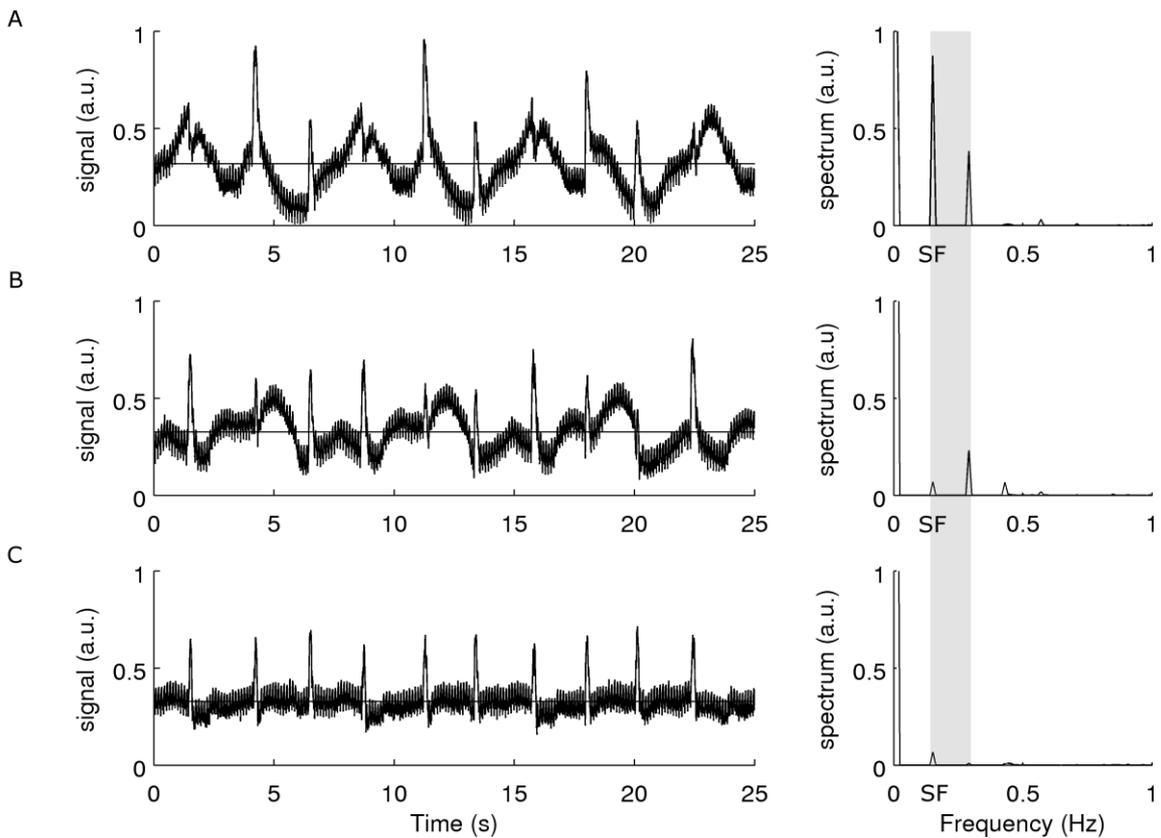

**Figure 60 Gating signals and respective spectra showing the modulation at the "sweep frequency" (SF): A) before any correction. B) after correction using magnitude of spokes (using the whole sweep). C) after correction using phase of projections (using the whole sweep).**

The gating signals are shown in Figure 60 for each correction scheme along with their power spectra. A significant reduction in components corresponding to the sweep frequency and its first harmonic is seen with each correction method, again with the phase correction performing better than the magnitude correction.

## 6.4 Discussion

The methods proposed for radial trajectory correction perform well in the difficult context of mouse cardiac imaging in free breathing. Corrections based on phase data were more effective than the magnitude data and that it is worthwhile correcting for the B0 effect. For retrospectively-gated sequences, the navigator signal is better with the phase correction method.

To measure k-space shifts, this experiment compared a method based on the magnitude of the spokes to one based on the phase of the projections. Not surprisingly, using the phase of the projections gave a more accurate estimate of gradient inaccuracies, due to the high sensitivity of this method. However, the method based on the magnitude achieved significant reduction in the image artifacts, and is still of clear benefit in cases where noise prohibits use of the phase data.

In order to improve the correction in the presence of cardiac and respiratory motion spokes sweeping a full circle were used. Although data was acquired with a small angle increment, any radial acquisition scheme could in theory be used (for example, a golden ratio scheme is often used (145)). In most cine MRI protocols each frame of the cardiac cycle is reconstructed independently, and opposite spokes acquired with this scheme are unlikely to fall within the same heart phase. In free breathing, if just two spokes are used to calculate the correction factors it is possible that they are acquired at peculiar instants in the cardiac or respiratory cycles that gives a poor fit to the average situation for the mouse. A greater number of spokes acquired as the heart moves leads to an averaging out of these types of effect making the correction scheme less likely to be influenced by abhorrent signals.

## 6.5 Chapter summary

Retrospective correction of free-breathing radial MRI data can be obtained using the imaging data, without requiring additional calibration or preparation. This correction can be used to improve image and navigator signal quality in radially acquired cine MRI. As seen in Chapter 5, this acquisition scheme is useful to obtain high speed-up with acceleration methods. This chapter concludes the section of this thesis focussing on global function. The next section will describe methods for tissue characterization with MRI in the context of mouse models of myocardial infarction.

# Section 3: Tissue characterization

**Chapter 7**

# A fast protocol for tissue viability assessment in mice

In this chapter a method to perform late gadolinium enhancement (LGE) imaging is presented, optimized for tissue viability assessment in interventional mouse models of myocardial infarction.

## *7.1 Introduction*

In preclinical studies assessing efficacy of novel treatments for myocardial infarction (MI), reliable quantification of infarct size is key in addressing the pathophysiology of ischaemia and predict long-term outcome (146). MRI is non-invasive and allows longitudinal assessment of the same animal from the acute stage to the subsequent weeks to monitor adverse remodelling (34, 147, 148). For this purpose, late gadolinium enhancement (LGE) imaging can be used to visualise tissue viability in vivo after injection of a contrast agent (149, 150). This protocol is routinely performed in patients with suspected MI via breath-hold segmented inversion recovery (IR), where images are typically acquired 10-30 minutes after intravenous injection of a gadolinium-based contrast agent (0.1-0.3 mmol/kg) (149). Optimisation of the inversion time is performed for each patient to null healthy myocardium signal in order to obtain maximum contrast in the late enhancement images (151).

In mice, the technique is more challenging due to the faster heart rate and smaller dimensions of the heart (38). Variability between subjects, both at baseline and in response to induced injuries, means that considerable numbers of mice can be required to establish treatment efficacy. Furthermore, long term doses of anaesthesia themselves cause cardiovascular effects confounding measurements (37). It is clear therefore that both fast and high-throughput protocols are needed for useful translational studies.

Studies of late gadolinium enhancement in rodents using IR at field strengths of 4.7T(89), 7T(92), and 9.4T(90, 91) have used segmented gradient-recalled echo (GRE) or fast low-angle shot (FLASH) sequences, acquiring several echo lines following an inversion. Recently, Price et al. (93) have proposed a LGE protocol for rodents based on a fast multi-slice approach. Contrast is optimised in each mouse by acquiring a Look-Locker sequence (22) to select the optimum TI to null healthy myocardium, a multi-slice IR-GRE sequence is then used for LGE imaging. Multi-slice acquisition in place of multi-echo allows for higher flip angles and therefore higher SNR efficiency (93).

In clinical examinations it is possible to perform the LGE acquisition during a breath-hold (151). In anaesthetised rodents, the reduced time and signal available prevents this kind of acquisition. To avoid motion artifacts, respiration signals are commonly used to trigger the inversion pulses (90, 92, 93). The variability of respiration rate between subjects leads to different saturation effects for each scan, reducing the reproducibility of the contrast between different subjects or sessions.

In an effort to accelerate acquisition, an IR sequence was developed and validated at 4.7T, providing high-quality LGE images as a three-minute extension to our standard imaging protocol for functional assessment. This strategy relies on acquiring multiple slices in every heartbeat interleaving slice packages between alternate TRs, and can be applied successfully at any point during the first 15 minutes after intravenous contrast agent injection. It was found that gating based solely on ECG gives artifact-free images and the sequence TR can be kept consistent within 550-750 ms. Choosing a suitable flip angle and TR, an optimum compromise was found between contrast and speed. With our sequence, contrast is consistent between animals for the same dose of gadolinium, and individual TI optimization is not crucial for reliable contrast.

## *7.2  Materials and methods*

### 7.2.1 Phantom Experiments

To select the parameters in our in vivo sequence, healthy and infarcted tissue were modelled with 7 vials containing different concentrations of Gadovist (Bayer Schering Pharma, Berlin, Germany) in water. Three of the concentrations used were chosen to model healthy myocardial tissue (T1>0.5 s) and four for a range of potential pathological enhancements (T1<0.5 s). Imaging was performed at 4.7 T with a Bruker BioSpec 47/40 system (Bruker Inc., Ettlingen, Germany). A birdcage coil of 12cm was used for both signal excitation and reception. T1 mapping was performed with an inversion-recovery Look-Locker sequence (FOV 8 cm, 128x128, TR/TE=10s/6.5 ms, BW 50 kHz, FA=20°, interecho time 30 ms, 50 echoes, 1 NEX). Data were fitted with an algorithm based on minimising $\chi2$ with Look-Locker correction factors as described in Deichmann et al (152).

IR sequences were then acquired with different flip angles and TRs (FOV 8 cm, 128x128, TE=2.8 ms, sequence TR 350-1550 ms, TI=280, BW 64 kHz, FA=10°-90°, 1 NEX, slice thickness 0.8 mm, 0.2 mm gap, selective inversion 0.8 mm thickness with 5 ms sech shaped pulse) to evaluate SNR and contrast for T1 and TR values covering the range of parameters that might be used in in vivo experiments. Like in the in vivo experiments, slices were acquired in two groups in alternate TRs to allow time for efficient T1 recovery following inversion (Figure 61).

Hyperenhancement (i.e. the ratio of the difference between enhanced and normal myocardial signal to normal myocardial signal, as described in Simonetti et al.(150)) and contrast to noise ratio (CNR, here defined as the ratio of the difference between enhanced and normal myocardial signal to the standard deviation of the noise (150)) were calculated in ROIs manually drawn in each vial. Hyperenhancements and CNR between pairs of vials were observed to find optimal scan parameters for *in vivo* scans. Optimal scan parameters would maximise both hyperenhancement and CNR.

### 7.2.2 In Vivo Experiments

Thirty-five C57bl6 male mice (8-10 weeks) from on-going drug assessment studies were selected to be scanned with our protocol 24 hours after induced acute MI.

### 7.2.3 Animal model of acute myocardial infarction

The animal model of acute myocardial infarction was obtained ligating the left coronary descending artery (LAD). Classically, the territory perfused by the human LAD involves the anterior aspect of the LV wall and the anterior two thirds of the septum. Ligations of the LAD in mice, instead, consistently cause infarction of the free wall of the LV extending to the apex, with sparing of the septum (153).

Surgery was performed by an independent operator (Dr Carmen Methner, Dept. of medicine). Myocardial infarction was induced in mice as previously described (154). Briefly, mice were anaesthetized with gaseous isoflurane, intubated endotracheally, and ventilated with 3 cmH2O positive end-expiratory pressure. A small thoracotomy was performed and the heart was exposed via stripping of the pericardium. The main branch of the LAD was identified and ligated for one hour using a 7.0 suture. The thoracic incision was closed and mice were monitored until recovery. Buprenorphine (0.05 mg/kg) was given subcutaneously just prior the end of the surgery and during the recovery period as necessary.

## 7.2.4 Image Acquisition

Animals were anaesthetised with gaseous isoflurane both for induction (3% in 1l/min O2) and maintenance (1.5-2% in 1l/min O2). A pressure sensor for respiration rate was used to monitor anaesthesia depth, rate was maintained in the range 30-45 breaths per minute. Prospective gating of the MRI sequences was achieved with ECG monitoring using paediatric EGC electrodes (3M Europe, Diegem, Belgium) on left and right forepaws. Body temperature was monitored using a rectal thermometer and a flowing-water heating blanket was used to maintain animal temperature at 37° C throughout the experiment.

A birdcage coil of 12cm was used for signal excitation and animals were positioned prone over a 2cm surface coil for signal reception. Depending on the heart rate of the animal, total scan time including functional evaluation before the LGE assessment ranged from 15 to 20 minutes.

## 7.2.5 Cine-MRI

After initial localization images, 4-chamber and 2-chamber views were acquired (FISP, TR/TE 7 ms/2.4 ms, 13-20 frames, 3.5 cm FOV, 256x256 matrix, 1 mm slice thickness, bandwidth 64 kHz, flip angle 20°, NEX 2). Using these scans as a reference, short axis slices were arranged perpendicularly to both the long-axis views to cover the left ventricle (LV) (FISP, TR/TE 7ms/2.4ms, 13-20 frames, 3.5 cm FOV, 256x256 matrix, 1 mm slice thickness, bandwidth 64 kHz, flip angle 20°, NEX 2). Full LV coverage was achieved with no slice gap with 8-10 slices.

## 7.2.6 Inversion Recovery Sequence

After the acquisition of the standard cine protocol, contrast agent was injected in situ (Gadovist, Bayer; 0.3 mmol/kg i.v.). Within the first 15 minutes after injection, IR images were acquired (FLASH, FOV 3.5 cm, 256x256, 0.8 mm slice thickness with 0.2 mm gap between adjacent slices TE= 2.8 ms, sequence TR=550-750 ms, FLASH TR 7 ms, bandwidth 64 kHz, flip angle 60°, 1 NEX, 0.8 mm, 0.2 mm gap, selective inversion 0.8 mm thickness with 5 ms sech shaped pulse). No respiration gating was used. For each mouse, a delay was calculated for the trigger such that after the inversion pulses and a fixed TI of 280 ms, acquisition would occur at end-diastole. The sequence TR was, as a result, an even multiple (4 or 6 times the heart period) between 550 and 750 ms. Slices were acquired in two groups in alternate TRs to allow time for efficient T1 recovery following inversion (Figure 61). Eight slices were usually required to cover the entire heart in a typical time of 3 minutes (depending on HR).

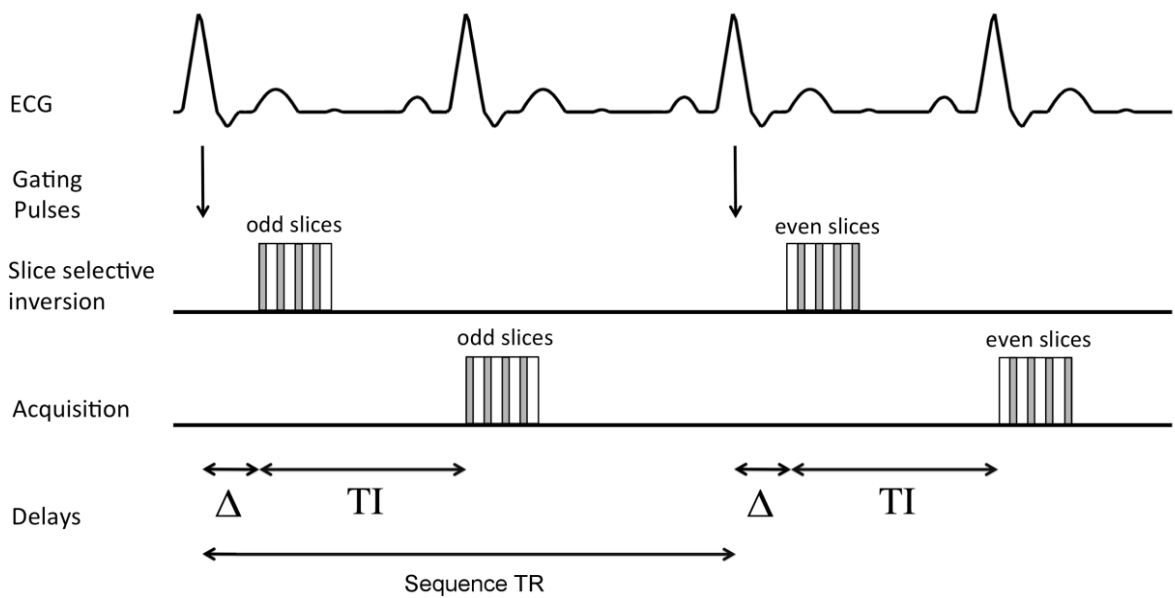

**Figure 61 Schematic representation of the acquisition strategy for our inversion recovery sequence. Odd and even slices are inverted and acquired in alternate blocks within each sequence TR. The delay Δ is calculated such that the inversion time TI is reached during end-diastole.**

### 7.2.7 Image Analysis

Delineation of the LV at each phase of the cardiac cycle excluded the papillary muscles and trabeculations throughout, which has been reported as improving reproducibility in LV mass measurements (155). The regions from each slice were combined using Simpson's rule to provide LV mass using Segment v1.9 (48).

To assess repeatability of these measurements, eight healthy male mice from the same batch were selected for repeated acquisition during the same session. Data were then analysed in a random order.

### 7.2.8 Viability Assessment

Infarcts were manually delineated on the inversion recovery images. Values are expressed as ratios of the LV mass as measured from the standard cine protocol. The hyperenhancement was measured for each mouse. To assess repeatability of infarct size measurements, eight mice were selected randomly and infarct size measurements were repeated twice on the same image stack.

### 7.2.9 Comparison with Histology

After completion of the protocol, six mice were randomly selected for histological assessment of infarct size to compare with the MRI measurements.

Histological measures were performed by an independent investigator (Dr. Carmen Methner, Dept. of Medicine). Hearts were excised, frozen, and then cut into 1 mm thick transverse slices. The slices were incubated in 1% triphenyltetrazolium chloride (TTC) in sodium phosphate buffer (pH 7.4) at 37°C for 20 min. This protocol stains the non-infarcted myocardium brick red. The slices were immersed in 10% formaldehyde solution to enhance the contrast between stained (viable) and unstained (necrotic) tissue. The infarct zone was determined by a planimetry measurement and was expressed as a percentage of LV.

## *7.3  Results*

### 7.3.1 Phantom Experiments

The T1 measurements of the vials were 0.25, 0.29, 0.36, 0.45, 0.60, 1.0, 2.80 s respectively (Figure 62). T1 values of 0.5-3 s were used to model healthy myocardium and vials with T1 < 0.5 s were used to model varying degrees of possible enhancement. The choice of 0.5 s as a threshold was made based on typical values observed during in-vivo experiments in our centre.

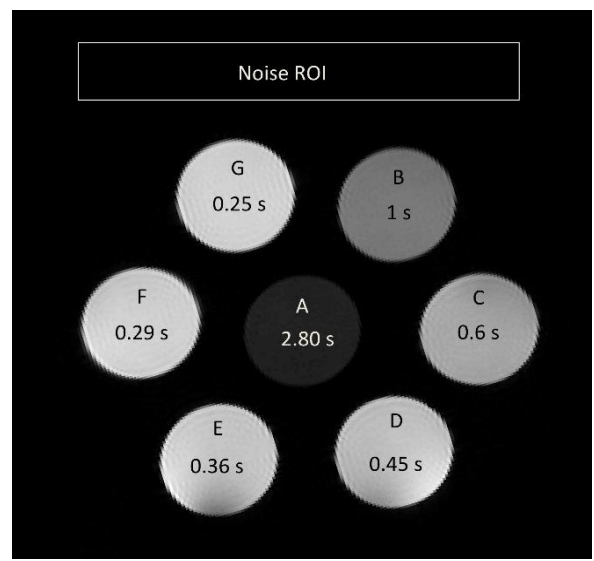

**Figure 62 Phantom used to optimize the IR sequence, T1 values are shown on each tube. For CNR and hyperenhancement calculations, values between 0.5-3 s (A-C) were used to model viable myocardium, values between 0-0.5 s (D-G) were used to model non-viable myocardium.**

The resulting hyperenhancements and CNRs between couples of vials can be observed in Figure 63. An area exists, corresponding to a sequence TR between 550-750 ms, a fixed TI of 280 ms, and a FA of 60° where the hyperenhancement was at least 200% in every case.

## 7.3.2 Late Gadolinium Enhancement

During the LGE protocol the heart rate and respiration rate were stable within 10% throughout each examination (intervals for heart rate were 400-500 bpm, respiration rate 30-45 bpm). The demonstrated protocol gave good contrast for each mouse, and hyperenhancement was at least 200% (apex 404 ± 265 [186-893], mid-ventricle 580 ± 366 [254-1291], base 540 ± 325 [268-1238], percentage mean ± s.d. [range], one slice per location, n=35) between infarcted and remote tissue for all the mice examined. No significant difference was observed between enhancement in apex, mid-ventricle or base. Intra-observer variability of LV mass measurement was found to be within 4%, in line with similar experiments (40, 43, 156). Infarct size intra-observer variability was within 4%. No artifacts arising from respiration were observed in any of the acquired images.

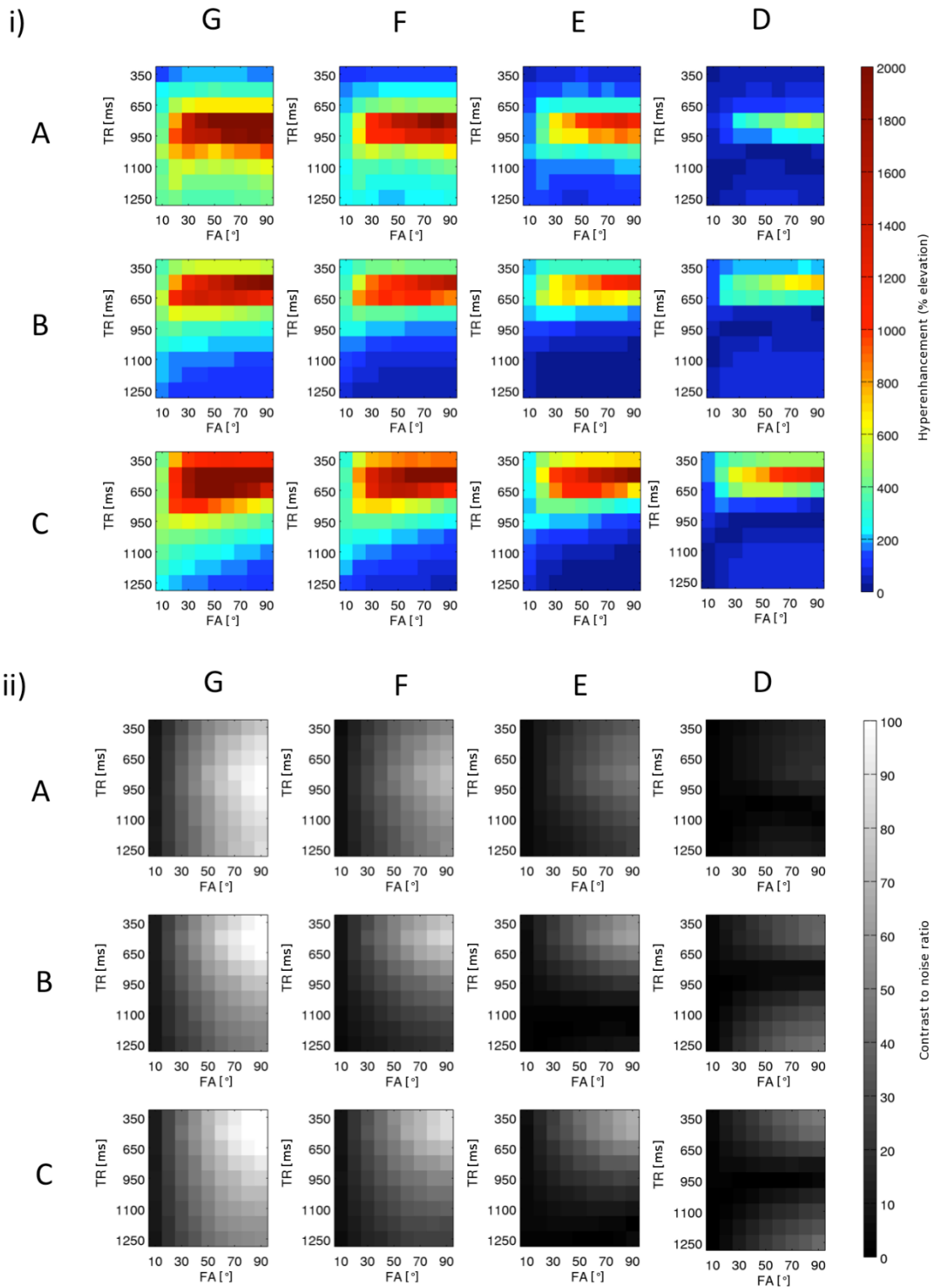

**Figure 63 i) Hyperenhancement between vials representing non-viable tissue and vials representing viable tissue as a function of FA and TR, from top left: Hyperenhancement between vial G and vial A, vial F and vial A, etc. ii) CNR between vials representing non-viable tissue and vials representing viable tissue as a function of FA and TR, from top left: CNR between vial G and vial A, vial F and vial A, etc. The noise ROI used to calculate CNR is shown in Figure 62.**

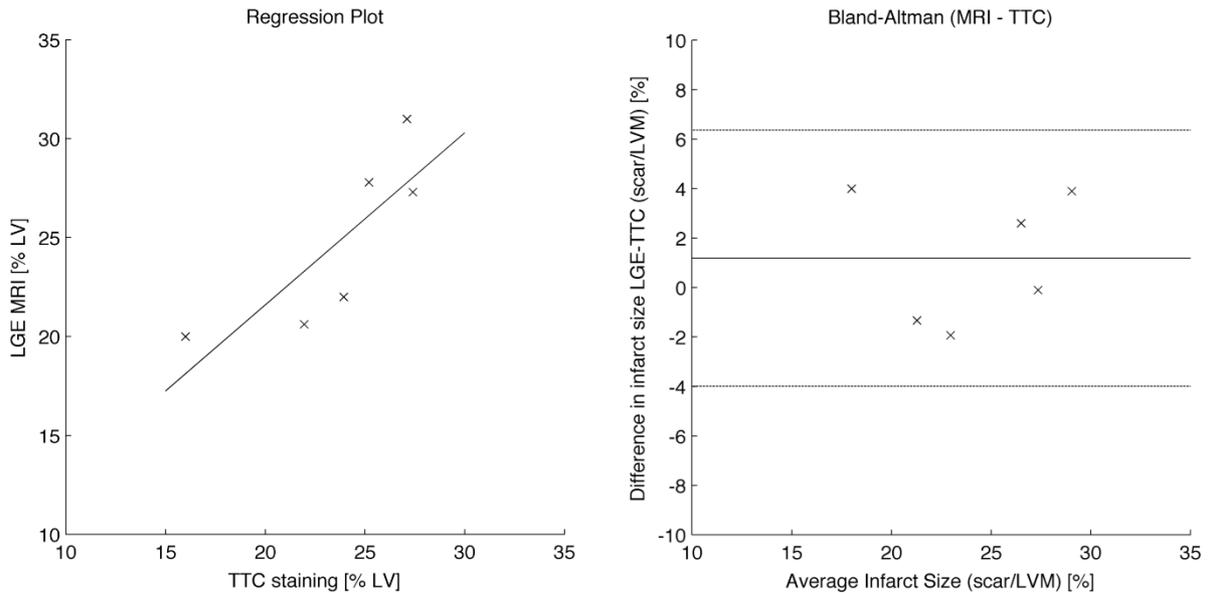

**Figure 64 Regression plot (slope=0.87, intercept=4.2, r2=0.67) and Bland-Altman plot comparing TTC staining with MRI. The graphs show excellent agreement between the two measurement techniques.**

## 7.3.3 Comparison with Histology

MRI measures and TTC results for the same animals are shown in Bland-Altman and correlation plots in Figure 64. Our data show excellent correlation and agreement between the techniques. Representative images of histology and MRI are shown in Figure 65.

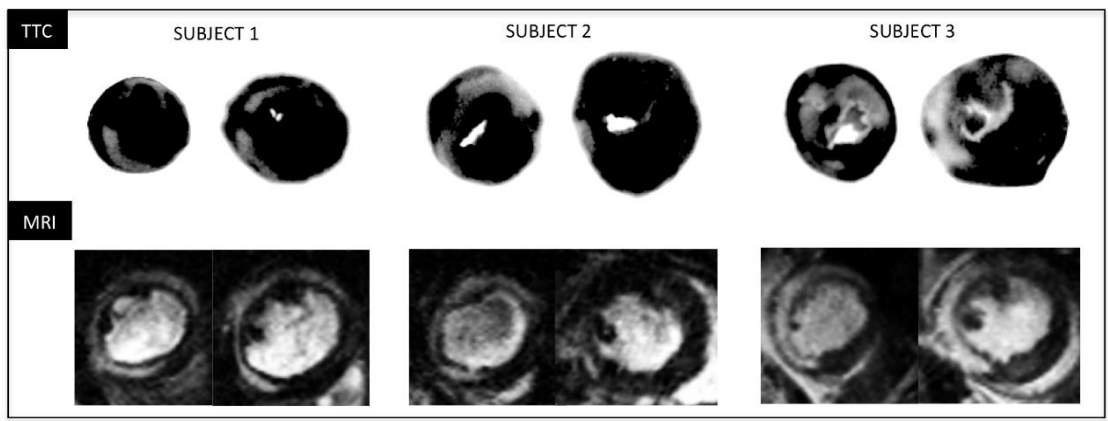

**Figure 65 Comparison of TTC histological stain (top) and MRI LGE imaging. In both images the infarcted regions appear hyperintense.**

## 7.4 Discussion

Late gadolinium enhancement imaging is a clear and accurate way to assess infarct size in vivo. While the protocols for performing this technique in humans are already well established, in rodents there is still a debate about the best method to use (92, 93). For instance, a recent study from Protti et al (92), has shown that in mice, TI-optimised standard IR FLASH does not perform better than a cine-MRI sequence without preparation at 7T.

In common with a recent study from Price et al. (93), this approach samples multiple slices rather than multiple echoes in each repetition. Most other LGE studies in rodents use repeated sampling of the same slices per TR (i.e. in Look-Locker sequences), which limits the flip angles that can be used. Since this sequence does not require this, larger flip angles can be used without saturating the slices and a greater SNR efficiency is therefore achieved.

The present study builds on the work of Price et al. by demonstrating in a phantom of typical parameters expected in healthy and infarcted myocardium that a subject-specific optimisation of TI is not required. This is further demonstrated in the consistency of contrast obtained between the animals in this study which all had the same TI. In pre-clinical LGE sequences, T1 saturation is usually exploited using sequence TRs similar to the T1 relaxation time of healthy myocardium (92, 93). In our protocol a similar principle is used, although our phantom experiments predicted that for a multi-slice experiment a shorter TR (550-750 ms) gives optimal contrast, higher speed and reliable hyperenhancement, confirmed by our in vivo results.

In small animals, the respiratory period is comparable with TR. A consequence of this is that if respiratory signals are used to gate the acquisition, their variability limits reproducibility. For this reason, to obtain higher repeatability of contrast, individual TI optimisation is usually performed in the most recent techniques using respiration signals to gate inversion pulses (92, 93). Our protocol does not use respiratory gating so that TR can be maintained in a confined range (550-750 ms) and contrast is reliable.

The present study demonstrates that a relatively low concentration of contrast agent can be used at 4.7T (0.3 mmol/kg, compared to 0.5-0.6 mmol/kg commonly used in mice (90, 91)), yielding good contrast which can be observed moments after injection. Furthermore, the contrast achieved is consistent between animals.

A multi-slice approach with a selective inversion (rather than a global inversion) may suffer from crosstalk between slices. In addition, the delay between inversion pulses being applied and the slices being acquired (i.e. 280 ms) allows time for cardiac motion, which can be deleterious to image quality. In this study, the effect of these potential issues was assessed by comparing hyperenhancement levels in neighbouring slices through the heart. There was no significant difference between slices, so these potential issues are not problematic in this sequence. This is further supported by the good agreement between the MRI assessment of infarct size and the TTC measurement of non-viable tissue.

None of the images showed motion artifacts, and even if the cardiac and respiratory rates throughout the scans were similar to typical literature values (90, 92), the relatively low breath rate was a consequence of the rather high (1.5-2%) isoflurane concentration used in this study. Under different anaesthetic regimes that do not achieve similar respiration rates to the present study, it may no longer be possible to achieve high quality images without appropriate gating.

While for studies requiring a complete assessment of the infarct on an individual basis a 3D method (i.e. Bohl et al. (90)) could be preferred, the protocol described here is suited to group analysis investigating infarct size. Serial, longitudinal scanning in mouse studies is expensive and time consuming, moreover it is a stressful procedure for animals in the acute stage following surgery. This protocol is fast and requires less time under anaesthesia, which makes it tailored to the testing of multiple experimental drugs with multiple dose levels in mouse models of acute MI.

In summary, this protocol consists in a multi-slice method with an efficient sampling scheme: multiple slices in successive TRs with an interleaved acquisition. The protocol achieves excellent CNR and accuracy in a short time for increased throughput and reliability. The strategy described here provides a fast viability assessment in mice one day after infarction, as an add-on at the end of the standard cine-imaging protocol. The total protocol time is 35 minutes from the start of anaesthesia to the end of MRI, so that more than 10 scans in a typical day are feasible. The protocol is easy to implement and does not require an additional sequence for TI optimization. In the context of drug testing in MI, a fast protocol achieves greater throughput and reduces mortality and adverse effects of anaesthesia.

## 7.5 Application: testing the efficacy of MitoSNO

The following paragraphs will present an experiment investigating the pharmaceutical efficacy of MitoSNO, a mitochondria-targeted drug (112). In this study, MRI was used to assess tissue viability at the acute stage and then to follow up heart function in the same mice 28 days after I/R injury. Mitochondria-targeted agents for protection in I/R injury were given intravenously before reperfusion. Importantly, a longitudinal study design was used which is ready for translation to clinical trials.

### 7.5.1 Background

Recently it has been shown that the mitochondria-targeted S-nitrosothiol MitoSNO protects against acute ischemia/reperfusion (IR) injury by inhibiting the reactivation of mitochondrial complex I in the first minutes of reperfusion of ischemic tissue, thereby preventing free radical formation that underlies IR injury (112). However, it remains unclear how this transient inhibition of mitochondrial complex I at reperfusion affects the long-term recovery of the heart following I/R injury.

In this experiment infarct size was assessed in vivo and heart function was assessed on the same mice at the chronic stage.

### 7.5.2 Methods

Mice were subjected to 30 min left coronary artery occlusion followed by reperfusion and recovery over 28 days. In seven mice, MitoSNO (100 ng/kg) was applied 5 min before the onset of reperfusion followed by 20 min infusion (1 ng/kg/min). Control mice (n=7) received a saline injection. Surgery was performed by an independent operator.

MRI was performed 24 hours after I/R injury at 4.7 T with a Bruker BioSpec 47/40 system (Bruker Inc., Ettlingen, Germany). A quadrature birdcage coil of 12cm was used for signal excitation and a four-channel cardiac receiver coil for signal reception. Animals were positioned prone. After initial localization images, 4-chamber and 2-chamber views were acquired. Using these scans as a reference, short axis slices were arranged perpendicularly to both the long-axis views to cover the LV (FISP, TR/TE 6 ms/2.1 ms, 13-20 frames, 3.5 cm FOV, 256x256 matrix, 1 mm slice thickness, bandwidth 78 kHz, flip angle 20°, NEX 1). Full LV coverage was achieved with no slice gap with 8-10 slices.

After the acquisition of the standard cine protocol, late gadolinium enhancement (LGE) was performed using a multi-slice inversion recovery sequence as described above. The MRI protocol excluding the LGE was repeated 28 days after I/R injury.

### 7.5.3 Results

MitoSNO treated mice exhibited reduced infarct size (2.1% ± 0.5% of total LV mass vs. 16.4 ± 2.2% in controls, p=0.001) and preserved function (64.4% ± 1.9% left ventricular ejection fraction vs. 51.3 ± 2.1% in controls, p=0.001). In addition, MitoSNO at reperfusion improved outcome measures 28 days post-IR, as preserved systolic function (63.7% ± 1.8% left ventricular ejection fraction vs. 53.7 ± 2.1% in controls, p=0.01).

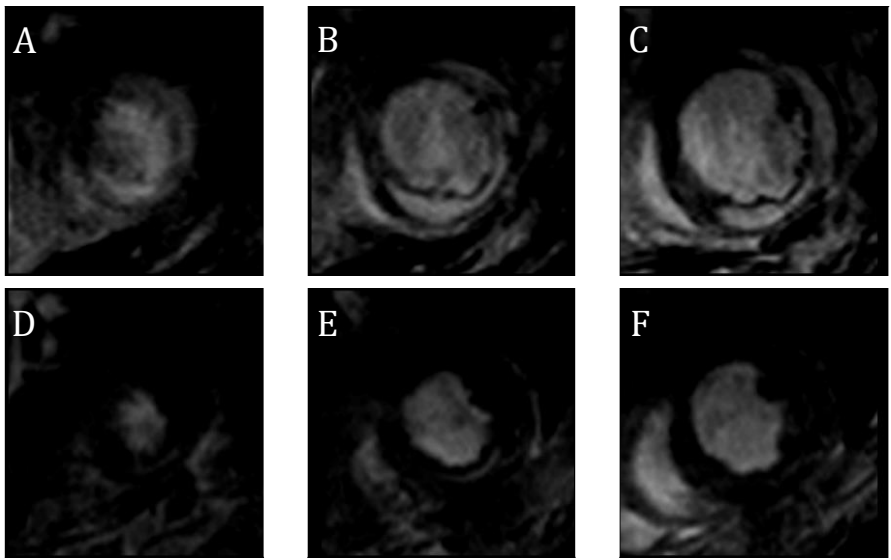

**Figure 66 Late gadolinium enhancement at 24 hours post I/R injury: A-C) Short axis slices in an untreated mouse: bright areas indicate non-viable tissue. D-F) Corresponding slices in a mouse treated with mitoSNO, the infarct area is significantly smaller.**

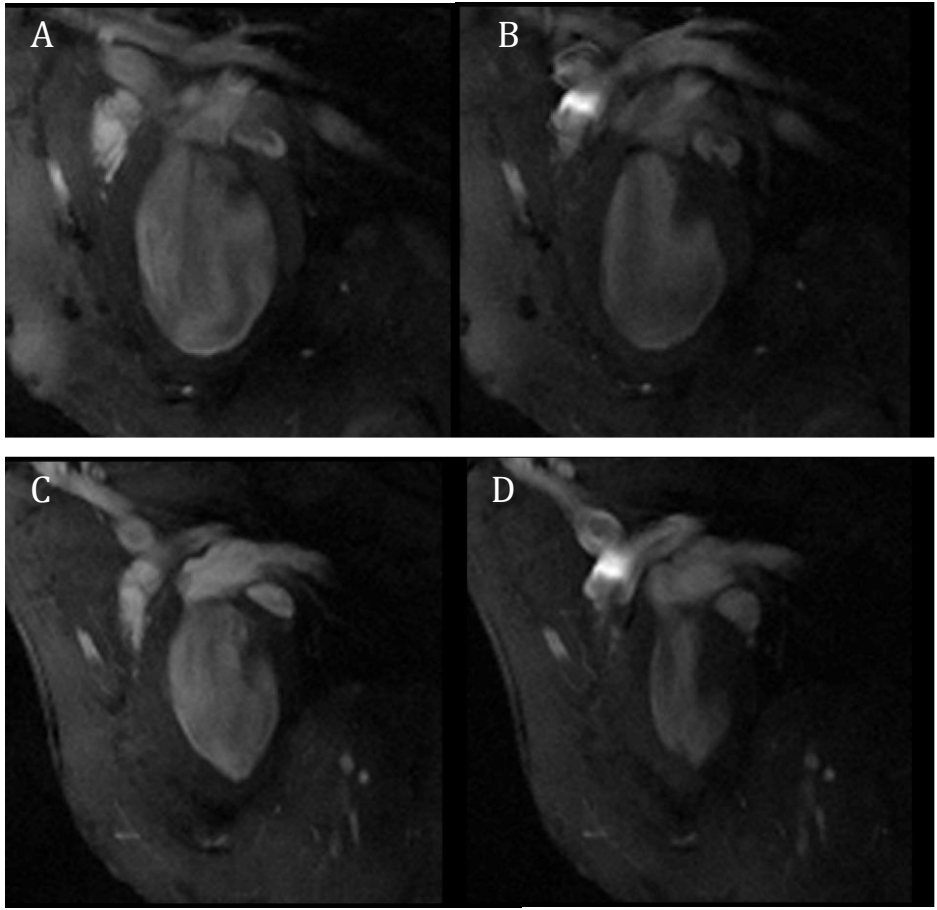

Figure 67 Long-axis 2 chamber view slices showing heart function in the follow-up 28 days after I/R injury: A) Untreated mouse in End diastole. B) Untreated mouse in end-systole. C) Mouse treated with MitoSNO in end diastole. D) Mouse treated with MitoSNO in end-systole. MitoSNO prevents the development of heart failure.

|                    | Controls (n=7) | MitoSNO (n=7)      |
|--------------------|----------------|--------------------|
| **LVM(µl)**        | $104 \pm 4$    | $88 \pm 5$ *       |
| **LVEDV(µl)**      | $65 \pm 6$     | $49 \pm 4$ *       |
| **LVESV (µl)**     | $32 \pm 3$     | $18 \pm 2$ **      |
| **LVSV(µl)**       | $34 \pm 3$     | $31 \pm 2$         |
| **LVEF(%)**        | $51 \pm 2$     | $64 \pm 2$ ***     |
| **Infarct size (% LV)** | $16 \pm 2$ | $2.1 \pm 0.5$*** |

Table 9 MRI-derived left ventricular volumes at the acute stage, data are mean ± SEM. * p <0.05, ** p<0.01, *** p<0.001

| | Controls (n=7) | MitoSNO (n=7) |
|---|---|---|
| **LVM(µl)** | 120 ± 7 | 101 ± 6 * |
| **LVEDV(µl)** | 76 ± 6 | 60 ± 6 * |
| **LVESV (µl)** | 36 ± 4 | 22 ± 3 * |
| **LVSV(µl)** | 40 ± 3 | 38 ± 3 |
| **LVEF(%)** | 54 ±2 | 64 ±1 ** |

Table 10 MRI-derived left ventricular volumes at the chronic stage, data are mean ± SEM. * p <0.05, ** p<0.01

### 7.5.4 Conclusions

LGE MRI was successfully used to investigate a novel pharmaceutical in-vivo, performing a follow-up on the same mice to assess remodelling. MitoSNO action acutely at reperfusion reduces infarct size and protects from post post-myocardial infarction heart failure. Therefore, targeted inhibition of mitochondrial complex I in the first minutes of reperfusion by MitoSNO is a rational therapeutic strategy for preventing subsequent heart failure in patients undergoing IR injury.

## 7.6 Chapter summary

This chapter has introduced a new method for fast assessment of tissue viability in the mouse heart. Comparing to existing methods, the protocol exposed here is faster and requires no TI optimization prior to scanning. The application of this method to a current problem in pharmacology demonstrates its potential in a practical situation. The protocol exposed here offers a new efficient tool for the measurement of infarct size in mouse models of heart disease. Although this method correlates well with histology and is highly sensitive in detecting scar, it lacks molecular specificity. Positron emission tomography, on the other hand, can be used for molecular imaging of specific metabolites but it lacks anatomical detail. The next chapter exposes a novel method for combining MRI and PET in the mouse heart.

# Chapter 8

# Multimodality assessment of the infarcted mouse heart

As seen in the precedent chapters, MRI achieves unparalleled anatomical detail and can be used to assess function and viability in vivo. However, all of the methods described so far rely on contrast between different tissues to derive volumetric measurements and not on a quantitative measure extracted from the signal itself. Although MRI gives excellent contrast and anatomical detail, the signal is non-specific. Positron emission tomography is a complementary technique highly specific to molecular binding but lacking the anatomical detail of MRI.

By combining these imaging techniques, damage and recovery can be assessed *in vivo* on the cellular, tissue and whole-organ scale. In this chapter, a novel method is presented using a combination of complementary measures of heart function on multiple levels using magnetic resonance imaging (MRI) and positron emission tomography (PET).

## 8.1 Introduction

Staging of the treatment in vivo is crucial to test novel pharmaceuticals for acute MI. Standard volumetric measurements of ventricle size at each phase of the heart (cine MRI) can be used to mesure volumes through systole. These parameters give the performance of the heart and provide a sensitive measure for heart failure (see 3.1).

Tissue perfusion can be measured by injecting a contrast agent which rapidly washes out of healthy tissue, though slowly accumulates in infarcted regions. Late gadolinium enhancement (LGE) imaging provides a measure of the infarcted region which compares well with direct histological measures of tissue viability (as described in Chapter 7).

By measuring the motion of the myocardium, muscular stress and contractility can be assessed to evaluate muscular performance throughout the left ventricle (LV) (see 3.1). Displacement encoding with stimulated echoes (DENSE) (78) is an MRI technique that provides these measures.

PET provides molecular imaging of tracers as they accumulate in metabolism. Here the glucose analogue 18F-fluorodeoxyglucose (FDG) was used as a direct marker of cellular viability (see Figure 68).

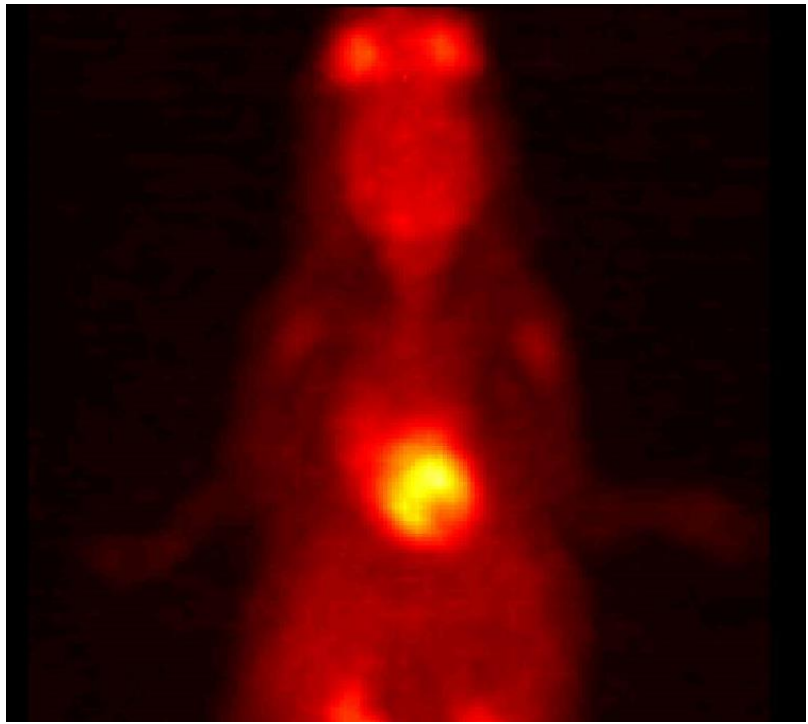

**Figure 68 Maximum intensity projection of a Mouse FDG-PET image.**

Used together, these different markers can give a holistic view of the heart: from cell metabolism to tissue perfusion, to muscle performance and finally to the overall function of the organ as a whole. Here a protocol to perform these techniques together in vivo is demonstrated.

## 8.2 Methods

Male C57Bl/6J mice (n=12) were used and the left anterior descending coronary artery (LAD) was occluded for 30 minutes to induce an ischaemic insult (see 7.2.3) (154, 157). Imaging was performed after 24 hours of reperfusion for every mouse.

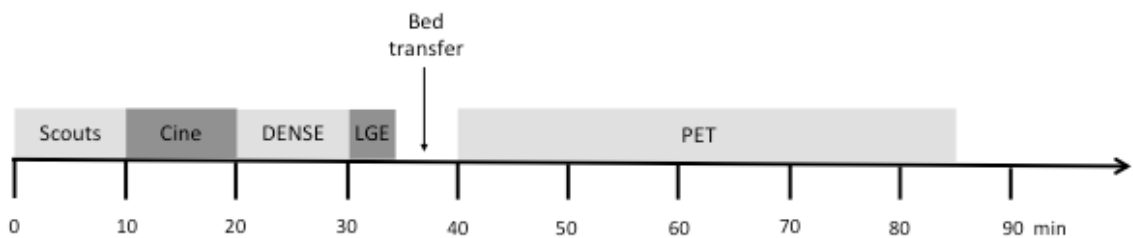

**Figure 69 Overview of the sequential multi-modality imaging protocol. The time spent in the MRI and the time spent in the PET are similar, suggesting that the same examination performed on a combined PET/MRI scanner would take half the time.**

A scheme of the imaging protocol is presented in Figure 69. MRI was performed with a 4.7T Bruker BioSpec system. Anaesthesia was induced with 3% and maintained with 1.25% isoflurane in oxygen. Temperature was monitored with a rectal probe and maintained constant via a heated water blanket, respiration was monitored using a pillow connected to a piezoelectric transducer. ECG signals, used for gating, were acquired with neonatal graphite (3M) electrodes placed over the forepaws.

A 12cm diameter birdcage was used to transmit the signal and a 2cm diameter surface coil was used for signal reception. The imaging protocol consisted of scout scans followed by ECG-gated FISP slices (TR/TE 7ms/2.4ms, 13-20 frames, 3.5 cm FOV, 256 matrix, 1 mm slice thickness, bandwidth 64kHz, flip angle 20°, 2 NEX), in the long-axis and in the short-axis to cover the whole heart. In post processing, volumes during different phases of the ECG were delineated and integrated over the whole heart using Simpson's rule to obtain global functional parameters.

Cine MRI was followed by DENSE imaging (3 short-axis slices encoding from end-diastole to end-systole, 1 mm thick, TR/TE ~200 ms/9.5ms, 3.5 cm FOV, 128 matrix, bandwidth 64kHz, flip angle 90°, 4 NEX with CANSEL (158). The DENSE images were analysed with in-house code using Matlab: phase images were extracted and unwrapped (159) to obtain displacement maps. The Green strain tensor (E) was calculated from the Jacobian matrix (**F**) by means of $\mathbf{E}=(\mathbf{F'F}\text{-}\mathbf{I})/2$, then decomposed into radial and circumferential strain components.

LGE MRI was then performed as described in Chapter 7. After 0.3 mmol/kg i.v. administration of Gd-DTPA, an ECG-gated FLASH IR-sequence was acquired with slices 0.8 mm thick with 0.2 mm gap (FOV 3.5 cm, 256x256, TE/TR 2.8/800 ms, bandwidth 64kHz, flip angle 60°, 1 NEX). Infarct size was measured by delineating the enhanced regions in the slices and integrating over the left ventricle.

After MRI, the imaging bed was transferred to the Focus 120 PET camera housed in the Cambridge PET/MRI scanner (160, 161) leaving the MRI receiver coil in place. The heart was positioned in the centre of the PET field of view moving the bed only axially. After injecting 10-30 MBq FDG, list mode gated PET was acquired for 45 minutes. PET data was corrected for detector efficiency, random and dead time, binned into 8 heart phases and reconstructed with a 3D filtered back projection algorithm (128x128x95 matrix 0.8 mm resolution). Standard uptake values were calculated over the last 15 minutes of the scan. For PET/MRI coregistration, a fixed matrix was used to adjust scaling and rotations (162). To derive translations, the slices from cine-MRI were stacked to obtain a 3D reconstruction and interpolated to obtain 0.3 mm isotropic resolution, PET images were interpolated to obtain the same resolution. The end-diastolic PET volume was registered to the end-diastolic MRI volume using SPM-Mouse coregistration tools (119).

## *8.3 Results*

During the imaging protocol heart rate was in the range of 400-500 bpm. LGE images 24 hours after I/R injury confirmed that a region of the myocardium had become necrotic in each mouse examined due to the LAD occlusion. The measurement of left ventricular EF revealed a marked decrease of function in mice with larger infarcts (Figure 72 $R^2$=0.76).

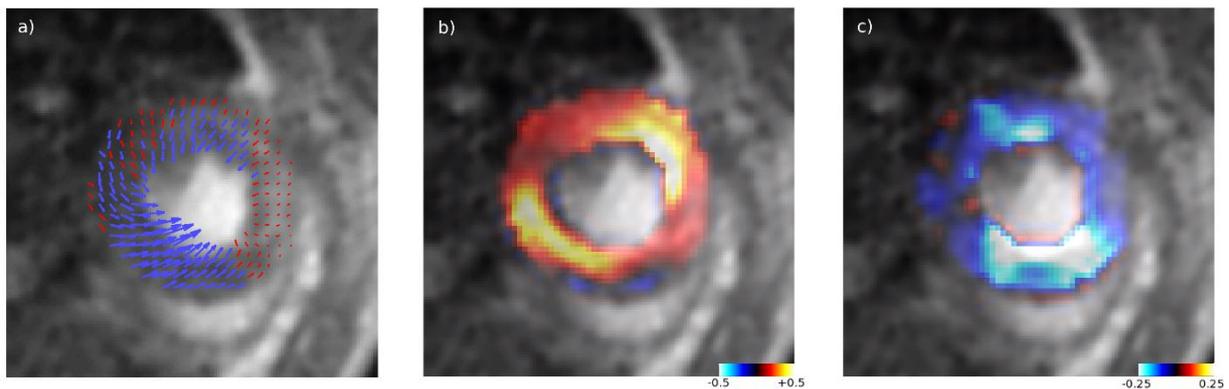

**Figure 70 End systolic short-axis slice from a single mouse: a) Voxel-wise displacements measured by DENSE MRI, areas of reduced contraction are indicated in red. b) Radial strain. c) Circumferential strain: areas of increased circumferential strain correspond to the hypokinetic areas in the displacement map.**

DENSE MRI images showed that the necrotic area and the tissue immediately close to the infarct had a reduced displacement (see Figure 71). Displacement images were processed to obtain radial and circumferential strains. As shown in Figure 70, regions of reduced displacement corresponded to areas with increased circumferential strain, indicating passive contraction.

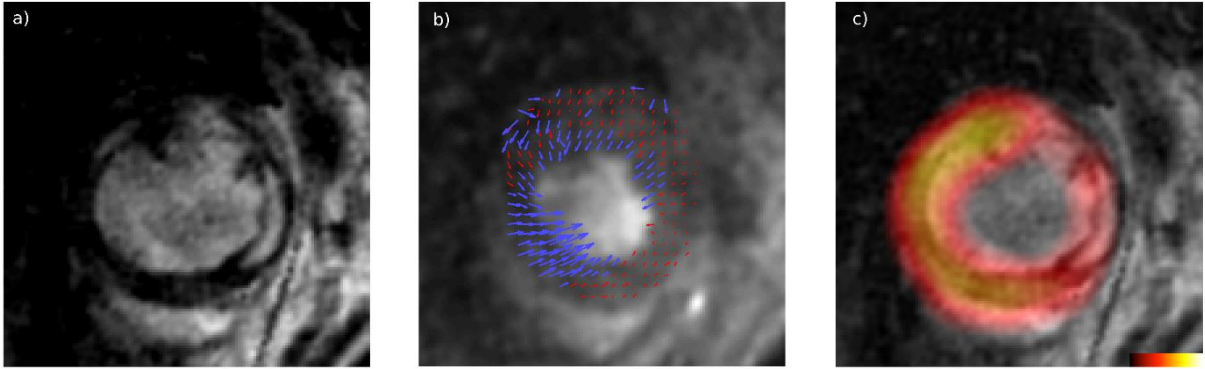

**Figure 71 One short-axis slice from a single mouse. a) End diastolic LGE image, areas of hyperenhancement correspond to non-viable tissue. b) End systolic DENSE-derived Displacement map: a hypokinetic area (marked in red) is present larger than the infarct. c) End diastolic FDG-PET uptake is reduced in the infarcted areas, although small infarcts are not visualised in the PET image.**

Global parameters plotted against LGE MRI infarct size are reported in Figure 72. Radial strain was globally reduced in mice with larger infarcts, due to an impairment in muscle thickening performance. On the other hand, circumferential strain values were globally increased in mice with bigger infarcts, with weakened circumferential shortening. Both strain components correlated well with infarct size (circumferential strain: $R^2$=0.88, radial strain: $R^2$=0.68).

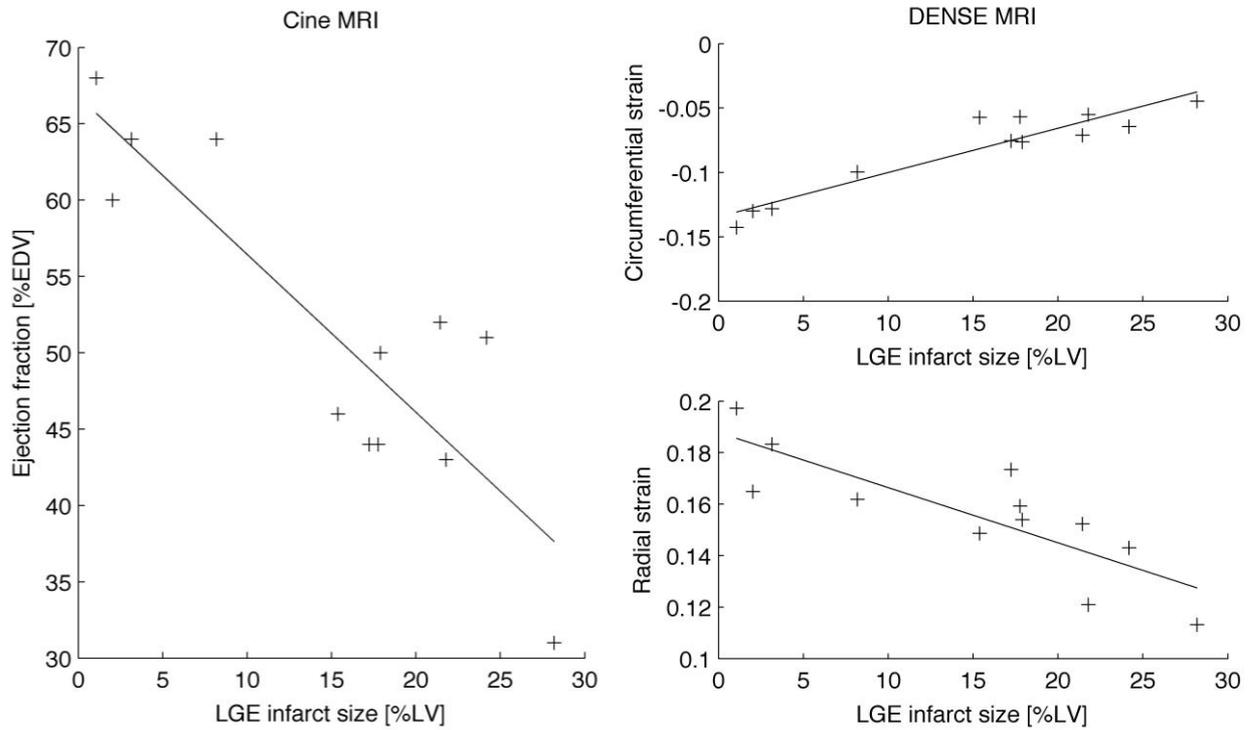

**Figure 72 Ejection fraction and DENSE-derived strains Vs. LGE MRI infarct size.**

Coregistration of PET and MRI was performed easily, requiring only an axial translation. FDG-PET uptake was reduced in the necrotic area, although non-transmural infarcts did not show an observable reduction in the signal (cfr Figure 71). There was good correlation between PET and MRI infarct size ($R^2$=0.57, see Figure 73).

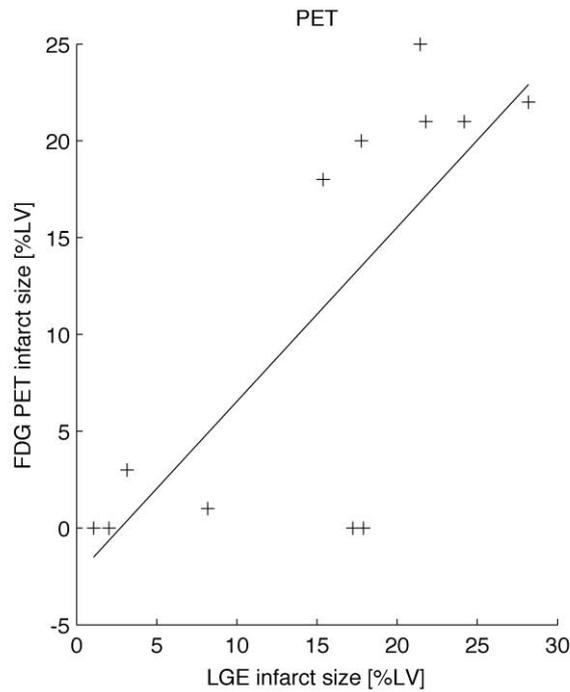

**Figure 73 PET infarct size Vs. MRI infarct size**

## 8.4 Discussion

In vivo imaging represents a non-invasive and translational approach for assessing efficacy of novel treatments targeting MI. In this work cine MRI, LGE MRI, DENSE MRI and FDG PET were performed for the first time in a single session. Each of these provides complementary information about heart disease and treatment to give a fuller picture of the response of the heart to therapy.

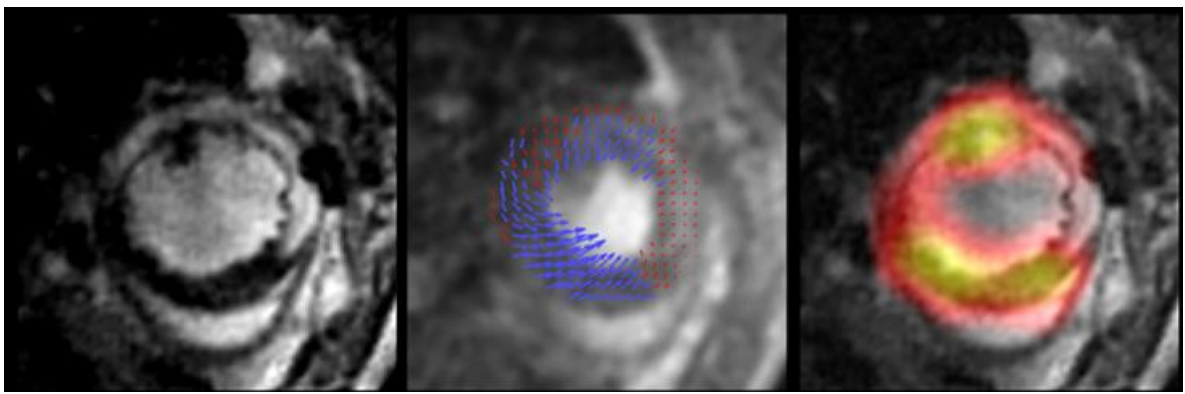

**Figure 74 LGE MRI, DENSE and FDG PET in the same heart, showing different aspects of acute disease.**

LGE MRI successfully identified the extent of the induced infarcts, by probing the tissue perfusion and washout characteristics of the myocardium after injecting a gadolinium contrast agent. EF measured with cine-MRI showed a reduction in function correlated with infarct size at 24 hours after I/R injury. This indicated a decreased efficiency of the whole organ in the acute stage of the disease. DENSE measurements showed that the infarcted as well as neighbouring tissue had become hypokinetic. The inability of these areas to contract efficiently, due to the recent I/R injury, will impact negatively on the disease progression, contributing to adverse remodelling.

On a molecular level, PET was used to assess tissue metabolism showing that glucose metabolism was reduced in the infarction. Despite the high correlation between data from different modalities, the quantities measured represent different aspects of the disease. The information derived from different metrics can be therefore used to stage disease and evaluate the time-course of treatment efficacy.

The acquisition of multi-modal data within one examination can be used to increase sensitivity and specificity of preclinical studies. Cine MRI is considered the gold standard for evaluating global heart function, outperforming echocardiogram, therefore requiring smaller group sizes (53). LGE, on the other hand, is a unique tool for measuring infarct size in vivo, providing information otherwise obtained with histological staining. These measures, performed on the same mice, can be used to refine and replace standard staining and echocardiographic measures, staging the disease in vivo with a longitudinal design. In addition, displacement maps derived using DENSE can measure muscle performance with higher resolution than other techniques. Possessing a spatial resolution almost ten times higher than standard MRI tagging techniques, DENSE strain maps can show regions of myocardium that are not functioning correctly despite having normal perfusion and thus having normal appearance in LGE or cine-MRI images. Within the same examination, the molecular mechanisms can be probed directly using radiolabelled compounds detected by PET. The main benefit of combining these techniques together is that it yields a fuller picture on multiple scales of recovery following damage. This is important as new therapeutic approaches are tested in the laboratory.

This protocol can be applied using standard equipment as found in most preclinical imaging facilities performing MRI and PET, moving the mouse on the same bed between scanners. However, the protocol would benefit substantially from simultaneity of PET and MRI (163). First, shorter imaging times mean less anaesthetic stress on the animals (therefore reduced mortality) as well as higher throughput. Secondly, using a combined PET/MRI scanner, the coregistration pipeline would be further improved. This would allow MRI data to be more readily used to improve PET reconstruction, for example by modelling partial volume effects and motion.

## 8.5 Application: protection in I/R injury with Riociguat

In the following, the application of this method to the evaluation a novel pharmaceutical for myocardial infarction is described.

### 8.5.1 Background

Recently, a new class of drugs, the so-called sGC stimulators, entered the clinical development for the treatment of pulmonary hypertension. The sGC stimulator Riociguat, has recently undergone Phase III clinical trials in patients with several subforms of pulmonary arterial hypertension (PAH) and with chronic thromboembolic pulmonary hypertension (CTEPH). However, The NO – sGC - PKG pathway is known to play an important role in the acute protection against cardiac reperfusion injury.

In this study we tested the effects of Riociguat on MI in a mouse model of ischemia/reperfusion.

## 8.5.2 Methods

Mice were subjected to 30 min ischaemia via ligation of the left main coronary artery to induce MI and either placebo or Riociguat (1.2 µmol/l) were given as a bolus 5 min before and 5 min after onset of reperfusion. Surgery and treatment was performed by an independent operator. After 24 hours the novel method described above was performed to assess viability, function and metabolism at the acute setting.

## 8.5.3 Results

In the Riociguat-treated mice, the resulting infarct size was smaller and LV systolic function analysed by MRI was better preserved. Strain measures were highly congruent. Results are reported in Table 11. Typical images are reported in Figure 75, Figure 76, and Figure 77.

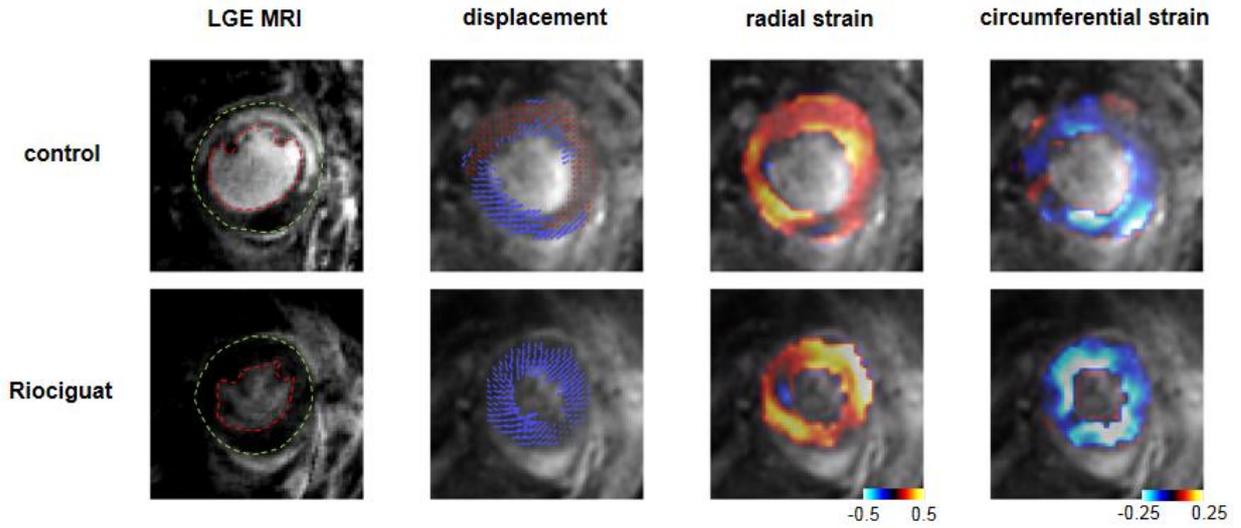

**Figure 75 Typical LGE-MRI and DENSE results for untreated and Riociguat treated mice.**

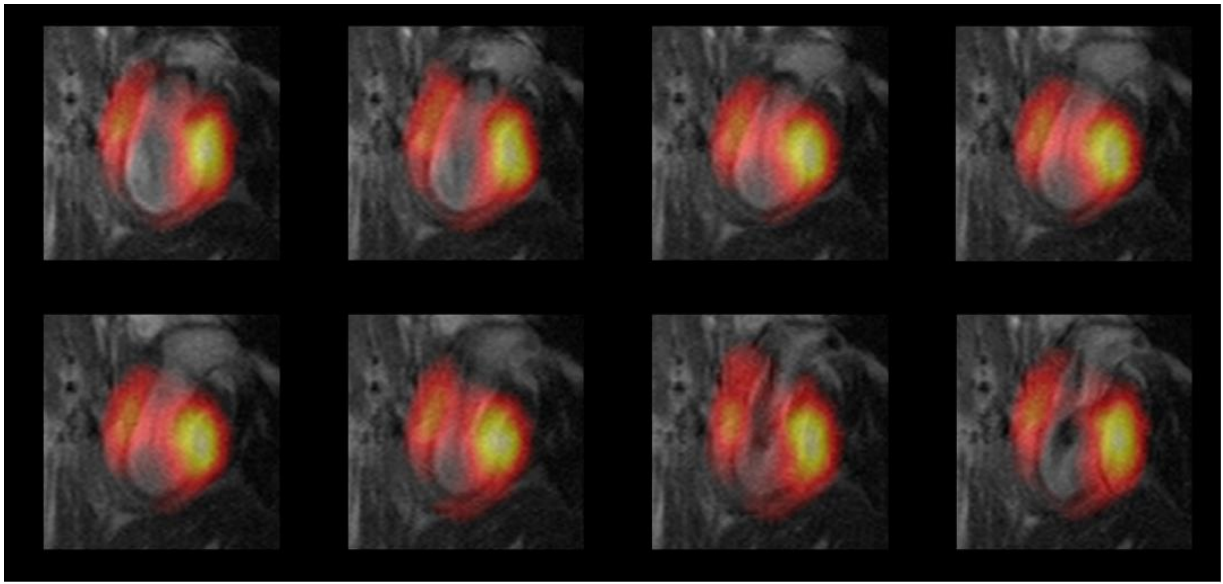

**Figure 76 Typical overlaid long axis cine-MRI and FDG PET frames from an untreated infarct.**

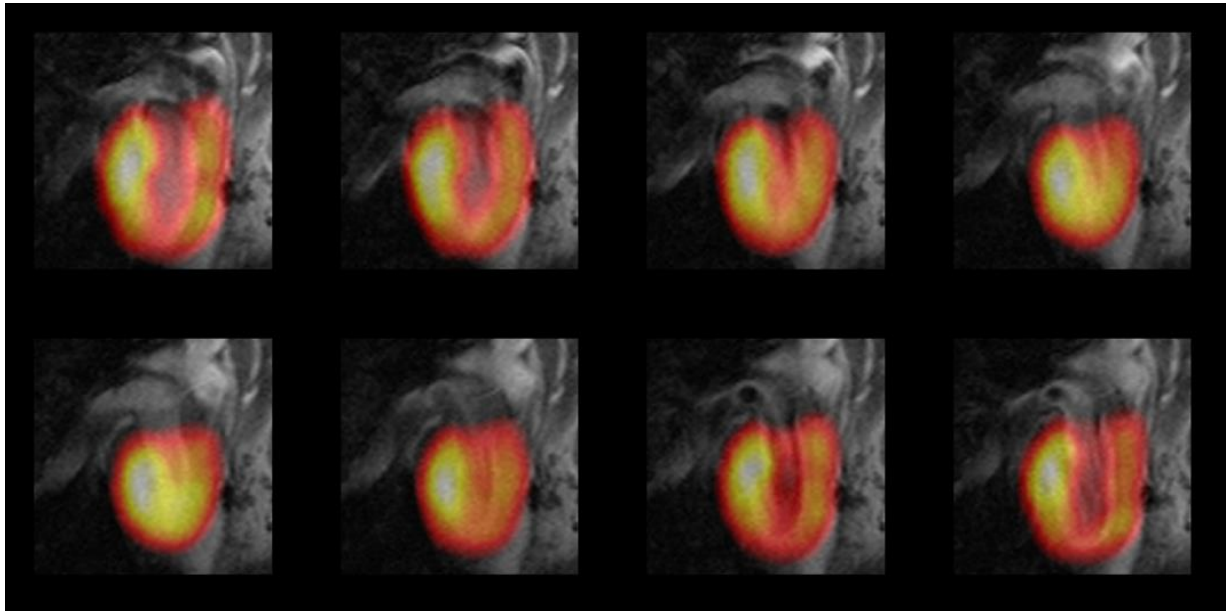

Figure 77 Typical overlaid long axis cine-MRI and FDG PET frames from a mouse treated with Riociguat.

| | control | Riociguat |
|---|---|---|
| **LVM (µl)** | 99 ± 4 | 92 ± 5 |
| **LVEDV (µl)** | 56 ± 3 | 46 ± 3* |
| **LVESV (µl)** | 32 ± 3 | 18 ± 1** |
| **LVSV (µl)** | 24 ± 1 | 27 ± 3 |
| **LVEF (%)** | 44 ± 3 | 60 ± 4** |
| **RVEDV (µl)** | 36 ± 2 | 41 ± 4 |
| **RVESV (µl)** | 13 ± 2 | 14 ± 1 |
| **RVSV (µl)** | 24 ± 1 | 27 ± 3 |
| **RVEF (%)** | 66 ± 3 | 66 ± 1 |
| **MRI - Infarct size (%LV)** | 22 ± 2 | 8 ± 3** |
| **Radial strain** | 0.144 ± 0.009 | 0.168 ± 0.007* |
| **Circumferential strain** | -0.061 ± 0.004 | -0.106 ± 0.013* |
| **PET - Infarct size (%LV)** | 18 ± 4 | 4 ± 3** |

Table 11 Key values obtained by MRI and PET 24 h after myocardial infarction from control and Riociguat treated mice. Values are mean ± sem. *$p$<0.05 **$p$<0.01

### 8.5.4 Conclusions

PET/MRI was successfully used to assess the effect of Riociguat in acute myocardial infarction. Mice treated acutely at the onset of reperfusion with the sGC stimulator Riociguat have smaller infarct size and better preserved function at the acute stage. These findings suggest that sGC stimulation after an acute MI may be a powerful therapeutic treatment strategy for preventing post-MI heart failure.

## *8.6 Chapter summary*

This chapter has shown that PET/MRI is an efficient tool for the assessment of disease and treatment in models of MI, achieving high accuracy and maintaining a high throughput. With PET/MRI technologies expanding at a fast pace, versatile protocols such as the one described here are a viable option for the wide-spread application of this technology in preclinical cardiac applications.

The potential of this method was demonstrated in a study investigating efficacy of a novel compound for myocardial infarction, finding that it reduces damage at the acute stge.

# Chapter 9

# Conclusions

In this thesis advanced methods for performing MRI in the mouse heart were developed and validated, outperforming current protocols for speed, accuracy and versatility. Each of these has been demonstrated in the context of current biomedical problems. In these contexts, the developed methods were not only proven accurate, but applicable to the practical situations where they are commonly needed. In each of the applications presented, the methods were able to determine differences between groups with a small number of animals, demonstrating high sensitivity to identify disease and treatment efficacy in preclinical cardiology.

## 9.1 Global heart function

Assessment of LV morphology throughout the cardiac cycle with MRI allowed for detection of abnormal LV chamber enlargement or increased wall thickness with high accuracy. Standard methods and common drawbacks were discussed in Chapter 4 and applied to current biological studies. Methods for semi-automatic segmentation were discussed to obtain functional values, obtaining high consistency both between and within observers, although high biases were measured between observers due to a subjective interpretation of anatomy.

Cine MRI techniques were applied to transgenic mice and myocardial infarction models. The technique was able to detect hypertrophic cardiomyopathy in the Complex I KO model. Preliminary observations were performed in the R6/2 mouse model of HD utilising a black-blood technique. However, the technique for black blood based on the saturation of signal in the major vessels was sub-optimal for an effective cancellation of the blood signal, and was therefore abandoned in future studies. A follow-up study in the same model used a pharmacological stress test with dobutamine, which unmasked right ventricular dysfunction in R6/2 mice. During measurement of heart response to inotropic stress, time limitations did not permit acquisition of the whole heart. Although better control of the stress could have been obtained  using continuous infusion through direct vascular access instead of intraperitoneal injections, counter indications of dobutamine are reduced when using lower doses (134, 164).  Scan durations remain a major limitation of standard cine MRI techniques (58, 59).

To address the problem of scan time, Chapter 5 investigated means of reconstructing partially-sampled data to accelerate acquisition. Cine MRI was made twelve times faster by applying a combination of parallel imaging and compressed sensing. In this experiment, retrospectively-gated techniques were used to extract functional parameters in free breathing for complete systolic assessment lasting just one minute. The technique was able to determine left and right ventricular function with preserved accuracy. Radial techniques were successfully used for the first time in accelerating cine MRI of the mouse heart, outperforming rectilinear sampling methods. When acquiring radial k-space, eddy currents generated artifacts in the images such as shading and halos. To correct these artifacts retrospectively, a novel method for trajectory correction was developed, described in Chapter 6. The combination of the acceleration methods and trajectory correction strategies can be implemented in imaging centres requiring high-throughput characterization of left or right ventricular function. Currently, the main drawback of the acceleration method is the processing time, which is prohibitive for online reconstruction of the data.

Further work advancing techniques to measure heart function can use undersampling techniques for achieving higher temporal resolution rather than speed. This is important for the assessment of diastolic dysfunction, which is currently difficult in mice due to the high temporal resolution required (165).

## 9.2 Tissue characterization

Contrast-enhanced MRI offers the unique feature of high-resolution characterization of tissue viability in mice *in vivo*, which can be then followed up longitudinally to assess different stages of heart failure for meaningful translational studies. A novel LGE method was developed for this purpose in Chapter 7, achieving results in good agreement with TTC staining. It was shown that in mice LGE imaging can be performed efficiently using multi-slice inversion recovery without TI optimization, simplifying the acquisition protocol and making faster protocols possible. The protocol achieves excellent CNR and accuracy in a short time for increased throughput and reliability, and can be performed as a three-minute "add-on" to the end of the cardiac examination. A drawback that remains when comparing this technique to histology is the absence in this setting of a method to measure area at risk. A T2-prepared scan before contrast agent injection would therefore represent an obvious extension of the current protocol (see 3.5).

As demonstrated here, techniques for assessing displacement and mechanical strains can be included in the imaging pipeline to identify hypokinetic areas, which often extend beyond the scar in infarcted tissue, and can be useful for assessment of acute disease. In addition, PET can be combined with MRI within a single examination, adding the capability of molecular imaging with a simplified coregistration pipeline. As shown in Chapter 8, FDG-PET performed one day after infarction shows glucose metabolism in the aftermath of the ischaemic insult, showing areas that are still metabolically functional.

This, combined with an assessment of global function, myocardial strain, and LGE imaging for viability, gives a complete evaluation of the heart in response to injury, from the macroscopic to the cellular level (Figure 78).

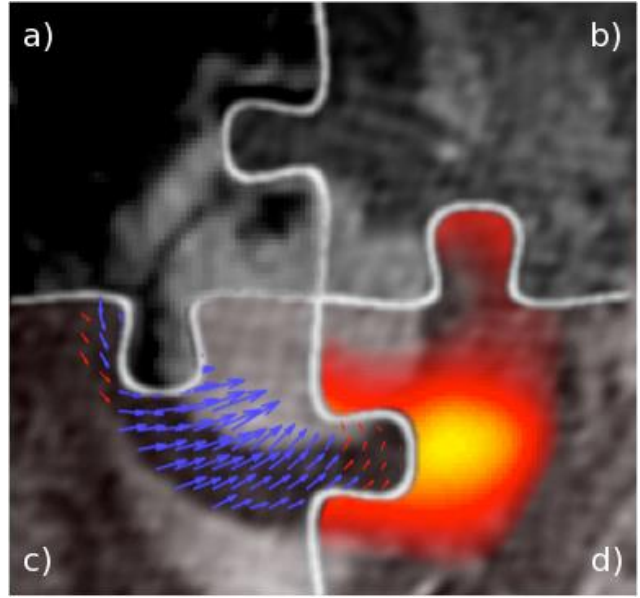

**Figure 78 Complementary information from multi-modality imaging can provide an accurate assessment of different aspects of a putative treatment: a) LGE MRI shows non-viable tissue. b) Cine MRI evaluates global heart function. c) DENSE MRI enquires muscle performance. d) PET is used for molecular imaging.**

The PET/MRI method demonstrated in this thesis is not specific to one tracer and can be used with different probes to answer specific questions in basic science or pharmacology. A substantial improvement to the methodology used here would derive blood plasma levels of tracer for full kinetic modelling thus directly quantifying metabolic rates. As well as for current investigations in preclinical cardiology, PET/MRI could be utilised for the implementation and cross-validation of diagnostic techniques with a direct comparison with histological methods. For instance, evaluation of tissue metabolism in the area at risk is important for clinical assessment of acute treatments. In addition, multi-modal approaches are a route to gain better knowledge of the mechanisms and limitations of current imaging examinations, such as gadolinium enhancement, as well as validating new diagnostic tools for clinical cardiology.

To summarise, MRI can be utilised to obtain a highly accurate assessment of heart function in mice non-invasively in a competitive time-scale. Highly-efficient methods to perform late gadolinium enhancement (LGE) can be utilised as a three-minute add-on to the cardiac MRI examination, in good agreement with tissue staining techniques. Further, DENSE and PET imaging can be added on to the imaging protocol in order to get a more detailed characterization of the heart, including respectively myocardial strains and metabolism. The methods exposed in this thesis have proven useful in the investigation of mechanisms behind disease using genetically modified mice, as well as the investigation of new compounds for acute myocardial infarction. Further investigation of these compounds in larger animal models and in clinical trials could confirm their efficacy and ultimately benefit patients. Consequently, the methodologies demonstrated in this thesis represent a significant advance for researchers in preclinical cardiology.

Guido Buonincontri

November 2013

# Publications resulting from this thesis

G Buonincontri, C Methner, T Krieg, TA Carpenter, SJ Sawiak - Functional assessment of the mouse heart with one-minute acquisition - *submitted*

G Buonincontri, C Methner, T Krieg, TA Carpenter, SJ Sawiak - An improved algorithm for trajectory correction in radial MRI – *submitted*

C Methner, ET Chouchani, G Buonincontri, A Logan, C Hu, SJ Sawiak, MP Murphy, T Krieg - Complex I deficiency due to selective loss of Ndufs4 in the mouse heart results in severe hypertrophic cardiomyopathy – *submitted*

C Methner, ET Chouchani, G Buonincontri, SJ Sawiak, MP Murphy, T Krieg - Mitochondria selective S-nitrosation by MitoSNO protects against post-infarct heart failure in mouse hearts – *submitted*

G Buonincontri, NI Wood, S Puttick, AO Ward, TA Carpenter SJ Sawiak, AJ Morton - Right ventricular dysfunction in the R6/2 transgenic mouse model of Huntington's disease is unmasked by dobutamine - *submitted*

C Methner, G Buonincontri, A Vujic, A Kretschmer, TA Carpenter, J Stasch, T Krieg - Riociguat reduces infarct size and post-infarct heart failure in mouse hearts: Insights from MRI/PET imaging – *PloS ONE, in press*

G Buonincontri, C Methner, T A Carpenter, S J Sawiak, T Krieg – PET and MRI in mouse models of myocardial infarction – *Journal of Visualized Experiments*, *in press*

G Buonincontri, C Methner, T Krieg, RC Hawkes, T A Carpenter, S J Sawiak – PET/MRI assessment of the infarcted mouse heart – *Nucl Instr and Meth in Physics Research Section A*, 2014, 734:152-155

G Buonincontri, SJ Sawiak, C Methner, T Krieg, RC Hawkes, TA Carpenter - PET/MRI in the infarcted mouse heart with the Cambridge split magnet - *Nucl Instr and Meth in Physics Research Section A,* 2013, 702:47–49

G Buonincontri, C Methner, T Krieg, TA Carpenter, SJ Sawiak – A fast protocol for infarct quantification in mice – *Journal of Magnetic Resonance Imaging,* 2013, 38(2):468-73

NI Wood, SJ Sawiak, G Buonincontri, Y Niu, AD Kane, TA Carpenter, DA Giussani, AJ Morton - Direct evidence of progressive cardiac dysfunction in a transgenic mouse model of Huntington's disease - *Journal of Huntington's Disease,* 2012, 1(1): 65–72

# Bibliography

1.      Murray CJ, Richards MA, Newton JN, Fenton KA, Anderson HR, Atkinson C, et al. UK health performance: findings of the Global Burden of Disease Study 2010. Lancet. 2013 Mar 23;381(9871):997-1020. PubMed PMID: 23668584.

2.      Patten RD, Hall-Porter MR. Small animal models of heart failure: development of novel therapies, past and present. Circulation Heart failure. 2009 Mar;2(2):138-44. PubMed PMID: 19808329.

3.      Guyton AC, Hall JE. Textbook of medical physiology. 11th ed. Philadelphia: Elsevier Saunders; 2006. xxxv, 1116 p. p.

4.      Nichols M, Townsend N, Scarborough P, Rayner M. Trends in age-specific coronary heart disease mortality in the European Union over three decades: 1980-2009. European heart journal. 2013 Oct;34(39):3017-27. PubMed PMID: 23801825. Pubmed Central PMCID: 3796269.

5.      Windecker S, Bax JJ, Myat A, Stone GW, Marber MS. Future treatment strategies in ST-segment elevation myocardial infarction. Lancet. 2013 Aug 17;382(9892):644-57. PubMed PMID: 23953388.

6.      Jessup M, Brozena S. Heart failure. The New England journal of medicine. 2003 May 15;348(20):2007-18. PubMed PMID: 12748317.

7.      Wessels A, Sedmera D. Developmental anatomy of the heart: a tale of mice and man. Physiological genomics. 2003 Nov 11;15(3):165-76. PubMed PMID: 14612588.

8.      Klocke R, Tian W, Kuhlmann MT, Nikol S. Surgical animal models of heart failure related to coronary heart disease. Cardiovascular research. 2007 Apr 1;74(1):29-38. PubMed PMID: 17188668.

9.      Pacher P, Nagayama T, Mukhopadhyay P, Bátkai S, Kass DA. Measurement of cardiac function using pressure-volume conductance catheter technique in mice and rats. Nature protocols. 2008;3:1422-34. PubMed PMID: 18772869.

10.     Larsen TS, Belke DD, Sas R, Giles WR, Severson DL, Lopaschuk GD, et al. The isolated working mouse heart: methodological considerations. Pflügers Archiv : European journal of physiology. 1999;437:979-85. PubMed PMID: 10370078.

11.     Buonincontri G, Methner C, Carpenter TA, Hawkes RC, Sawiak SJ, Krieg T. MRI and PET in mouse models of myocardial infarction. Journal of visualized experiments : JoVE. 2013;in press(82):e50806. PubMed PMID: 24378323.

12.     Wood NI, Sawiak SJ, Buonincontri G, Niu Y, Kane AD, Carpenter TA, et al. Direct Evidence of Progressive Cardiac Dysfunction in a Transgenic Mouse Model of Huntington's Disease - Journal of Huntington's Disease - Volume 1, Number 1 / 2012 - IOS Press. Journal of Huntington's Disease. 2012;1(1):65-72.

13.     Buonincontri G, Methner C, Krieg T, Carpenter TA, Sawiak SJ. A fast protocol for infarct quantification in mice. J Magn Reson Imaging. 2013 Aug;38(2):468-73. PubMed PMID: 23292906.

14.     Buonincontri G, Methner C, Krieg T, Hawkes RC, Adrian Carpenter T, Sawiak SJ. PET/MRI assessment of the infarcted mouse heart. Nuclear Instruments and Methods in Physics Research Section A: Accelerators, Spectrometers, Detectors and Associated Equipment. 2014;734(0):152-5.

15.     Haacke EM. Magnetic resonance imaging : physical principles and sequence design. New York: Wiley; 1999. xxvii, 914 p. p.

16.     McRobbie DW. MRI from picture to proton. 2nd ed. Cambridge, UK ; New York: Cambridge University Press; 2007. xii, 394 p. p.

17.     Bloch F. Nuclear Induction. Physical Review. 1946;70(7-8):460-74.

18.     Hahn E. Spin Echoes. Physical Review. 1950;80(4): 580-94.

19.     Haase A, Frahm J, Matthaei D, Hanicke W, Merboldt KD. FLASH imaging: rapid NMR imaging using low flip-angle pulses. 1986. J Magn Reson. 2011 Dec;213(2):533-41. PubMed PMID: 22152368.

20.     Haase A. FLASH MR imaging: a success story since 25 years. J Magn Reson. 2011 Dec;213(2):542-3. PubMed PMID: 22152369.

21.     Hawkes RC, Patz S. Rapid Fourier imaging using steady-state free precession. Magnetic resonance in medicine : official journal of the Society of Magnetic Resonance in Medicine / Society of Magnetic Resonance in Medicine. 1987 Jan;4(1):9-23. PubMed PMID: 3821484.

22.     Look DC. Time Saving in Measurement of NMR and EPR Relaxation Times. Review of Scientific Instruments. 1970;41:250.

23.     Gudbjartsson H, Patz S. The Rician distribution of noisy MRI data. Magnetic resonance in medicine : official journal of the Society of Magnetic Resonance in Medicine / Society of Magnetic Resonance in Medicine. 1995 Dec;34(6):910-4. PubMed PMID: 8598820. Pubmed Central PMCID: 2254141.

24.     Andersen AH. On the Rician distribution of noisy MRI data. Magnetic resonance in medicine : official journal of the Society of Magnetic Resonance in Medicine / Society of Magnetic Resonance in Medicine. 1996 Aug;36(2):331-3. PubMed PMID: 8843389.

25.     Herman GT. Image reconstruction from projections : the fundamentals of computerized tomography. San Francisco: Academic Press; 1980. xiv, 316 p. p.

26.     O'Sullivan JD. A fast sinc function gridding algorithm for fourier inversion in computer tomography. IEEE Trans Med Imaging. 1985;4(4):200-7. PubMed PMID: 18243972.

27.     Sedarat H, Nishimura DG. On the optimality of the gridding reconstruction algorithm. Ieee T Med Imaging. 2000 Apr;19(4):306-17. PubMed PMID: WOS:000088106800006. English.

28.     Fessler JA, Sutton BP. Nonuniform fast Fourier transforms using min-max interpolation. Ieee T Signal Proces. 2003 Feb;51(2):560-74. PubMed PMID: ISI:000181323700023. English.

29.     Moussavi A, Untenberger M, Uecker M, Frahm J. Correction of gradient-induced phase errors in radial MRI. Magnetic resonance in medicine : official journal of the Society of Magnetic Resonance in Medicine / Society of Magnetic Resonance in Medicine. 2013 Feb 25. PubMed PMID: 23440722. Epub 2013/02/27. Eng.

30.     Deshmane A, Gulani V, Griswold MA, Seiberlich N. Parallel MR imaging. J Magn Reson Imaging. 2012 Jul;36(1):55-72. PubMed PMID: 22696125.

31.     Tsao J, Kozerke S. MRI temporal acceleration techniques. J Magn Reson Imaging. 2012 Sep;36(3):543-60. PubMed PMID: 22903655.

32.     Lustig M, Donoho D, Pauly JM. Sparse MRI: The application of compressed sensing for rapid MR imaging. Magnetic resonance in medicine : official journal of the Society of Magnetic Resonance in Medicine / Society of Magnetic Resonance in Medicine. 2007 Dec;58(6):1182-95. PubMed PMID: 17969013.

33.     Lustig M, Pauly JM. SPIRiT: Iterative self-consistent parallel imaging reconstruction from arbitrary k-space. Magnetic resonance in medicine : official journal of the Society of Magnetic Resonance in Medicine / Society of Magnetic Resonance in Medicine. 2010 Aug;64(2):457-71. PubMed PMID: 20665790. Pubmed Central PMCID: 2925465. Epub 2010/07/29. eng.

34.     Dall'Armellina E, Karamitsos TD, Neubauer S, Choudhury RP. CMR for characterization of the myocardium in acute coronary syndromes. Nature reviews Cardiology. 2010 Nov;7(11):624-36. PubMed PMID: 20856263.

35.     Price AN, Cheung KK, Cleary JO, Campbell AE, Riegler J, Lythgoe MF. Cardiovascular magnetic resonance imaging in experimental models. The open cardiovascular medicine journal. 2010;4:278-92. PubMed PMID: 21331311. Pubmed Central PMCID: 3040459.

36.     Siri FM, Jelicks LA, Leinwand LA, Gardin JM. Gated magnetic resonance imaging of normal and hypertrophied murine hearts. The American journal of physiology. 1997 May;272(5 Pt 2):H2394-402. PubMed PMID: 9176310.

37.     Kober F, Iltis I, Cozzone PJ, Bernard M. Cine-MRI assessment of cardiac function in mice anesthetized with ketamine/xylazine and isoflurane. Magma (New York, NY). 2004;17:157-61. PubMed PMID: 15609036.

38.     Epstein FH. MR in mouse models of cardiac disease. NMR in biomedicine. 2007;20:238-55. PubMed PMID: 17451182.

39.     Schneider JE, Wiesmann F, Lygate CA, Neubauer S. How to perform an accurate assessment of cardiac function in mice using high-resolution magnetic resonance imaging. Journal of cardiovascular magnetic resonance : official journal of the Society for Cardiovascular Magnetic Resonance. 2006;8(5):693-701. PubMed PMID: 16891228. Epub 2006/08/08. eng.

40.     Heijman E, de Graaf W, Niessen P, Nauerth A, van Eys G, de Graaf L, et al. Comparison between prospective and retrospective triggering for mouse cardiac MRI. Nmr in Biomedicine. 2007 Jun;20(4):439-47. PubMed PMID: ISI:000246767000005. English.

41.     Larson AC, White RD, Laub G, McVeigh ER, Li DB, Simonetti OP. Self-gated cardiac cine MRI. Magnet Reson Med. 2004 Jan;51(1):93-102. PubMed PMID: ISI:000188041500013. English.

42.     Hiba B, Richard N, Thibault H, Janier M. Cardiac and respiratory self-gated cine MRI in the mouse: comparison between radial and rectilinear techniques at 7T. Magnetic resonance in medicine : official journal of the Society of Magnetic Resonance in Medicine / Society of Magnetic Resonance in Medicine. 2007 Oct;58(4):745-53. PubMed PMID: 17899593.

43.     Ruff J, Wiesmann F, Hiller K-H, Voll S, von Kienlin M, Bauer WR, et al. Magnetic resonance microimaging for noninvasive quantification of myocardial function and mass in the mouse. Magnet Reson Med. 1998;40:43-8.

44.     Streif JU, Herold V, Szimtenings M, Lanz TE, Nahrendorf M, Wiesmann F, et al. In vivo time-resolved quantitative motion mapping of the murine myocardium with phase contrast MRI. Magnetic resonance in medicine : official journal of the Society of Magnetic Resonance in Medicine / Society of Magnetic Resonance in Medicine. 2003 Feb;49(2):315-21. PubMed PMID: 12541252.

45.     Wiesmann F, Ruff J, Engelhardt S, Hein L, Dienesch C, Leupold A, et al. Dobutamine-stress magnetic resonance microimaging in mice : acute changes of cardiac geometry and function in normal and failing murine hearts. Circulation research. 2001;88:563-9. PubMed PMID: 11282889.

46.     Miraux S, Calmettes G, Massot P, Lefrancois W, Parzy E, Muller B, et al. 4D retrospective black blood trueFISP imaging of mouse heart. Magnetic resonance in medicine : official journal of the Society of Magnetic Resonance in Medicine / Society of Magnetic Resonance in Medicine. 2009 Nov;62(5):1099-105. PubMed PMID: 19780163.

47.     Yu X, Tesiram YA, Towner RA, Abbott A, Patterson E, Huang S, et al. Early myocardial dysfunction in streptozotocin-induced diabetic mice: a study using in vivo magnetic resonance imaging (MRI). Cardiovascular diabetology. 2007;6:6. PubMed PMID: 17309798. Pubmed Central PMCID: 1805425.

48.     Heiberg E, Sjögren J, Ugander M, Carlsson M, Engblom H, Arheden H. Design and validation of Segment--freely available software for cardiovascular image analysis. BMC medical imaging. 2010;10:1. PubMed PMID: 20064248.

49.     Heiberg E, Wigstrom L, Carlsson M, Bolger AF, Karlsson M. Time resolved three-dimensional automated segmentation of the left ventricle. Comput Cardiol. 2005;32:599-602. PubMed PMID: WOS:000236378300151. English.

50.     Heiberg E, Engblom H, Engvall J, Hedstrom E, Ugander M, Arheden H. Semi-automatic quantification of myocardial infarction from delayed contrast enhanced magnetic resonance imaging. Scandinavian cardiovascular journal : SCJ. 2005 Oct;39(5):267-75. PubMed PMID: 16269396.

51.     Franco F, Thomas GD, Giroir B, Bryant D, Bullock MC, Chwialkowski MC, et al. Magnetic resonance imaging and invasive evaluation of development of heart failure in transgenic mice with myocardial expression of tumor necrosis factor-alpha. Circulation. 1999 Jan 26;99(3):448-54. PubMed PMID: 9918534.

52.     Jacoby C, Molojavyi A, Flogel U, Merx MW, Ding Z, Schrader J. Direct comparison of magnetic resonance imaging and conductance microcatheter in the evaluation of left ventricular function in mice. Basic research in cardiology. 2006 Jan;101(1):87-95. PubMed PMID: 16132173.

53.     Stuckey DJ, Carr CA, Tyler DJ, Clarke K. Cine-MRI versus two-dimensional echocardiography to measure in vivo left ventricular function in rat heart. Nmr in Biomedicine. 2008 Aug;21(7):765-72. PubMed PMID: 18457349. Epub 2008/05/07. eng.

54.     Tyrankiewicz U, Skorka T, Jablonska M, Petkow-Dimitrow P, Chlopicki S. Characterization of the cardiac response to a low and high dose of dobutamine in the mouse model of dilated cardiomyopathy by MRI in vivo. Journal of magnetic resonance imaging : JMRI. 2013;37:669-77. PubMed PMID: 23125067.

55.     ten Hove M, Lygate CA, Fischer A, Schneider JE, Sang AE, Hulbert K, et al. Reduced inotropic reserve and increased susceptibility to cardiac ischemia/reperfusion injury in phosphocreatine-deficient guanidinoacetate-N-methyltransferase-knockout mice. Circulation. 2005 May 17;111(19):2477-85. PubMed PMID: 15883212.

56.     Slawson SE, Roman BB, Williams DS, Koretsky AP. Cardiac MRI of the normal and hypertrophied mouse heart. Magnetic resonance in medicine : official journal of the Society of Magnetic Resonance in Medicine / Society of Magnetic Resonance in Medicine. 1998 Jun;39(6):980-7. PubMed PMID: 9621922.

57.     Franco F, Dubois SK, Peshock RM, Shohet RV. Magnetic resonance imaging accurately estimates LV mass in a transgenic mouse model of cardiac hypertrophy. The American journal of physiology. 1998 Feb;274(2 Pt 2):H679-83. PubMed PMID: 9486274.

58.     Schneider JE, Lanz T, Barnes H, Medway D, Stork LA, Lygate CA, et al. Ultra-fast and accurate assessment of cardiac function in rats using accelerated MRI at 9.4 Tesla. Magnetic resonance in medicine : official journal of the Society of Magnetic Resonance in Medicine / Society of Magnetic Resonance in Medicine. 2008 Mar;59(3):636-41. PubMed PMID: 18306411. Epub 2008/02/29. eng.

59.     Ratering D, Baltes C, Dorries C, Rudin M. Accelerated cardiovascular magnetic resonance of the mouse heart using self-gated parallel imaging strategies does not compromise accuracy of structural and functional measures. Journal of cardiovascular magnetic resonance : official journal of the Society for Cardiovascular Magnetic Resonance. 2010;12:43. PubMed PMID: 20663156. Pubmed Central PMCID: 2918602. Epub 2010/07/29. eng.

60.     Wech T, Lemke A, Medway D, Stork LA, Lygate CA, Neubauer S, et al. Accelerating cine-MR imaging in mouse hearts using compressed sensing. J Magn Reson Imaging. 2011 Nov;34(5):1072-9. PubMed PMID: 21932360. Pubmed Central PMCID: 3261377. Epub 2011/09/21. eng.

61.     Bland JM, Altman DG. Statistical methods for assessing agreement between two methods of clinical measurement. Lancet. 1986 Feb 8;1(8476):307-10. PubMed PMID: 2868172.

62.     Dinkel J, Bartling SH, Kuntz J, Grasruck M, Kopp-Schneider A, Iwasaki M, et al. Intrinsic gating for small-animal computed tomography: a robust ECG-less paradigm for deriving cardiac phase information and functional imaging. Circulation Cardiovascular imaging. 2008 Nov;1(3):235-43. PubMed PMID: 19808548. Epub 2009/10/08. eng.

63.     Goetschalckx K, Rademakers F, Bogaert J. Right ventricular function by MRI. Current opinion in cardiology. 2010;25:451-5. PubMed PMID: 20543681.

64.     Voelkel NF, Quaife RA, Leinwand LA, Barst RJ, McGoon MD, Meldrum DR, et al. Right ventricular function and failure: report of a National Heart, Lung, and Blood Institute working group on cellular and molecular mechanisms of right heart failure. Circulation. 2006;114:1883-91. PubMed PMID: 17060398.

65.     Haddad F, Hunt SA, Rosenthal DN, Murphy DJ. Right ventricular function in cardiovascular disease, part I: Anatomy, physiology, aging, and functional assessment of the right ventricle. Circulation. 2008;117:1436-48. PubMed PMID: 18347220.

66.     Haddad F, Doyle R, Murphy DJ, Hunt SA. Right ventricular function in cardiovascular disease, part II: pathophysiology, clinical importance, and management of right ventricular failure. Circulation. 2008;117:1717-31. PubMed PMID: 18378625.

67.     Wiesmann F, Frydrychowicz A, Rautenberg J, Illinger R, Rommel E, Haase A, et al. Analysis of right ventricular function in healthy mice and a murine model of heart failure by in vivo MRI. American journal of physiology Heart and circulatory physiology. 2002;283:H1065-71. PubMed PMID: 12181136.

68.     Stuckey DJ, Carr CA, Camelliti P, Tyler DJ, Davies KE, Clarke K. In vivo MRI Characterization of Progressive Cardiac Dysfunction in the mdx Mouse Model of Muscular Dystrophy. PloS one. 2012;7:e28569. PubMed PMID: 22235247.

69.     van Nierop BJ, van Assen HC, van Deel ED, Niesen LB, Duncker DJ, Strijkers GJ, et al. Phenotyping of left and right ventricular function in mouse models of compensated hypertrophy and heart failure with cardiac MRI. PloS one. 2013;8(2):e55424. PubMed PMID: 23383329. Pubmed Central PMCID: 3562232.

70.     Zerhouni EA, Parish DM, Rogers WJ, Yang A, Shapiro EP. Human heart: tagging with MR imaging--a method for noninvasive assessment of myocardial motion. Radiology. 1988 Oct;169(1):59-63. PubMed PMID: 3420283.

71.     Axel L, Dougherty L. MR imaging of motion with spatial modulation of magnetization. Radiology. 1989 Jun;171(3):841-5. PubMed PMID: 2717762.

72.     de Crespigny AJ, Carpenter TA, Hall LD. Cardiac tagging in the rat using a DANTE sequence. Magnetic resonance in medicine : official journal of the Society of Magnetic Resonance in Medicine / Society of Magnetic Resonance in Medicine. 1991 Sep;21(1):151-6. PubMed PMID: 1943673.

73.     Osman NF, McVeigh ER, Prince JL. Imaging heart motion using harmonic phase MRI. Ieee T Med Imaging. 2000 Mar;19(3):186-202. PubMed PMID: WOS:000087569300004. English.

74.     Heijman E, Strijkers GJ, Habets J, Janssen B, Nicolay K. Magnetic resonance imaging of regional cardiac function in the mouse. Magn Reson Mater Phy. 2004 Dec;17(3-6):170-8. PubMed PMID: WOS:000227620700011. English.

75.     Dall'Armellina E, Jung BA, Lygate CA, Neubauer S, Markl M, Schneider JE. Improved method for quantification of regional cardiac function in mice using phase-contrast MRI. Magnetic resonance in medicine : official journal of the Society of Magnetic Resonance in Medicine / Society of Magnetic Resonance in Medicine. 2012 Feb;67(2):541-51. PubMed PMID: 21674616. Pubmed Central PMCID: 3378699.

76.     Aletras AH, Ding S, Balaban RS, Wen H. DENSE: displacement encoding with stimulated echoes in cardiac functional MRI. J Magn Reson. 1999 Mar;137(1):247-52. PubMed PMID: 10053155. Pubmed Central PMCID: 2887318.

77.     Gilson WD, Yang Z, French BA, Epstein FH. Complementary displacement-encoded MRI for contrast-enhanced infarct detection and quantification of myocardial function in mice. Magnetic resonance in medicine : official journal of the Society of Magnetic Resonance in Medicine / Society of Magnetic Resonance in Medicine. 2004 Apr;51(4):744-52. PubMed PMID: 15065247. Epub 2004/04/06. eng.

78.     Gilson WD, Yang Z, French BA, Epstein FH. Measurement of myocardial mechanics in mice before and after infarction using multislice displacement-encoded MRI with 3D motion encoding. Am J Physiol Heart Circ Physiol. 2005 Mar;288(3):H1491-7. PubMed PMID: 15513963.

79.     Zhong X, Gibberman LB, Spottiswoode BS, Gilliam AD, Meyer CH, French BA, et al. Comprehensive cardiovascular magnetic resonance of myocardial mechanics in mice using three-dimensional cine DENSE. Journal of cardiovascular magnetic resonance : official journal of the Society for Cardiovascular Magnetic Resonance. 2011;13:83. PubMed PMID: 22208954. Pubmed Central PMCID: 3278394.

80.     Gilliam AD, Epstein FH. Automated motion estimation for 2-D cine DENSE MRI. IEEE Trans Med Imaging. 2012 Sep;31(9):1669-81. PubMed PMID: 22575669.

81.     Hung CL, Verma A, Uno H, Shin SH, Bourgoun M, Hassanein AH, et al. Longitudinal and circumferential strain rate, left ventricular remodeling, and prognosis after myocardial infarction. J Am Coll Cardiol. 2010 Nov 23;56(22):1812-22. PubMed PMID: 21087709.

82.     Hsu LY, Rhoads KL, Holly JE, Kellman P, Aletras AH, Arai AE. Quantitative myocardial perfusion analysis with a dual-bolus contrast-enhanced first-pass MRI technique in humans. J Magn Reson Imaging. 2006 Mar;23(3):315-22. PubMed PMID: 16463299.

83.     Ichihara T, Ishida M, Kitagawa K, Ichikawa Y, Natsume T, Yamaki N, et al. Quantitative analysis of first-pass contrast-enhanced myocardial perfusion MRI using a Patlak plot method and blood saturation correction. Magnetic resonance in medicine : official journal of the Society of Magnetic Resonance in Medicine / Society of Magnetic Resonance in Medicine. 2009 Aug;62(2):373-83. PubMed PMID: 19353669.

84.     van Nierop BJ, Coolen BF, Dijk WJ, Hendriks AD, de Graaf L, Nicolay K, et al. Quantitative first-pass perfusion MRI of the mouse myocardium. Magnetic resonance in medicine : official journal of the Society of Magnetic Resonance in Medicine / Society of Magnetic Resonance in Medicine. 2013 Jun;69(6):1735-44. PubMed PMID: 22907879.

85.     Streif JU, Nahrendorf M, Hiller KH, Waller C, Wiesmann F, Rommel E, et al. In vivo assessment of absolute perfusion and intracapillary blood volume in the murine myocardium by spin labeling magnetic resonance imaging. Magnetic resonance in medicine : official journal of the Society of Magnetic Resonance in Medicine / Society of Magnetic Resonance in Medicine. 2005 Mar;53(3):584-92. PubMed PMID: 15723416.

86.     Vandsburger MH, French BA, Helm PA, Roy RJ, Kramer CM, Young AA, et al. Multi-parameter in vivo cardiac magnetic resonance imaging demonstrates normal perfusion reserve despite severely attenuated beta-adrenergic functional response in neuronal nitric oxide synthase knockout mice. European heart journal. 2007 Nov;28(22):2792-8. PubMed PMID: 17602202.

87.     Kober F, Iltis I, Cozzone PJ, Bernard M. Myocardial blood flow mapping in mice using high-resolution spin labeling magnetic resonance imaging: influence of ketamine/xylazine and isoflurane anesthesia. Magnetic resonance in medicine : official journal of the Society of Magnetic Resonance in Medicine / Society of Magnetic Resonance in Medicine. 2005 Mar;53(3):601-6. PubMed PMID: 15723407.

88.     Wagner A, Mahrholdt H, Holly TA, Elliott MD, Regenfus M, Parker M, et al. Contrast-enhanced MRI and routine single photon emission computed tomography (SPECT) perfusion imaging for detection of subendocardial myocardial infarcts: an imaging study. Lancet. 2003 Feb 1;361(9355):374-9. PubMed PMID: 12573373.

89.     Thomas D, Dumont C, Pickup S, Misselwitz B, Zhou R, Horowitz J, et al. T1-weighted cine FLASH is superior to IR imaging of post-infarction myocardial viability at 4.7T. Journal of cardiovascular magnetic resonance : official journal of the Society for Cardiovascular Magnetic Resonance. 2006;8:345-52. PubMed PMID: 16669177.

90.     Bohl S, Lygate CA, Barnes H, Medway D, Stork L-A, Schulz-Menger J, et al. Advanced methods for quantification of infarct size in mice using three-dimensional high-field late gadolinium enhancement MRI. American journal of physiology Heart and circulatory physiology. 2009;296:H1200-8. PubMed PMID: 19218501.

91.     Chapon C, Herlihy AH, Bhakoo KK. Assessment of myocardial infarction in mice by late gadolinium enhancement MR imaging using an inversion recovery pulse sequence at 9.4T. Journal of cardiovascular magnetic resonance : official journal of the Society for Cardiovascular Magnetic Resonance. 2008;10:6. PubMed PMID: 18272007.

92.     Protti A, Sirker A, Shah AM, Botnar R. Late gadolinium enhancement of acute myocardial infarction in mice at 7T: cine-FLASH versus inversion recovery. Journal of magnetic resonance imaging : JMRI. 2010;32:878-86. PubMed PMID: 20882618.

93.     Price AN, Cheung KK, Lim SY, Yellon DM, Hausenloy DJ, Lythgoe MF. Rapid assessment of myocardial infarct size in rodents using multi-slice inversion recovery late gadolinium enhancement CMR at 9.4T. Journal of cardiovascular magnetic resonance : official journal of the Society for Cardiovascular Magnetic Resonance. 2011;13:44. PubMed PMID: 21892953. en.

94.     Schneider JE, Cassidy PJ, Lygate C, Tyler DJ, Wiesmann F, Grieve SM, et al. Fast, high-resolution in vivo cine magnetic resonance imaging in normal and failing mouse hearts on a vertical 11.7 T system. J Magn Reson Imaging. 2003 Dec;18(6):691-701. PubMed PMID: 14635154.

95.     Beyers RJ, Smith RS, Xu Y, Piras BA, Salerno M, Berr SS, et al. T(2) -weighted MRI of post-infarct myocardial edema in mice. Magnetic resonance in medicine : official journal of the Society of Magnetic Resonance in Medicine / Society of Magnetic Resonance in Medicine. 2012 Jan;67(1):201-9. PubMed PMID: 21630350. Pubmed Central PMCID: 3188362.

96.     Bun SS, Kober F, Jacquier A, Espinosa L, Kalifa J, Bonzi MF, et al. Value of in vivo T2 measurement for myocardial fibrosis assessment in diabetic mice at 11.75 T. Investigative radiology. 2012 May;47(5):319-23. PubMed PMID: 22488510.

97.     O'Regan DP, Ariff B, Neuwirth C, Tan Y, Durighel G, Cook SA. Assessment of severe reperfusion injury with T2* cardiac MRI in patients with acute myocardial infarction. Heart. 2010 Dec;96(23):1885-91. PubMed PMID: 20965977.

98.     Pennell DJ. T2* magnetic resonance and myocardial iron in thalassemia. Annals of the New York Academy of Sciences. 2005;1054:373-8. PubMed PMID: 16339685.

99.     Shea SM, Fieno DS, Schirf BE, Bi X, Huang J, Omary RA, et al. T2-prepared steady-state free precession blood oxygen level-dependent MR imaging of myocardial perfusion in a dog stenosis model. Radiology. 2005 Aug;236(2):503-9. PubMed PMID: 16040907.

100.     Aguor EN, Arslan F, van de Kolk CW, Nederhoff MG, Doevendans PA, van Echteld CJ, et al. Quantitative T 2* assessment of acute and chronic myocardial ischemia/reperfusion injury in mice. Magma. 2012 Oct;25(5):369-79. PubMed PMID: 22327962. Pubmed Central PMCID: 3458196.

101.     Ruff J, Wiesmann F, Lanz T, Haase A. Magnetic resonance imaging of coronary arteries and heart valves in a living mouse: techniques and preliminary results. J Magn Reson. 2000 Oct;146(2):290-6. PubMed PMID: 11001845.

102.     Strijkers GJ, Bouts A, Blankesteijn WM, Peeters TH, Vilanova A, van Prooijen MC, et al. Diffusion tensor imaging of left ventricular remodeling in response to myocardial infarction in the mouse. NMR Biomed. 2009 Feb;22(2):182-90. PubMed PMID: 18780284.

103.     Akki A, Gupta A, Weiss RG. Magnetic resonance imaging and spectroscopy of the murine cardiovascular system. Am J Physiol Heart Circ Physiol. 2013 Mar 1;304(5):H633-48. PubMed PMID: 23292717. Pubmed Central PMCID: 3602757.

104.     Chacko VP, Aresta F, Chacko SM, Weiss RG. MRI/MRS assessment of in vivo murine cardiac metabolism, morphology, and function at physiological heart rates. Am J Physiol Heart Circ Physiol. 2000 Nov;279(5):H2218-24. PubMed PMID: 11045956.

105.     Stegger L, Heijman E, Schafers KP, Nicolay K, Schafers MA, Strijkers GJ. Quantification of left ventricular volumes and ejection fraction in mice using PET, compared with MRI. J Nucl Med. 2009 Jan;50(1):132-8. PubMed PMID: 19091898.

106.     Buscher K, Judenhofer MS, Kuhlmann MT, Hermann S, Wehrl HF, Schafers KP, et al. Isochronous Assessment of Cardiac Metabolism and Function in Mice Using Hybrid PET/MRI. J Nucl Med. 2010 Aug 1;51(8):1277-84. PubMed PMID: WOS:000280645600026. English.

107.     Lee WW, Marinelli B, van der Laan AM, Sena BF, Gorbatov R, Leuschner F, et al. PET/MRI of inflammation in myocardial infarction. J Am Coll Cardiol. 2012 Jan 10;59(2):153-63. PubMed PMID: 22222080. Pubmed Central PMCID: 3257823.

108.     Heijman E, Aben JP, Penners C, Niessen P, Guillaume R, van Eys G, et al. Evaluation of manual and automatic segmentation of the mouse heart from CINE MR images. J Magn Reson Imaging. 2008 Jan;27(1):86-93. PubMed PMID: 18050352.

109.     Ingraham CA, Burwell LS, Skalska J, Brookes PS, Howell RL, Sheu SS, et al. NDUFS4: creation of a mouse model mimicking a Complex I disorder. Mitochondrion. 2009 Jun;9(3):204-10. PubMed PMID: 19460290. Pubmed Central PMCID: 2783808.

110.     Quintana A, Morgan PG, Kruse SE, Palmiter RD, Sedensky MM. Altered anesthetic sensitivity of mice lacking Ndufs4, a subunit of mitochondrial complex I. PloS one. 2012;7(8):e42904. PubMed PMID: 22912761. Pubmed Central PMCID: 3422219.

111.     Sterky FH, Hoffman AF, Milenkovic D, Bao B, Paganelli A, Edgar D, et al. Altered dopamine metabolism and increased vulnerability to MPTP in mice with partial deficiency of mitochondrial complex I in dopamine neurons. Human molecular genetics. 2012 Mar 1;21(5):1078-89. PubMed PMID: 22090423. Pubmed Central PMCID: 3277308.

112.     Chouchani ET, Methner C, Nadtochiy SM, Logan A, Pell VR, Ding S, et al. Cardioprotection by S-nitrosation of a cysteine switch on mitochondrial complex I. Nature medicine. 2013 Jun;19(6):753-9. PubMed PMID: 23708290.

113.     Sathasivam K, Hobbs C, Turmaine M, Mangiarini L, Mahal A, Bertaux F, et al. Formation of Polyglutamine Inclusions in Non-CNS Tissue. Human molecular genetics. 1999;8:813-22.

114.     Chiu E, Alexander L. Causes of death in Huntington's disease. The Medical journal of Australia. 1982;1:153. PubMed PMID: 6210834.

115.     Lanska DJ, Lanska MJ, Lavine L, Schoenberg BS. Conditions associated with Huntington's disease at death. A case-control study. Archives of neurology. 1988;45:878-80. PubMed PMID: 2969233.

116.     Lanska DJ, Lavine L, Lanska MJ, Schoenberg BS. Huntington's disease mortality in the United States. Neurology. 1988;38:769-.

117.     van der Burg JMM, Björkqvist M, Brundin P. Beyond the brain: widespread pathology in Huntington's disease. Lancet neurology. 2009;8:765-74. PubMed PMID: 19608102.

118.     Aziz NA, Anguelova GV, Marinus J, van Dijk JG, Roos RAC. Autonomic symptoms in patients and pre-manifest mutation carriers of Huntington's disease. European journal of neurology : the official journal of the European Federation of Neurological Societies. 2010;17:1068-74. PubMed PMID: 20192977.

119.     Sawiak SJ, Wood NI, Williams GB, Morton AJ, Carpenter TA. Voxel-based morphometry in the R6/2 transgenic mouse reveals differences between genotypes not seen with manual 2D morphometry. Neurobiology of disease. 2009;33:20-7. PubMed PMID: 18930824.

120.     Sawiak SJ, Wood NI, Williams GB, Morton AJ, Carpenter TA. Use of magnetic resonance imaging for anatomical phenotyping of the R6/2 mouse model of Huntington's disease. Neurobiology of disease. 2009;33:12-9. PubMed PMID: 18930823.

121.     Wood NI, Glynn D, Morton AJ. "Brain training" improves cognitive performance and survival in a transgenic mouse model of Huntington's disease. Neurobiology of disease. 2011;42:427-37. PubMed PMID: 21324361.

122.     Carter RJ, Lione LA, Humby T, Mangiarini L, Mahal A, Bates GP, et al. Characterization of Progressive Motor Deficits in Mice Transgenic for the Human Huntington's Disease Mutation. J Neurosci. 1999;19:3248-57.

123.     Ciamei A, Morton AJ. Rigidity in social and emotional memory in the R6/2 mouse model of Huntington's disease. Neurobiology of learning and memory. 2008;89:533-44. PubMed PMID: 18069020.

124.     Ciamei A, Morton AJ. Progressive imbalance in the interaction between spatial and procedural memory systems in the R6/2 mouse model of Huntington's disease. Neurobiology of learning and memory. 2009;92:417-28. PubMed PMID: 19524696.

125.     Lione LA, Carter RJ, Hunt MJ, Bates GP, Morton AJ, Dunnett SB. Selective discrimination learning impairments in mice expressing the human Huntington's disease mutation. The Journal of neuroscience : the official journal of the Society for Neuroscience. 1999;19:10428-37. PubMed PMID: 10575040.

126.     Wood NI, Carta V, Milde S, Skillings EA, McAllister CJ, Ang YLM, et al. Responses to environmental enrichment differ with sex and genotype in a transgenic mouse model of Huntington's disease. PloS one. 2010;5:e9077. PubMed PMID: 20174443.

127.     Morton AJ, Wood NI, Hastings MH, Hurelbrink C, Barker RA, Maywood ES. Disintegration of the sleep-wake cycle and circadian timing in Huntington's disease. The Journal of neuroscience : the official journal of the Society for Neuroscience. 2005;25:157-63. PubMed PMID: 15634777.

128.     Maywood ES, Fraenkel E, McAllister CJ, Wood N, Reddy AB, Hastings MH, et al. Disruption of peripheral circadian timekeeping in a mouse model of Huntington's disease and its restoration by temporally scheduled feeding. The Journal of neuroscience : the official journal of the Society for Neuroscience. 2010;30:10199-204. PubMed PMID: 20668203.

129.     Kudo T, Schroeder A, Loh DH, Kuljis D, Jordan MC, Roos KP, et al. Dysfunctions in circadian behavior and physiology in mouse models of Huntington's disease. Experimental neurology. 2011;228:80-90. PubMed PMID: 21184755.

130.     Mangiarini L, Sathasivam K, Seller M, Cozens B, Harper A, Hetherington C, et al. Exon 1 of the HD gene with an expanded CAG repeat is sufficient to cause a progressive neurological phenotype in transgenic mice. Cell. 1996;87:493-506. PubMed PMID: 8898202.

131.     Duzdevich D, Li J, Whang J, Takahashi H, Takeyasu K, Dryden DTF, et al. Unusual structures are present in DNA fragments containing super-long Huntingtin CAG repeats. PloS one. 2011;6:e17119. PubMed PMID: 21347256.

132.     van Rugge FP, van der Wall EE, de Roos A, Bruschke AV. Dobutamine stress magnetic resonance imaging for detection of coronary artery disease. Journal of the American College of Cardiology. 1993;22:431-9. PubMed PMID: 8335812.

133.     Berr SS, Roy RJ, French BA, Yang Z, Gilson W, Kramer CM, et al. Black blood gradient echo cine magnetic resonance imaging of the mouse heart. Magnetic resonance in medicine : official journal of the Society of Magnetic Resonance in Medicine / Society of Magnetic Resonance in Medicine. 2005 May;53(5):1074-9. PubMed PMID: 15844138.

134.     Krahwinkel W, Ketteler T, Godke J, Wolfertz J, Ulbricht LJ, Krakau I, et al. Dobutamine stress echocardiography. European heart journal. 1997 Jun;18 Suppl D:D9-15. PubMed PMID: 9183605.

135.     Motaal AG, Coolen BF, Abdurrachim D, Castro RM, Prompers JJ, Florack LM, et al. Accelerated high-frame-rate mouse heart cine-MRI using compressed sensing reconstruction. NMR Biomed. 2012 Oct 29. PubMed PMID: 23109290.

136.     Montesinos P, Abascal JF, Cusso L, Vaquero JJ, Desco M. Application of the compressed sensing technique to self-gated cardiac cine sequences in small animals. Magnetic resonance in medicine : official journal of the Society of Magnetic Resonance in Medicine / Society of Magnetic Resonance in Medicine. 2013 Sep 16. PubMed PMID: 24105815.

137.     Bauschke HH, Borwein JM. Projection algorithms for solving convex feasibility problems. Siam Rev. 1996 Sep;38(3):367-426. PubMed PMID: ISI:A1996VH11600001. English.

138.     Nam S, Akcakaya M, Basha T, Stehning C, Manning WJ, Tarokh V, et al. Compressed sensing reconstruction for whole-heart imaging with 3D radial trajectories: a graphics processing unit implementation. Magnetic resonance in medicine : official journal of the Society of Magnetic Resonance in Medicine / Society of Magnetic Resonance in Medicine. 2013 Jan;69(1):91-102. PubMed PMID: 22392604. Pubmed Central PMCID: 3371294. Epub 2012/03/07. eng.

139.     Constantine G, Shan K, Flamm SD, Sivananthan MU. Role of MRI in clinical cardiology. Lancet. 2004 Jun 26;363(9427):2162-71. PubMed PMID: 15220041. Epub 2004/06/29. eng.

140.     Bellenger NG, Davies LC, Francis JM, Coats AJ, Pennell DJ. Reduction in sample size for studies of remodeling in heart failure by the use of cardiovascular magnetic resonance. Journal of cardiovascular magnetic resonance : official journal of the Society for Cardiovascular Magnetic Resonance. 2000;2(4):271-8. PubMed PMID: 11545126. Epub 2001/09/08. eng.

141.    Zhang S, Uecker M, Voit D, Merboldt KD, Frahm J. Real-time cardiovascular magnetic resonance at high temporal resolution: radial FLASH with nonlinear inverse reconstruction. Journal of cardiovascular magnetic resonance : official journal of the Society for Cardiovascular Magnetic Resonance. 2010;12:39. PubMed PMID: 20615228. Pubmed Central PMCID: 2911425.

142.    Vanvaals JJ, Bergman AH. Optimization of Eddy-Current Compensation. J Magn Reson. 1990 Oct 15;90(1):52-70. PubMed PMID: ISI:A1990EF60300005. English.

143.    Jehenson P, Syrota A. Correction of Distortions Due to the Pulsed Magnetic-Field Gradient-Induced Shift in B0 Field by Postprocessing. Magnet Reson Med. 1989 Nov;12(2):253-6. PubMed PMID: ISI:A1989AX06500011. English.

144.    Block KT, Uecker M, editors. Simple Method for Adaptive Gradient-Delay Compensation in Radial MRI. proceedings of the 19th ISMRM 2011; 2011; Montreal.

145.    Winkelmann S, Schaeffter T, Koehler T, Eggers H, Doessel O. An optimal radial profile order based on the golden ratio for time-resolved MRI. Ieee T Med Imaging. 2007 Jan;26(1):68-76. PubMed PMID: ISI:000243286800006. English.

146.    Borst O, Ochmann C, Schönberger T, Jacoby C, Stellos K, Seizer P, et al. Methods employed for induction and analysis of experimental myocardial infarction in mice. Cellular physiology and biochemistry : international journal of experimental cellular physiology, biochemistry, and pharmacology. 2011;28:1-12. PubMed PMID: 21865843.

147.    Geelen T, Paulis LE, Coolen BF, Nicolay K, Strijkers GJ. Contrast-enhanced MRI of murine myocardial infarction - part I. NMR Biomed. 2012 Aug;25(8):953-68. PubMed PMID: 22308108.

148.    Coolen BF, Paulis LE, Geelen T, Nicolay K, Strijkers GJ. Contrast-enhanced MRI of murine myocardial infarction - part II. NMR Biomed. 2012 Aug;25(8):969-84. PubMed PMID: 22311260.

149.    Kim RJ, Albert TS, Wible JH, Elliott MD, Allen JC, Lee JC, et al. Performance of delayed-enhancement magnetic resonance imaging with gadoversetamide contrast for the detection and assessment of myocardial infarction: an international, multicenter, double-blinded, randomized trial. Circulation. 2008 Feb 5;117(5):629-37. PubMed PMID: 18212288. Epub 2008/01/24. eng.

150.    Simonetti OP, Kim RJ, Fieno DS, Hillenbrand HB, Wu E, Bundy JM, et al. An improved MR imaging technique for the visualization of myocardial infarction. Radiology. 2001;218:215-23. PubMed PMID: 11152805. en.

151.    Kim RJ, Shah DJ, Judd RM. How we perform delayed enhancement imaging. Journal of cardiovascular magnetic resonance : official journal of the Society for Cardiovascular Magnetic Resonance. 2003;5:505-14. PubMed PMID: 12882082.

152.    Deichmann R, Haase A. Quantification of T1 values by SNAPSHOT-FLASH NMR imaging. Journal of Magnetic Resonance (1969). 1992;96:608-12.

153.    Salto-Tellez M, Yung Lim S, El-Oakley RM, Tang TP, ZA AL, Lim SK. Myocardial infarction in the C57BL/6J mouse: a quantifiable and highly reproducible experimental model. Cardiovascular pathology : the official journal of the Society for Cardiovascular Pathology. 2004 Mar-Apr;13(2):91-7. PubMed PMID: 15033158.

154.    Methner C, Schmidt K, Cohen MV, Downey JM, Krieg T. Both A2a and A2b adenosine receptors at reperfusion are necessary to reduce infarct size in mouse hearts. American journal of physiology Heart and circulatory physiology. 2010;299:H1262-4. PubMed PMID: 20709859.

155.    Weinsaft JW, Cham MD, Janik M, Min JK, Henschke CI, Yankelevitz DF, et al. Left ventricular papillary muscles and trabeculae are significant determinants of cardiac MRI volumetric measurements: effects on clinical standards in patients with advanced systolic dysfunction. International journal of cardiology. 2008;126:359-65. PubMed PMID: 17698216.

156.    Schneider JE, Cassidy PJ, Lygate C, Tyler DJ, Wiesmann F, Grieve SM, et al. Fast, high-resolution in vivo cine magnetic resonance imaging in normal and failing mouse hearts on a vertical 11.7 T system. Journal of magnetic resonance imaging : JMRI. 2003;18:691-701. PubMed PMID: 14635154.

157.    Methner C, Lukowski R, Grube K, Loga F, Smith RA, Murphy MP, et al. Protection through postconditioning or a mitochondria-targeted S-nitrosothiol is unaffected by cardiomyocyte-selective ablation of protein kinase G. Basic research in cardiology. 2013 Mar;108(2):337. PubMed PMID: 23423145. Epub 2013/02/21. eng.

158.    Epstein FH, Gilson WD. Displacement-encoded cardiac MRI using cosine and sine modulation to eliminate (CANSEL) artifact-generating echoes. Magnetic resonance in medicine : official journal of the Society of Magnetic Resonance in Medicine / Society of Magnetic Resonance in Medicine. 2004 Oct;52(4):774-81. PubMed PMID: 15389939. Epub 2004/09/25. eng.

159.    Flynn TJ. Two-dimensional phase unwrapping with minimum weighted discontinuity. J Opt Soc Am A. 1997 Oct;14(10):2692-701. PubMed PMID: WOS:A1997XY74200013. English.

160.     Lucas AJ, Hawkes RC, Guerra P, Ansorge RE, Nutt RE, Clark JC, et al. Development of a combined microPET (R)-MR system. Ieee Nucl Sci Conf R. 2006:2345-8. PubMed PMID: ISI:000288875602088. English.

161.     Hawkes RC, Fryer TD, Siegel S, Ansorge RE, Carpenter TA. Preliminary evaluation of a combined microPET-MR system. Technology in cancer research & treatment. 2010 Feb;9(1):53-60. PubMed PMID: 20082530. Epub 2010/01/20. eng.

162.     Sawiak SJ, Hawkes RC, Ansorge RE, Carpenter TA. Reliability of using a fixed matrix in coregistration of combined PET-MRI in a split magnet design. Nucl Instrum Meth A. 2013 Feb 21;702:54-5. PubMed PMID: ISI:000314682300016. English.

163.     Buonincontri G, Sawiak SJ, Methner C, Krieg T, Hawkes RC, Carpenter TA. PET/MRI in the infarcted mouse heart with the Cambridge split magnet. Nucl Instrum Meth A. 2013 Feb 21;702:47-9. PubMed PMID: WOS:000314682300014. English.

164.     Mertes H, Sawada SG, Ryan T, Segar DS, Kovacs R, Foltz J, et al. Symptoms, adverse effects, and complications associated with dobutamine stress echocardiography. Experience in 1118 patients. Circulation. 1993 Jul;88(1):15-9. PubMed PMID: 8319327.

165.     Coolen BF, Abdurrachim D, Motaal AG, Nicolay K, Prompers JJ, Strijkers GJ. High frame rate retrospectively triggered Cine MRI for assessment of murine diastolic function. Magnetic resonance in medicine : official journal of the Society of Magnetic Resonance in Medicine / Society of Magnetic Resonance in Medicine. 2013 Mar 1;69(3):648-56. PubMed PMID: 22517471.